\documentclass[iop,numberedappendix]{emulateapj}

\usepackage[utf8]{inputenc}
\usepackage{amsmath}
\usepackage{amssymb}
\usepackage{graphicx}
\usepackage{hyperref}
\usepackage{times}

\newcommand{\dd}{\,\mathrm{d}\,}
\newcommand{\pp}{\partial}
\newcommand{\D}{\displaystyle}
\newcommand{\bhat}[1]{\hat{\boldsymbol{#1}}}

\begin{document}
\title{Stability of Hall equilibria in neutron star crusts}
\author{Pablo Marchant}
\affil{Instituto de Astrofísica, Facultad de Física, Pontificia Universidad Católica de Chile\\
	Av. Vicuña Mackenna 4860, Macul 7820436, Santiago, Chile}
\affil{Argelander-Institut für Astronomie, Universität Bonn\\
	Auf dem Hügel 71, 53121 Bonn, Germany}
\email{pablo@astro.uni-bonn.de}
\author{Andreas Reisenegger}
\affil{Instituto de Astrofísica, Facultad de Física, Pontificia Universidad Católica de Chile\\
	Av. Vicuña Mackenna 4860, Macul 7820436, Santiago, Chile}
\author{Juan Alejandro Valdivia}
\affil{Departamento de Física, Facultad de Ciencias, Universidad de Chile\\
	Casilla 653, Santiago, Chile}
\author{Jaime H. Hoyos}
\affil{Departamento de Ciencias Básicas, Universidad de Medellín\\
	Cra. 87 \# 30-65, Medellín, Colombia}

\begin{abstract}
In the solid crusts of neutron stars, the advection of the magnetic field by the current-carrying electrons, an effect known as Hall drift, should play a very important role as the ions remain essentially fixed (as long as the solid does not break). Although Hall drift preserves the magnetic field energy, it has been argued that it m{}ay drive a turbulent cascade to scales at which Ohmic dissipation becomes effective, allowing a much faster decay in objects with very strong fields. On the other hand, it has been found that there are ``Hall equilibria'', i.e., field configurations that are unaffected by Hall drift. Here, we address the crucial question of the stability of these equilibria through axially symmetric (2D) numerical simulations of Hall drift and Ohmic diffusion, with the simplifying assumption of uniform electron density and conductivity. We demonstrate the 2D-stability of a purely poloidal equilibrium, for which Ohmic dissipation makes the field evolve towards an attractor state through adjacent stable configurations, around which damped oscillations occur. For this field, the decay scales with the Ohmic timescale. We also study the case of an unstable equilibrium consisting of both poloidal and toroidal field components that are confined within the crust. This field evolves into a stable configuration, which undergoes damped oscillations superimposed on a slow evolution towards an attractor, just as the purely poloidal one.
\end{abstract}
\section{Introduction}\label{ch:intro}
Hall drift, the effect of advection of magnetic flux by the current-carrying electrons, has long been thought to be important for the evolution of the crustal fields of neutron stars. \citet{jon+88} initially hypothesized that this effect could transport flux towards the outer regions of the crust, where Ohmic dissipation is much more effective than in the inner regions. On the other hand, \citet{golrei+92} provided an order-of-magnitude estimate that implies that magnetar-like fields could evolve in less than a million years. Since Hall drift is a conservative effect, their proposed mechanism for the field decay was based on the production of a turbulent cascade. This would drive the field to small-scale structures upon which Ohmic dissipation could act effectively.

In order to study the possible presence of instabilities, \citet{rhegep+02} studied a homogeneous plane-parallel model and performed a perturbation analysis against a background field that lies on the plane, discovering unstable modes with exponential growth rates. Using a similar model, \citet{vai+00} showed that the evolution through Hall drift is governed by Burgers' equation, which develops strong discontinuities corresponding to current sheets. The inclusion of Ohmic dissipation results in rapid decay of magnetic energy in these regions. \citet{rei+07} extended this result to the case of a purely toroidal (azimuthal) and axially symmetric magnetic field, showing in this case that the evolution through Hall drift can again be described with Burgers' equation, with analogous implications. They also argued that any small poloidal perturbation to the toroidal field would end up being amplified.

Although these analytical developments are a big step forward in the understanding of the evolution of the magnetic field in a neutron star, a full comprehension of this process seems to require the use of numerical simulations due to its complex non-linear character. Early simulations performed by \citet{shaurp+91} for the case of a spatially homogenous resistivity and electron density, show that in fact purely toroidal fields can develop strong discontinuities and experience fast Ohmic dissipation. Also with homogenous models, but dealing mostly with predominantly poloidal (meridional) fields, \citet{shaurp+97} and \citet{holrud+02} observed oscillatory phenomena, without a strong enhancement of the decay rates through Hall drift, compared to the case of pure Ohmic dissipation. Taking into account a range of models covering predominantly toroidal and predominantly poloidal fields, \citet{kojkis+12} showed that the decay rates were consistently higher as the toroidal field increased in importance. In models that included a simple stratified crust \citep{holrud+04} or with realistic values for electron density and resistivity \citep{geppon+07}, efficient decay is observed associated with the formation of strong current sheets, and not due to displacement of magnetic field to areas of lower conductivity as proposed by \citet{jon+88}. Also, as an ordered component remains for long times, the magnetic field does not appear to undergo the Hall cascade of \citet{golrei+92} either.

Recently, elaborate models have been produced that take into account the full magneto-thermal evolution of the neutron star (though allowing for field evolution only in the crust). \citet{pon+09} studied the coupling of the magnetic field to the thermal evolution using only Ohmic decay, while \citet{vig+13} included Hall drift into these models. Their results show the critical importance of using a fully consistent model in order to produce a realistic evolution, as feedback between the magnetic field and the thermal structure is very strong in both ways.

We intend to study the magnetic field stability in axial symmetry for a couple of Hall equilibrium configurations that can be expressed in closed analytical forms. \citet{gou+13b} already studied the evolution due to Hall drift using different initial conditions which included a poloidal equilibrium field. Although the early evolution differed from case to case, they noted that on long times the field evolved towards similar configurations that consist of a poloidal field with a dipole and a counter-aligned octupole (meaning that the dipolar and octupolar components have opposite signs at the poles), coupled through a weak quadrupolar toroidal field. \citet{gou+13c} further developed this concept referring to this final state as an ``attractor''.

The structure of this paper is as follows: \S \ref{ch::methods} briefly describes the methods used in our work. \S \ref{ch::poltor} deals with the effects of having initial conditions that are initially dominated by either a poloidal or a toroidal field, essentially reproducing the results of \citet{kojkis+12} but with a consistent normalization of the field that allows proper comparisons between simulations. \S \ref{ch::equil} provides a study of the stability of two different equilibrium fields, one purely poloidal extending to the vacuum outside the star, and another one that is a mix of a poloidal and a toroidal field, artificially constrained to remain confined to the crust. Finally, \S \ref{ch:conclusions} provides our conclusions and discussion of the results.

\section{Methods}\label{ch::methods}
\subsection{Hall drift and Ohmic decay}\label{intro::h}
We will restrict ourselves to the magnetic field evolution in the crusts of neutron stars, ignoring possible effects involving the fluid core. In the crust, ions are locked into a crystal lattice, and the only freely moving charged species are the electrons. This electron fluid should have a negligible acceleration, which in turn implies that the Lorentz force should be equal to the time-averaged momentum loss through collisions. Under these approximations, the evolution of the magnetic field $\boldsymbol{B}$ can be described by \citep[e.g.,][]{golrei+92}
\begin{eqnarray}
\frac{\pp \boldsymbol{B}}{\pp t}=-\nabla\times\left(\frac{c}{4\pi ne}[\nabla\times\boldsymbol{B}]\times\boldsymbol{B}+\eta\nabla\times\boldsymbol{B}\right),\label{intro::timeeq}
\end{eqnarray}
where $\eta$ is the magnetic diffusivity and $n$ the electron density. For simplicity, in this work we only consider models for which $n$ and $\eta$ are constant both in space and time. This equation contains two different effects that act on two distinct timescales, which can be estimated as
\begin{eqnarray}
t_{Hall}\equiv\frac{4\pi ne L^2}{cB_0},\quad t_{Ohm}\equiv\frac{L^2}{\eta},
\end{eqnarray}
where we introduced the characteristic value $B_0$ for the magnetic field, and $L$ is a characteristic length scale of variation, which for this work we take to be the thickness of the neutron star crust. The ratio of these two quantities defines the so-called magnetization parameter $R_B$ (a close analog to the Reynolds number of fluid mechanics),
\begin{eqnarray}
 R_B=\frac{t_{Ohm}}{t_{Hall}}=\frac{cB_0}{4\pi \eta ne},
\end{eqnarray}
which quantifies the relative importance of both effects. As the value of $R_B$ is fundamentally dependent on the meaning we give to the characteristic field $B_0$, it is desirable to choose this value in a physically unambiguous way. For the purpose of this work, we will define the characteristic field $B_0$ as
\begin{eqnarray}
B_0^2\equiv \frac{8\pi E}{V_{crust}} \label{anal::normB0}
\end{eqnarray}
where $V_{crust}$ is the volume of the crust and $E$ is the total magnetic energy (including the external vacuum field, if present).
\subsection{Axially symmetric fields}\label{intro::asf}
We restrict ourselves to axially symmetric fields, in which case $\boldsymbol{B}$ can be written in terms of two scalar functions as \citep[see, for instance,][]{chapre+56}
\begin{eqnarray}
\boldsymbol{B}=\nabla\alpha(r,\theta)\times\nabla\phi+\beta(r,\theta)\nabla\phi,
\end{eqnarray}
where $r$, $\theta$, and $\phi$ are the conventional spherical coordinates.

Using this, and defining $\chi\equiv c/(4\pi en\varpi^2)$ and $\varpi\equiv r\sin\theta$, Eq. (\ref{intro::timeeq}) can be decomposed into a purely poloidal and a purely toroidal part, from which two scalar equations for the time derivatives of $\alpha$ and $\beta$ are obtained \citep{rei+07}, namely

\begin{eqnarray}
\begin{aligned}
\frac{\pp\alpha}{\pp t}=&\varpi^2\chi[\nabla\alpha\times\nabla\beta]\cdot\nabla\phi+R_B^{-1}\eta\Delta^*\alpha\\
\frac{\pp \beta}{\pp t}=&\varpi^2\nabla\cdot\left(\chi\nabla\phi\times[\Delta^*\alpha\nabla\alpha+\beta\nabla\beta]+R_B^{-1}\frac{\eta\nabla\beta}{\varpi^2}\right),
\end{aligned}\label{intro::eqs}
\end{eqnarray}
where $t_{Hall}$ and $B_0$ are, respectively, the units of time and magnetic field. Also, $\Delta^*$ is the Grad-Shafranov operator, defined as
\begin{eqnarray}
\Delta^*\equiv\varpi^2\nabla\cdot(\varpi^{-2}\nabla)=\pp_r^2+\frac{\sin\theta}{r^2}\pp_\theta\left(\frac{\pp_\theta}{\sin\theta}\right).
\end{eqnarray}

\subsection{Boundary conditions}\label{intro::bc}
In order to avoid the complications of magnetic field evolution in the core \citep[e.g.,][]{golrei+92,hoy+08,hoy+10}, we assume that the core is a superconductor with a perfect Meissner effect, and the radius of the crust-core interface is defined as $r_{min}$, which we choose as $r_{min}=0.75 R$ where $R$ is the radius of the star. The conditions used at this boundary are then the continuity of the radial component of the magnetic field and the continuity of the tangential electric field, which we call ``Meissner boundary conditions''.

In order to compare our simulations with those of \citet{holrud+02} and \citet{kojkis+12}, we will also simulate the case with ``zero boundary conditions'' where both $\alpha=0$ and $\beta=0$ are forced at the inner boundary. This approximation is usually justified by saying that, since $R_B$ is a large number for high field neutron stars, resistive terms can be ignored, in which case the Meissner boundary conditions reduce to that. However, even in this limit, there are no physical reasons to impose this condition, since it implies that there is no surface current in the $\theta$-direction, whereas surface currents in the $\phi$-direction are allowed.

For the boundary condition at the surface, we take the exterior of the star to be a vacuum, with all 3 field components continuous across the interface, as surface currents should dissipate very efficiently.

For a comprehensive description of the boundary conditions used, refer to Appendix \ref{appendix::boundary}.

\subsection{Numerical methods}\label{ch::num}
For the purpose of studying the evolution of an axially symmetric field in a neutron star crust, we have developed a FTCS (forward time centered space) code to solve Eqs. (\ref{intro::eqs}), where the temporal discretization is done via the forward Euler method, which is first order in time, while spatial derivatives are solved with a central difference scheme that is second order in space. The precise details of the implementation are described in Appendix \ref{appendix::num}. This code is freely available for download at \url{https://github.com/orlox/hall_evolution}.

We also had access to the spectral code developed by \citet{hol+00}, against which we made comparisons. The result of these are described in Appendix \ref{appendix::holcom}.

\section{Fields with dominant poloidal or toroidal components}\label{ch::poltor}
As shown by the models of constant electron density and resistivity of \citet{kojkis+12}, evolution due to Hall drift is significantly different depending on whether the poloidal or the toroidal component is dominant. However, their definition of the characteristic field $B_0$ is simply the strongest value it takes, which means that their simulations of mixed fields at equal $R_B$ do not share the same total magnetic energy. Here we perform a similar analysis, but we compare the evolution of different field configurations that do share the same energy.

Defining $\boldsymbol{B}_{11p}$ and $\boldsymbol{B}_{11t}$, the fundamental poloidal and toroidal Ohmic modes for zero boundary conditions (see Appendix \ref{appendix::ohmmodes}), such that both are normalized to the same energy, we study combinations of the form
\begin{eqnarray}
\boldsymbol{B}=\sqrt{E_P/E}\boldsymbol{B}_{11p}+\sqrt{1-E_P/E}\boldsymbol{B}_{11t},\label{anal::bcomb}
\end{eqnarray}
all of which have the same total energy (and thus, the same characteristic magnetic field $B_0$ as given by Eq. (\ref{anal::normB0})), and for which the relevance of each component is given by the ratio of poloidal to total energy $E_P/E$.

For most of this section we use zero boundary conditions at the crust-core interface, and compare at the end for some cases how the results are modified by switching to Meissner boundary conditions. This is mainly because the Meissner boundary conditions as implemented here are computationally expensive to study, running into numerical problems for the case of large $R_B$. In the normalization used here, the maximum value of $\boldsymbol{B}_{11t}$ is approximately $1.75B_0$, which can be compared for instance with the normalization used by \citet{holrud+02}, that chooses $B_{max}=B_0$. The practical meaning of this is that our simulations with predominantly toroidal fields done with $R_B=100$ should be comparable to their simulations with $R_B=200$.

\begin{figure}
\begin{center}

\includegraphics[scale=0.22]{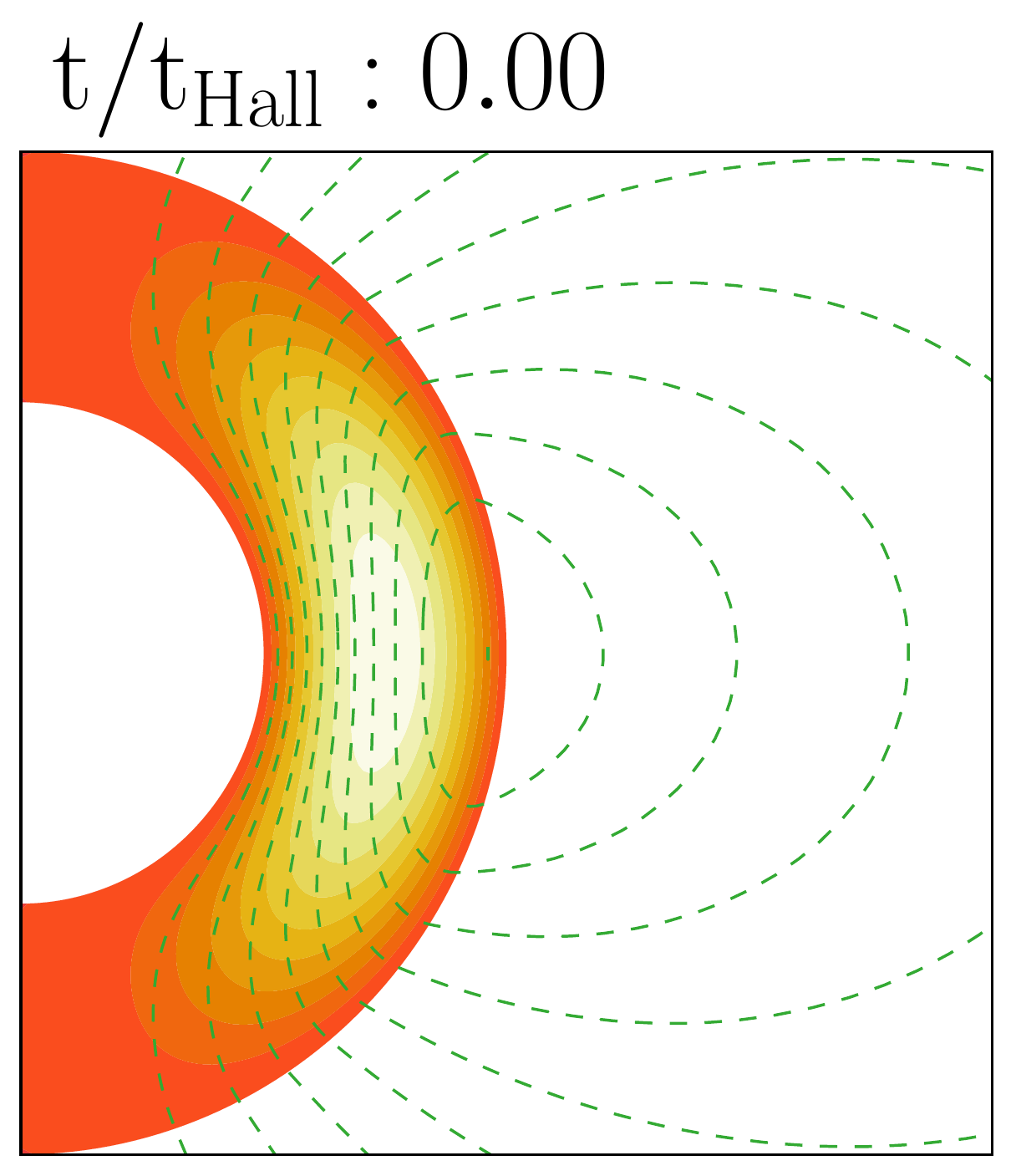}
\includegraphics[scale=0.22]{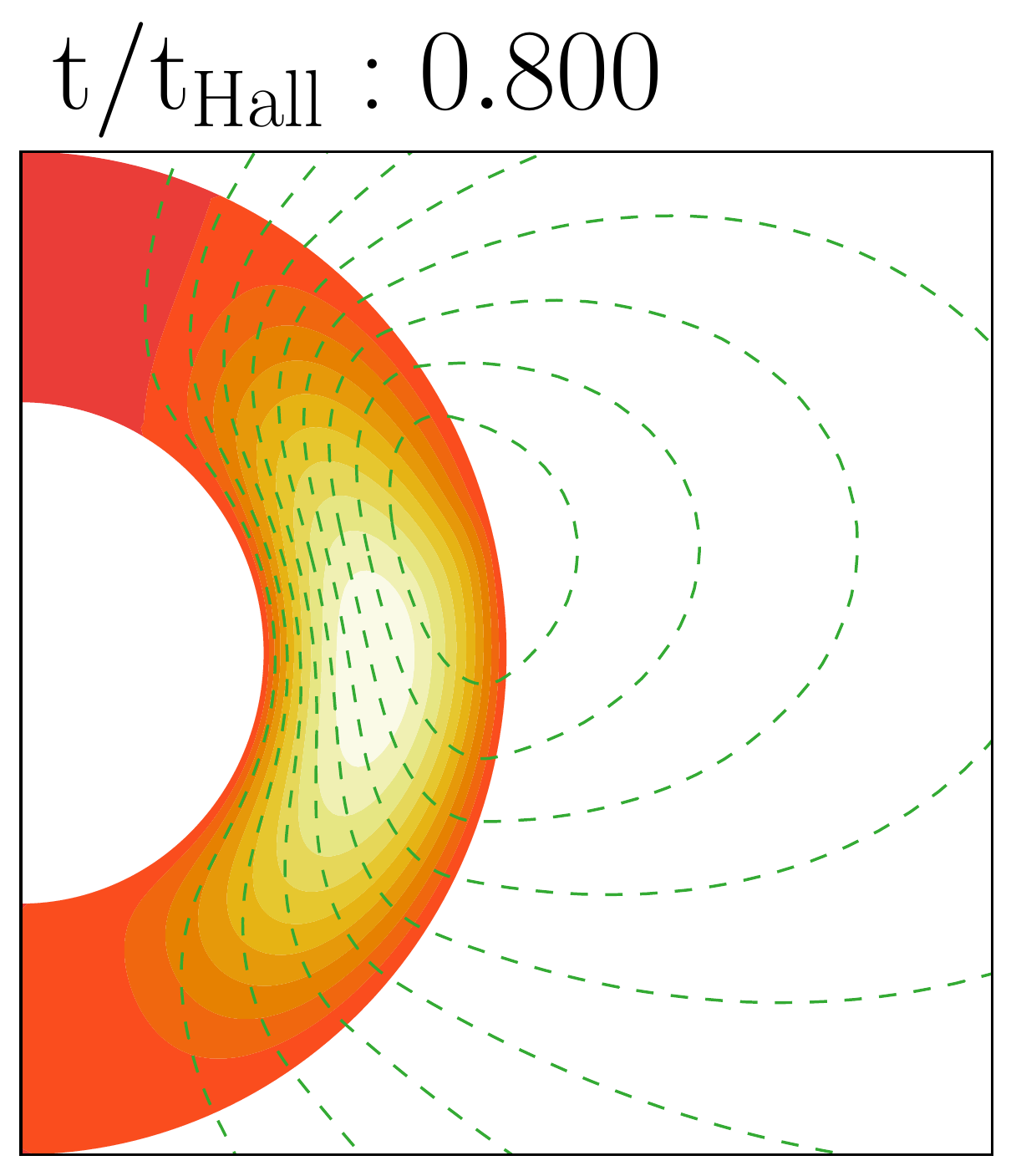}
\includegraphics[scale=0.22]{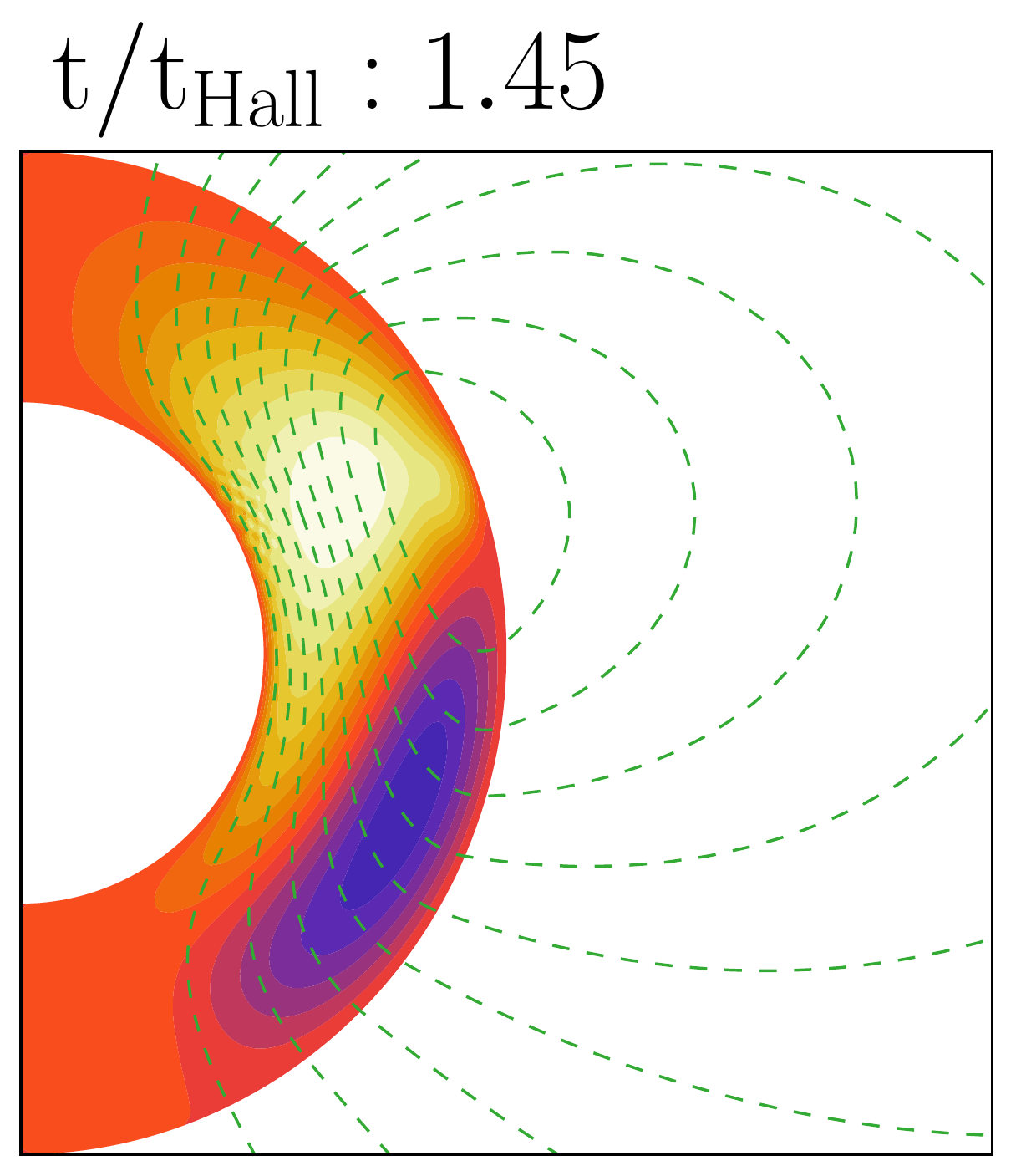}
\includegraphics[scale=0.22]{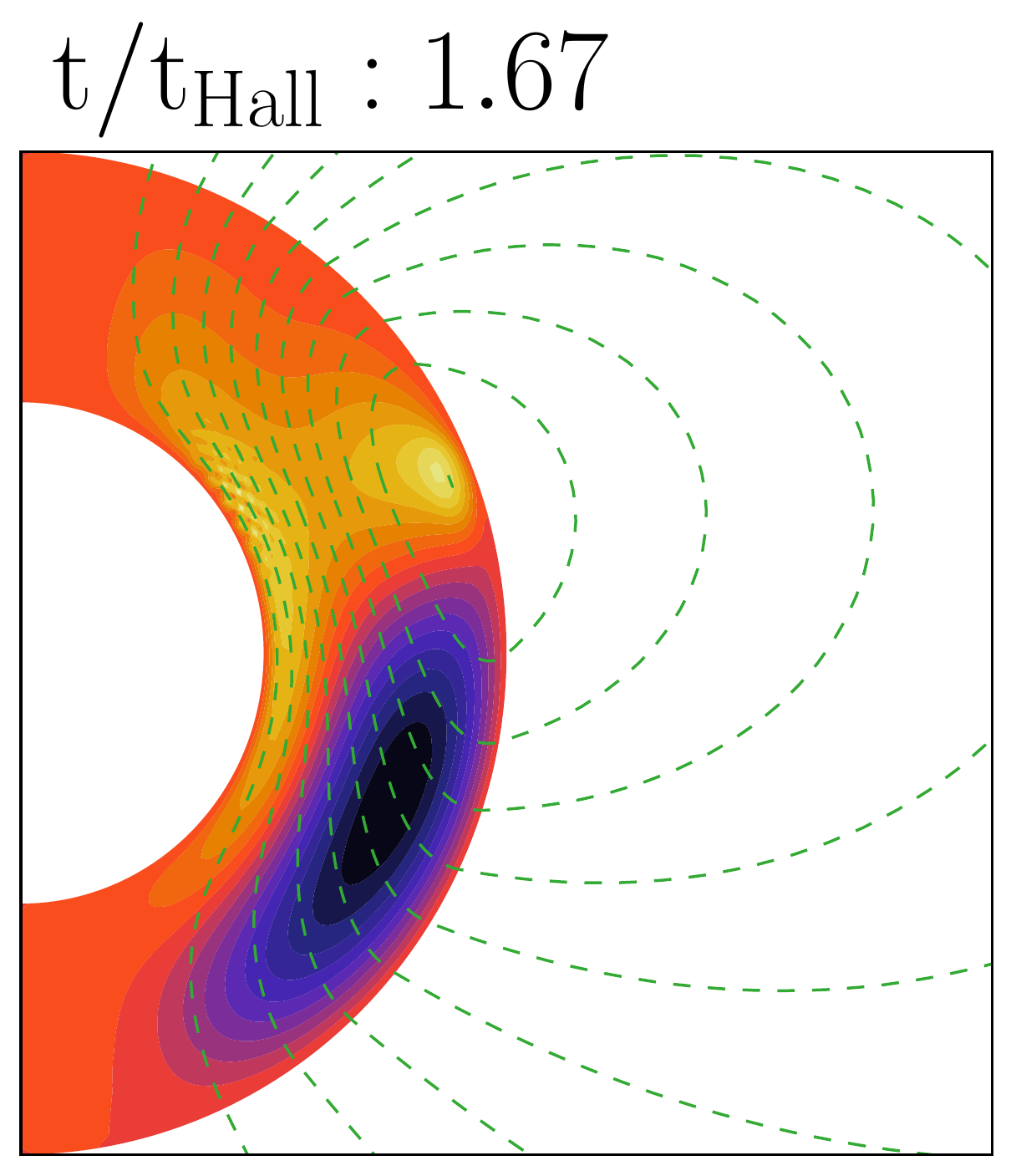}
\includegraphics[scale=0.22]{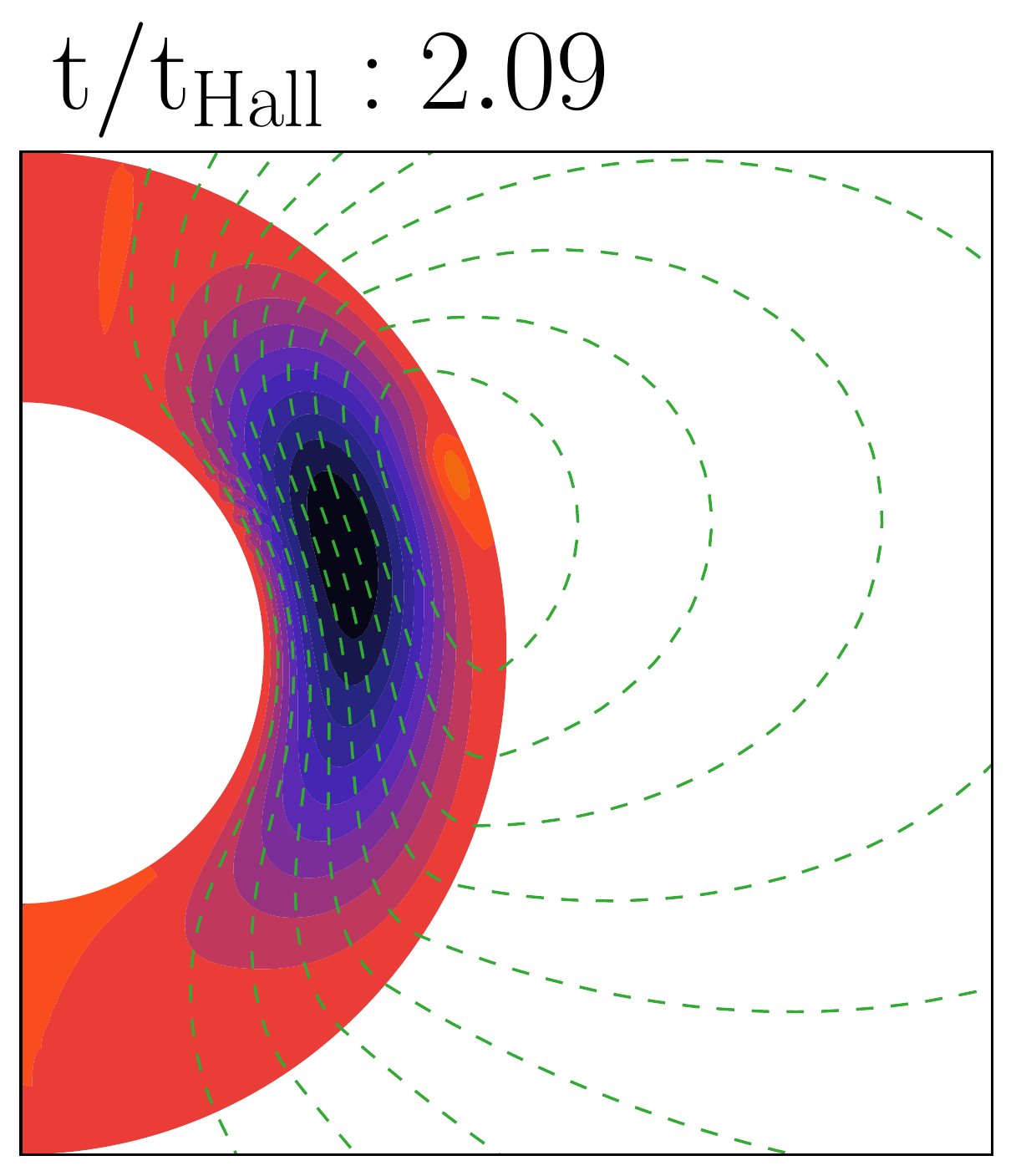}
\includegraphics[scale=0.22]{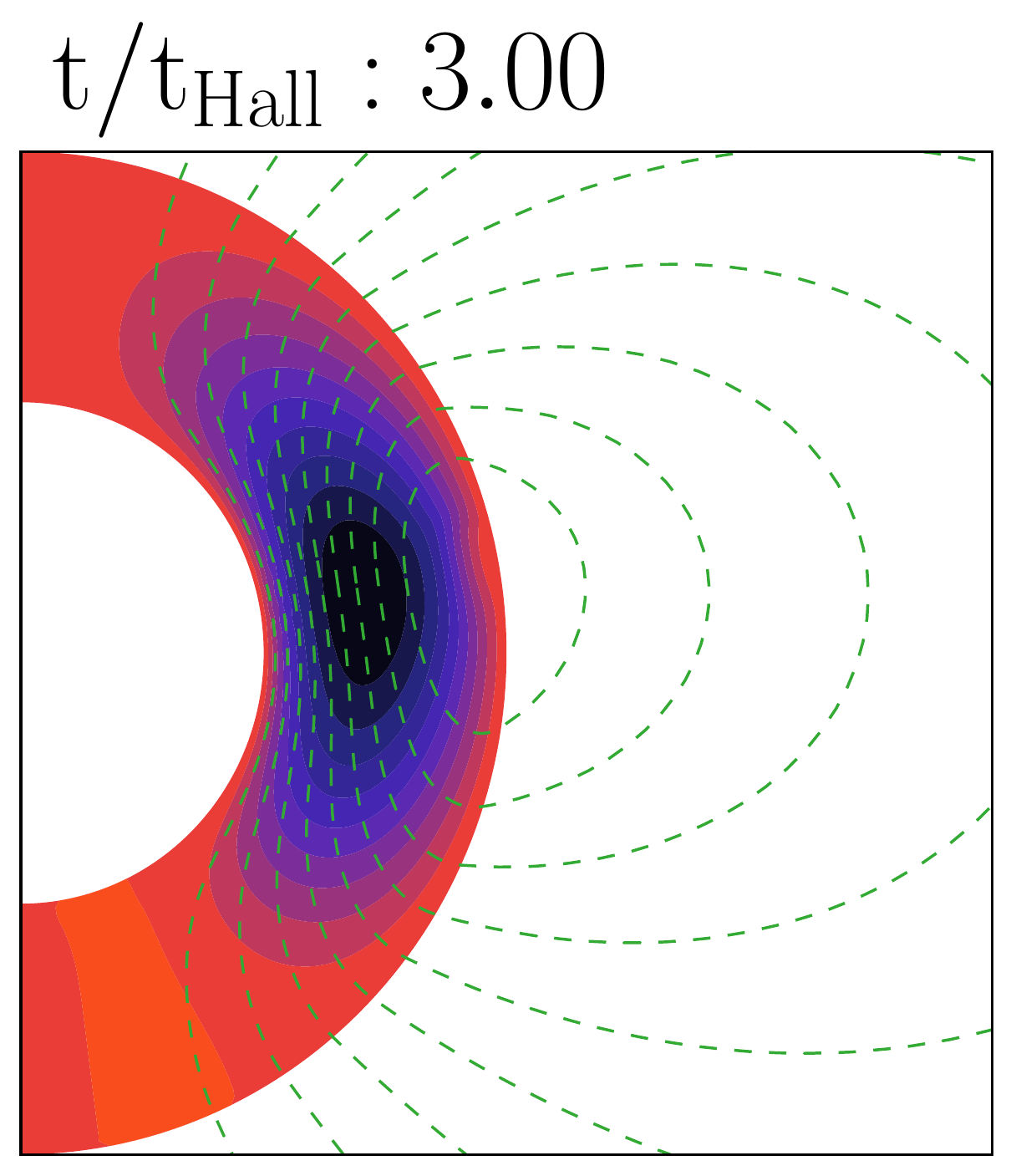}
\includegraphics[scale=0.22]{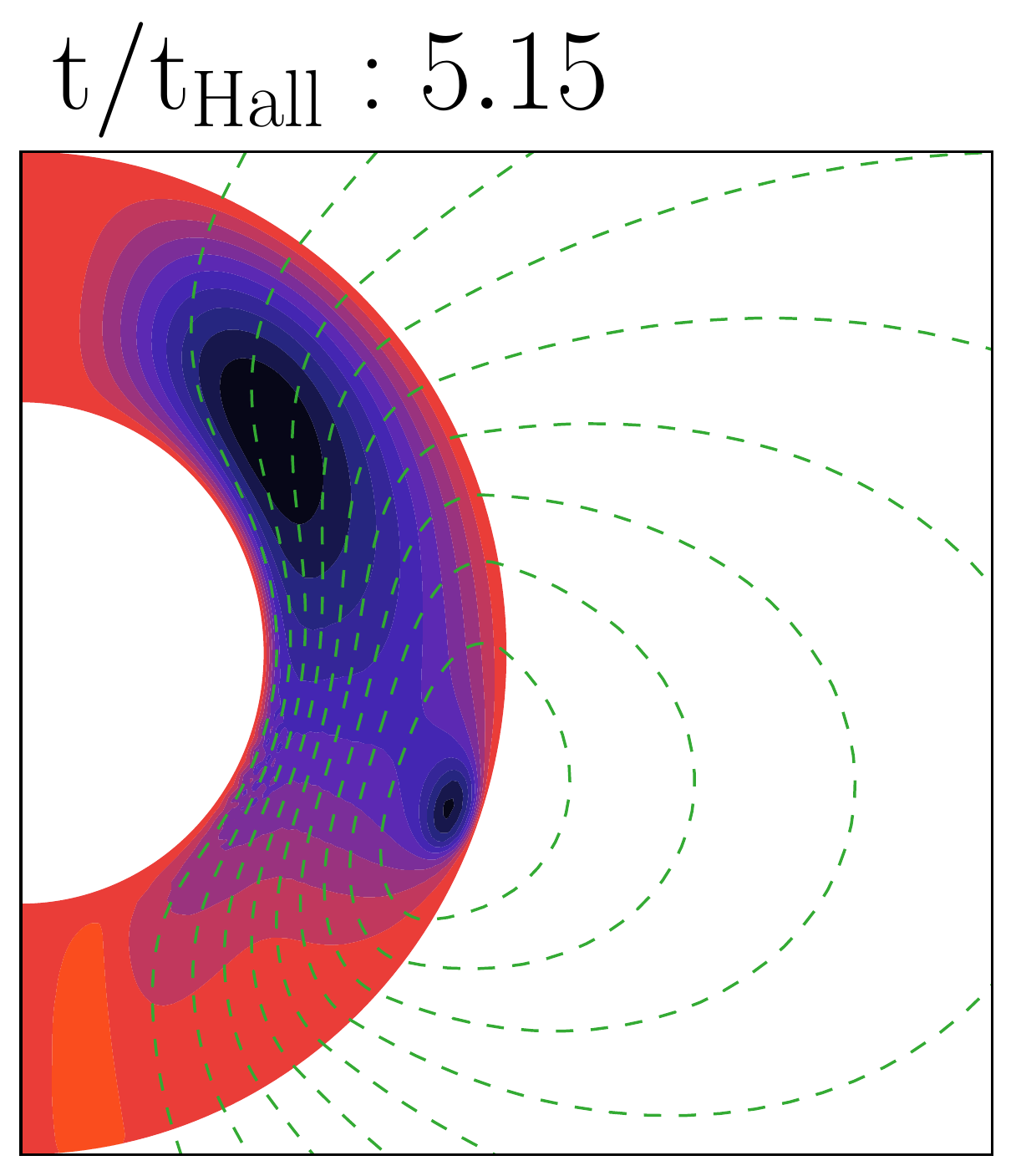}
\includegraphics[scale=0.22]{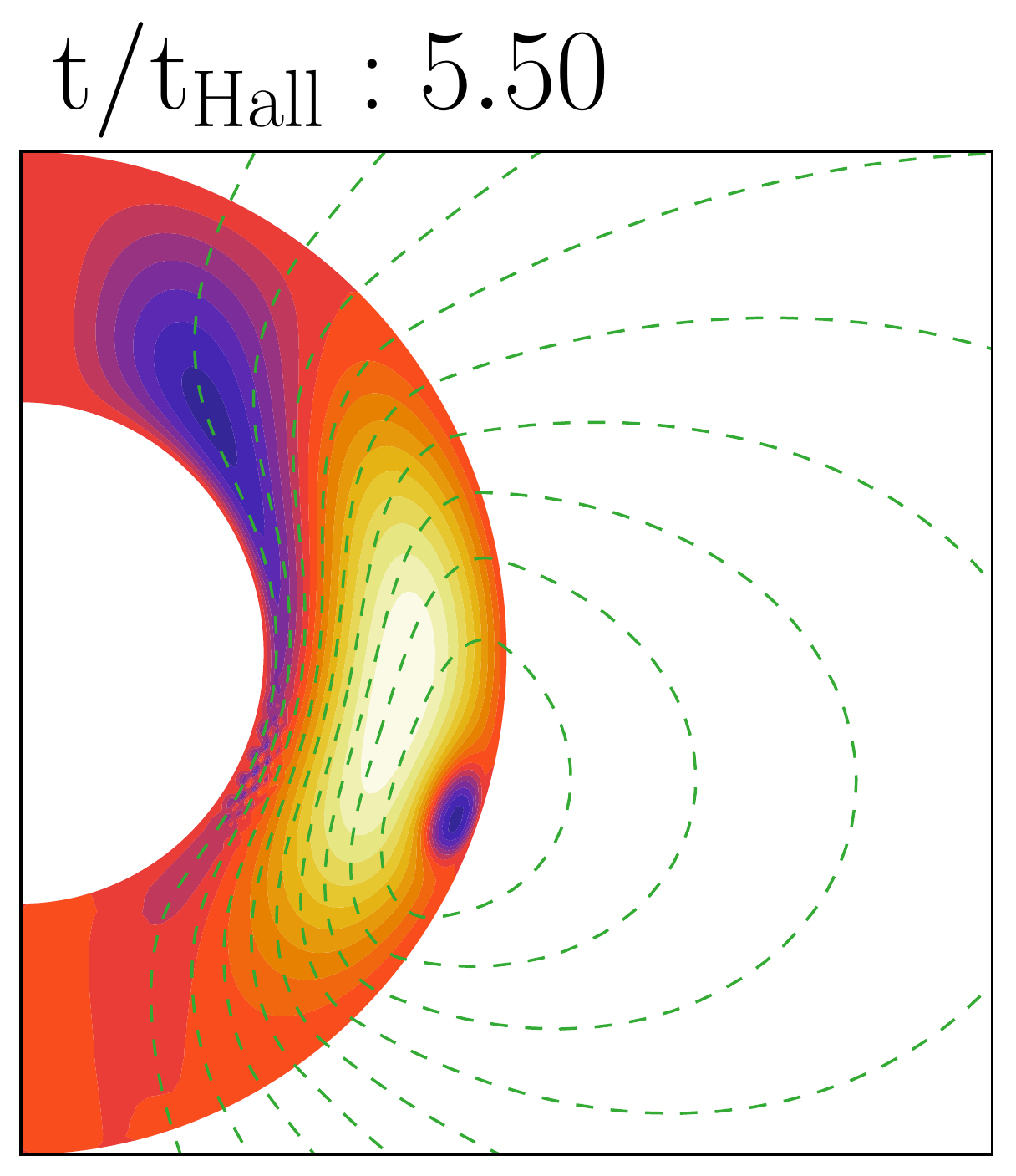}
\includegraphics[scale=0.22]{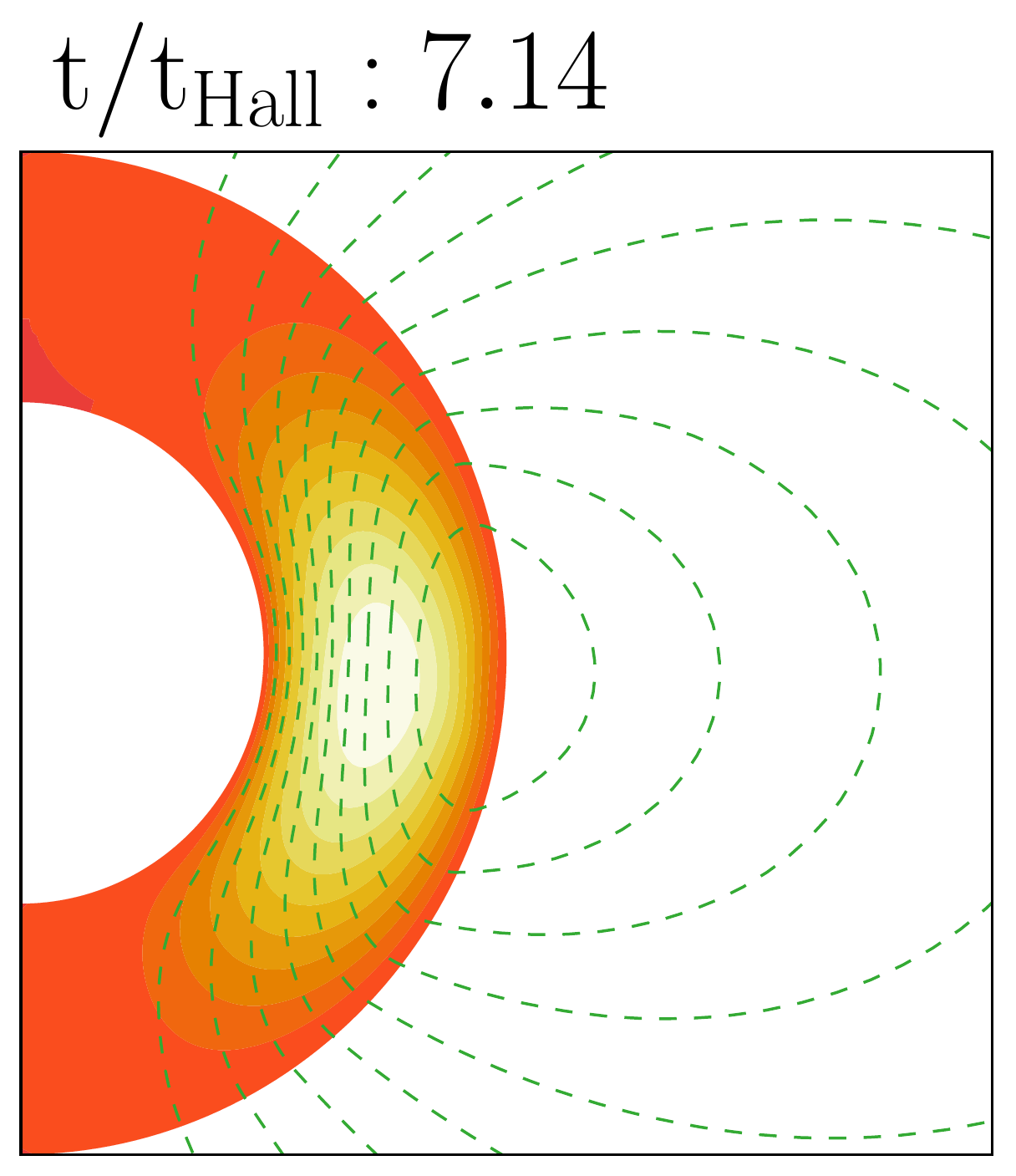}
\includegraphics[scale=0.22]{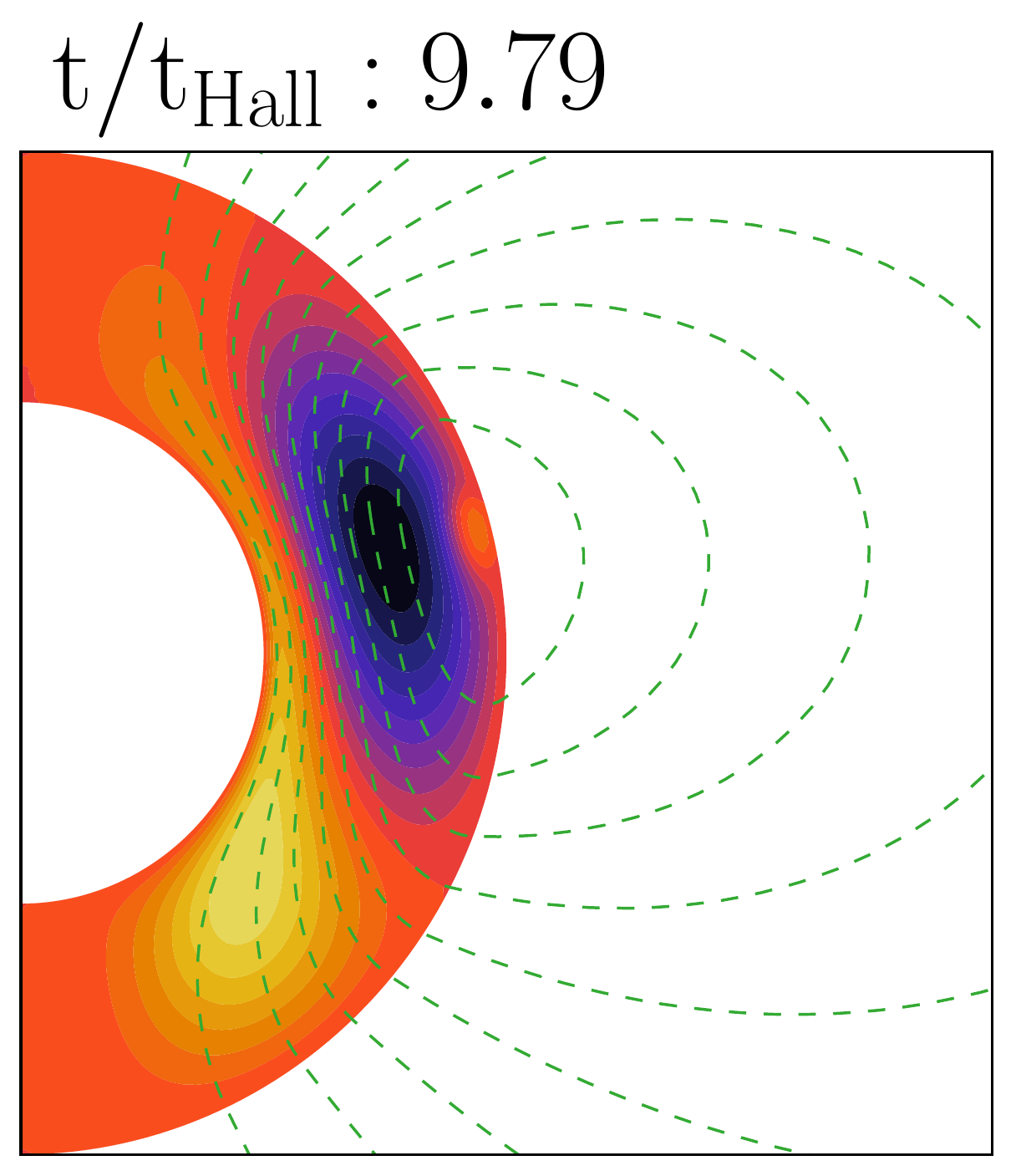}
\includegraphics[scale=0.22]{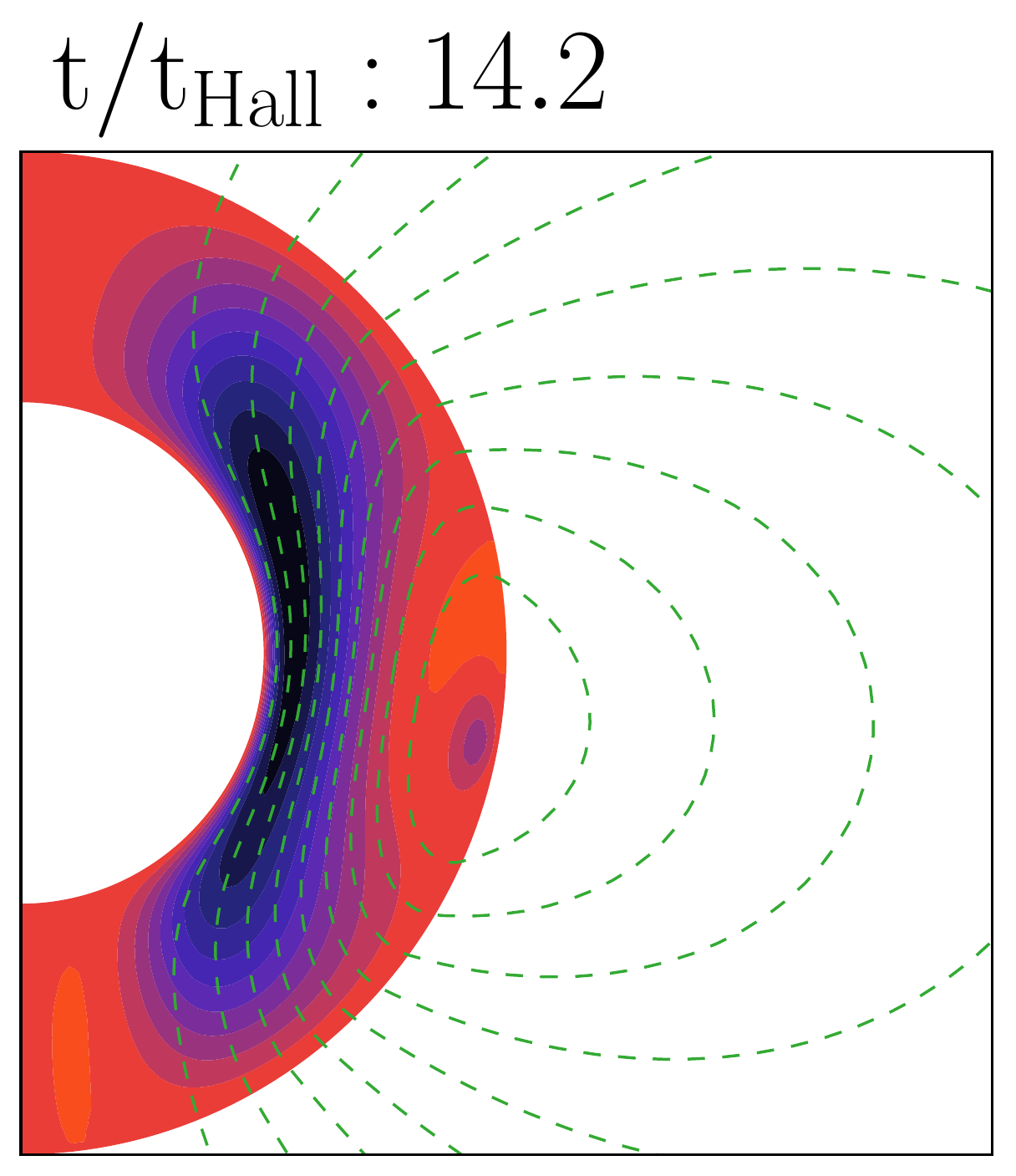}
\includegraphics[scale=0.22]{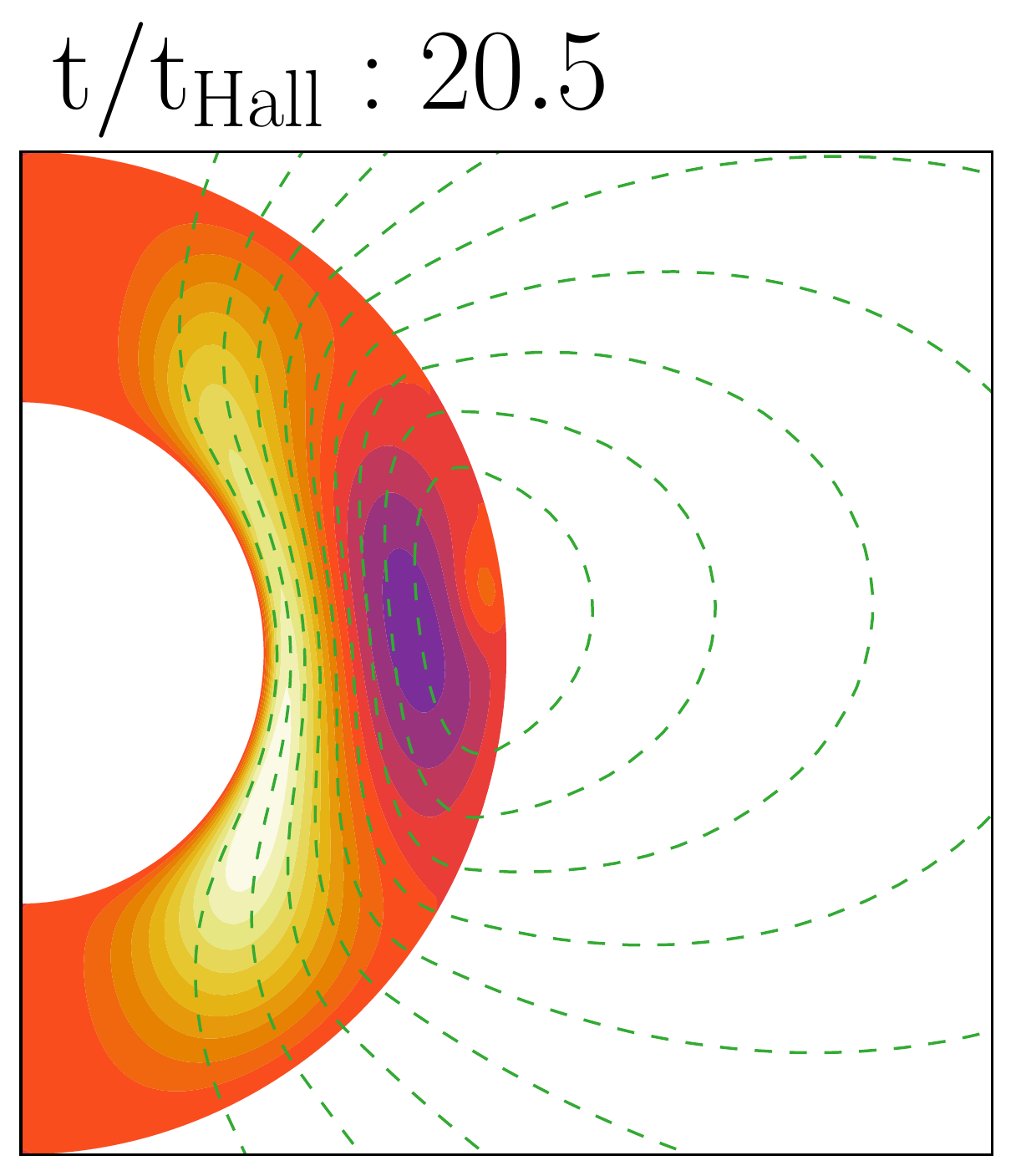}
\includegraphics[scale=0.22]{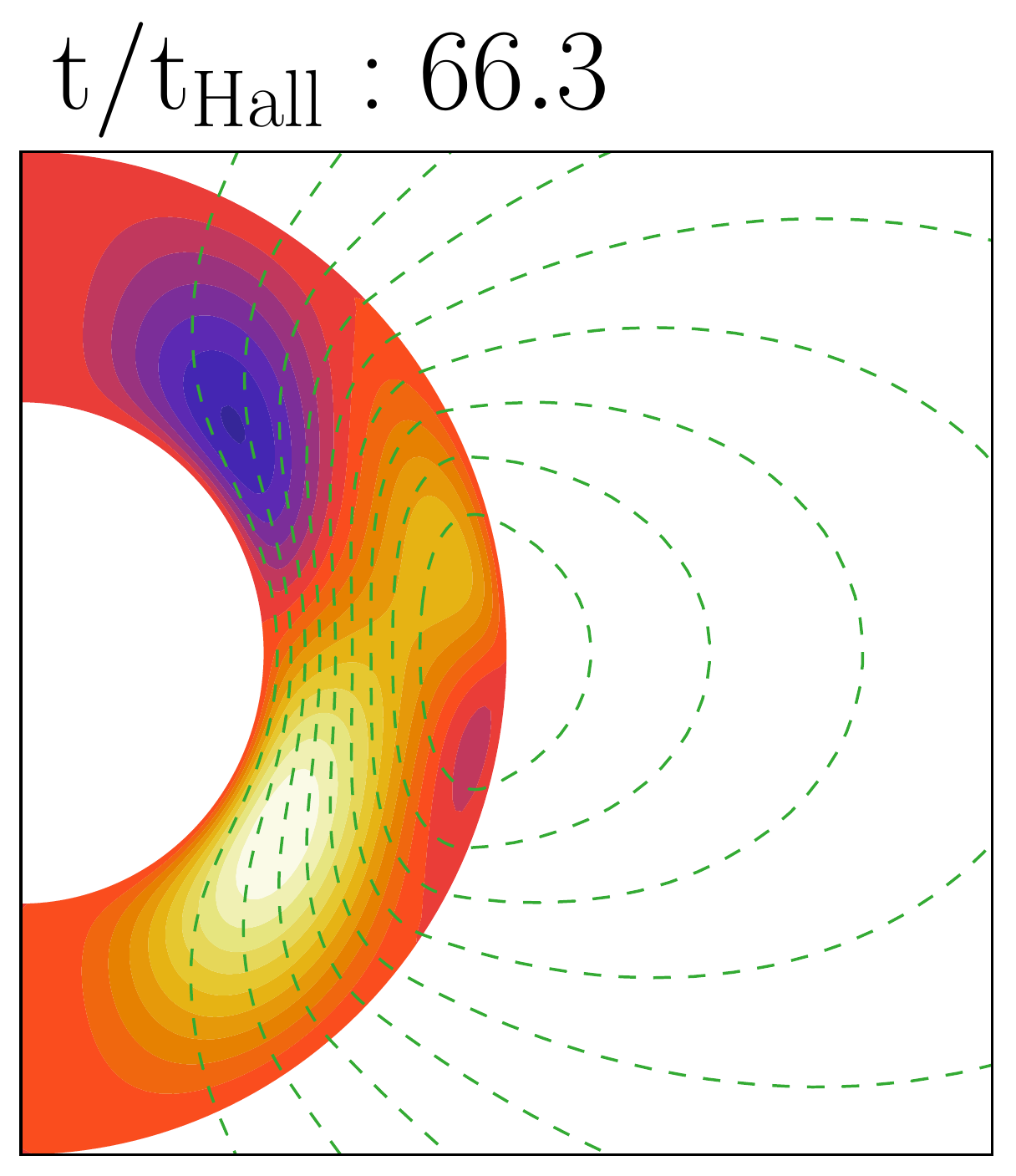}
\includegraphics[scale=0.22]{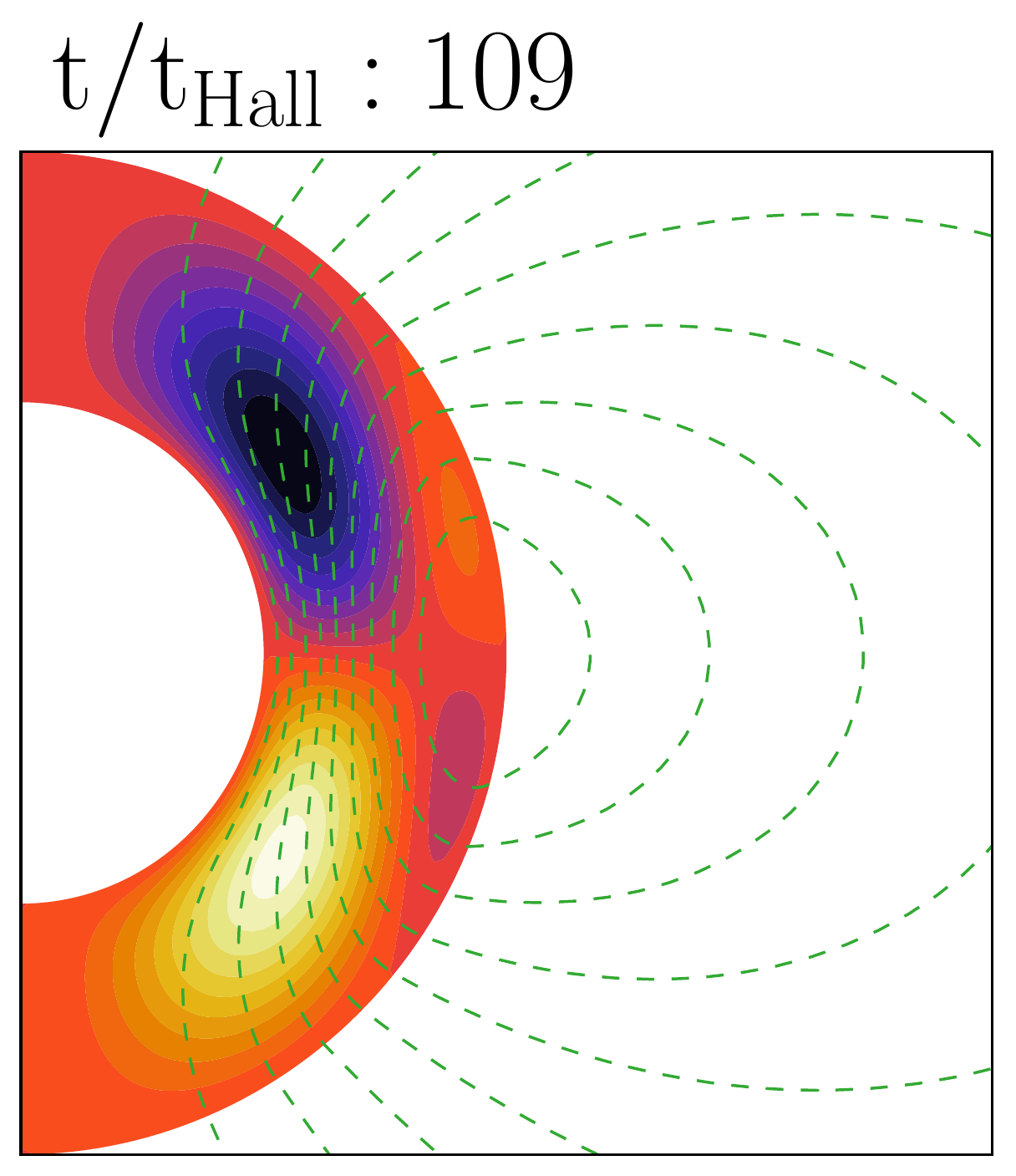}
\includegraphics[scale=0.22]{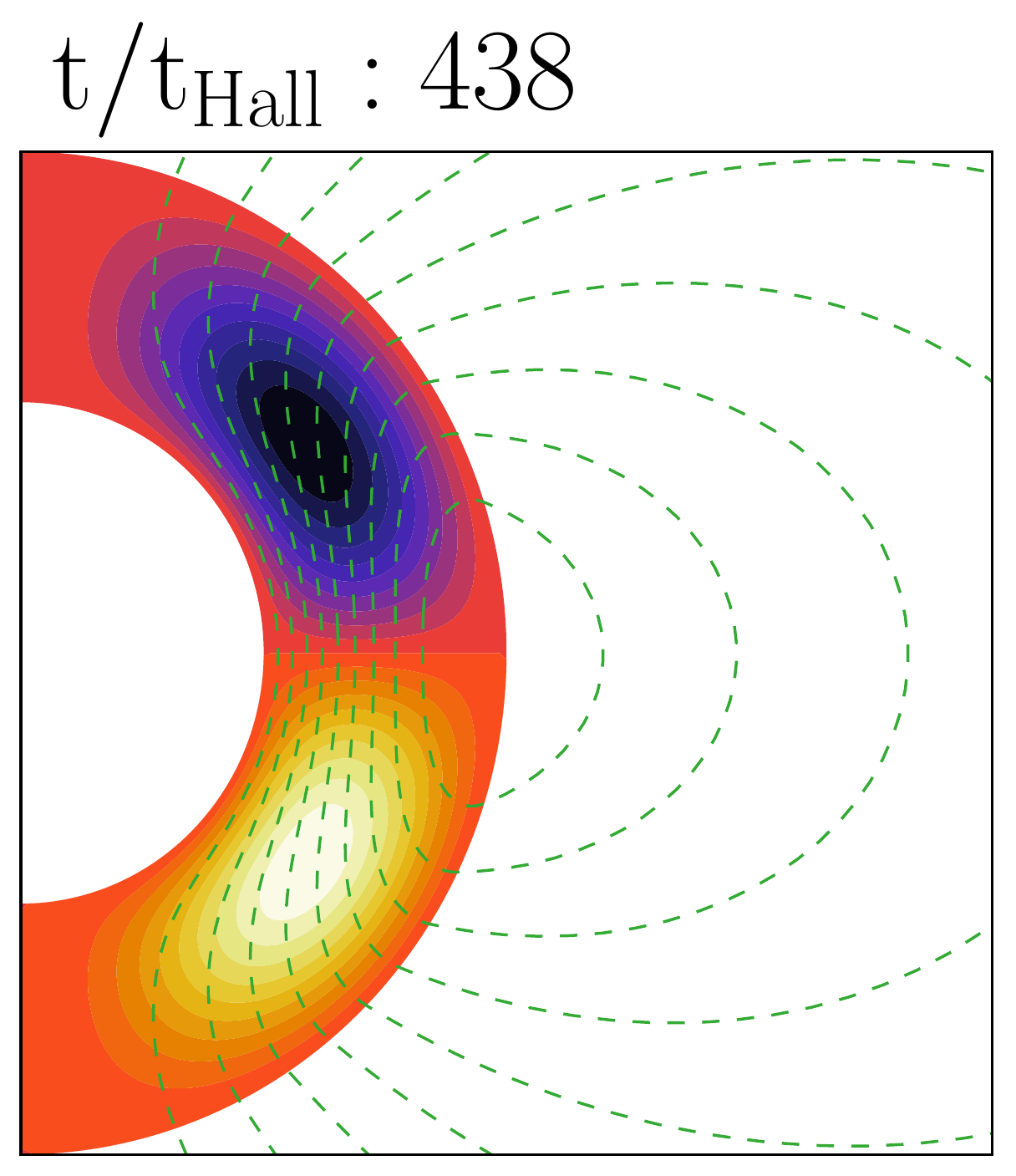}
\includegraphics[scale=0.8]{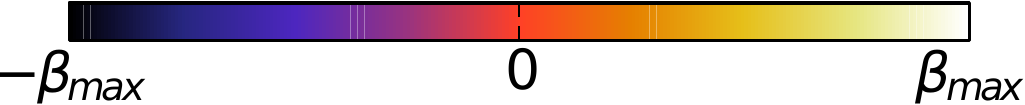}
\end{center}
\caption{Evolution of a poloidally dominated field given by Eq. (\ref{anal::bcomb}) with $E_P/E=0.9$. The color plot shows the intensity of $\beta$, with the color scale ranging from $-\beta_{max}$ to $\beta_{max}$, where $\beta_{max}=\max(|\beta|)$ is computed independently for each snapshot, and the contours are lines of constant $\alpha$ (corresponding to poloidal field lines). In these and subsequent plots the thickness of the crust is doubled to ease visualization and the exterior is also rescaled so poloidal field lines do not look discontinuous. We use a resolution of $40$ radial and $120$ angular steps, and a factor of the critical timestep $k_c=0.005$ (refer to Appendix \ref{appendix::timestep} for a definition of $k_c$)}\label{anal::poldomsim}
\end{figure}

In order to properly explore how the poloidally dominated regime is separated from the toroidally dominated one, we perform simulations using the field of Eq. (\ref{anal::bcomb}) with $E_P/E=0.9,0.7,0.5,0.3,0.1$ and $R_B=100$. Fig. \ref{anal::poldomsim} shows the simulation with $E_P/E=0.9$, from which it can be seen that the current associated to the toroidal field drags poloidal field lines closer to one of the poles, after which the bending of poloidal field lines changes the orientation of the toroidal field. The poloidal field lines are then dragged towards the opposite pole, where the process is repeated in what appears to be stable oscillations until the evolution is dominated by Ohmic dissipation with a mixture of the fundamental poloidal Ohm mode and the $n=1,l=2$ toroidal Ohm mode.

\begin{figure}
\begin{center}
\includegraphics[scale=0.22]{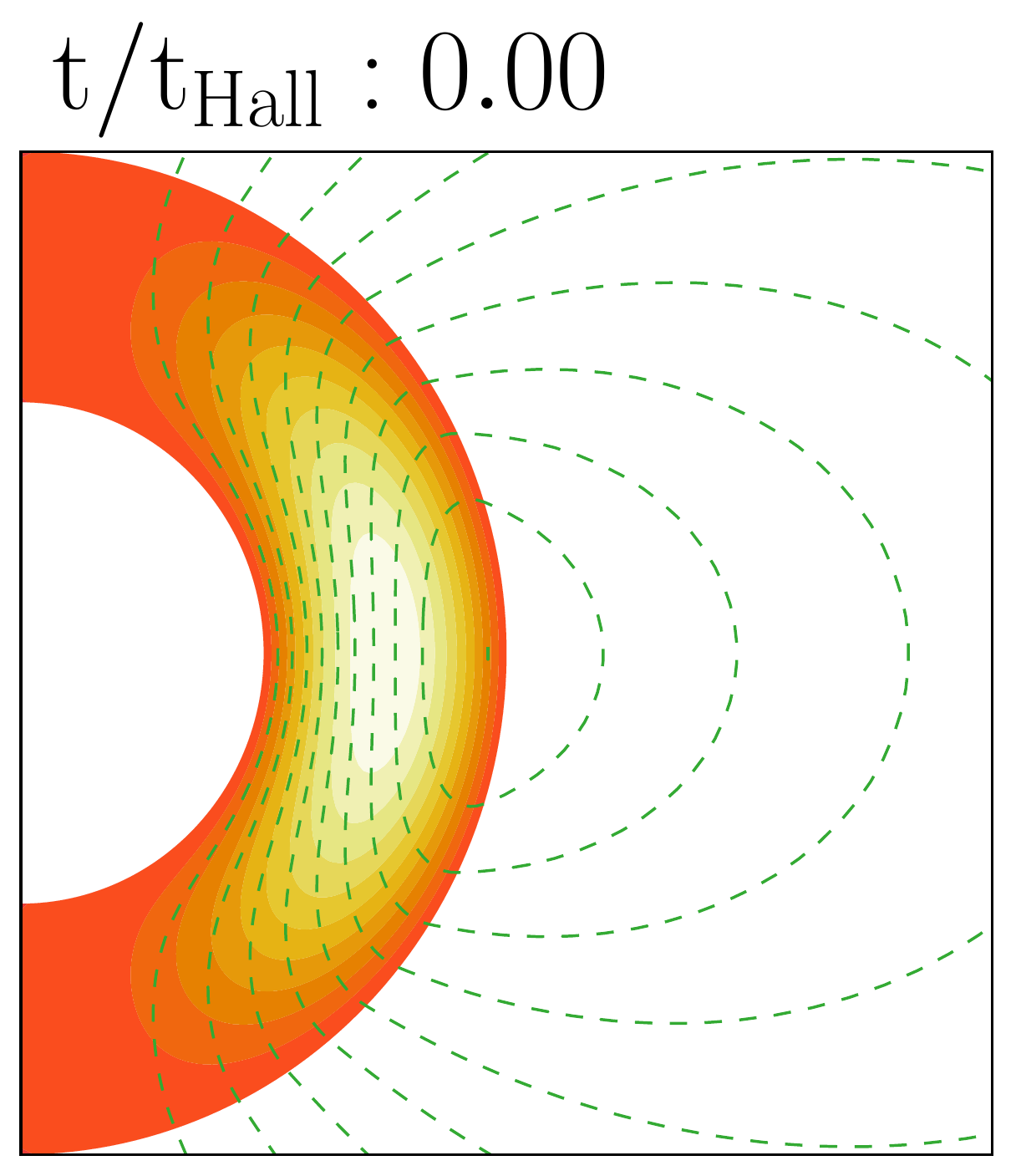}
\includegraphics[scale=0.22]{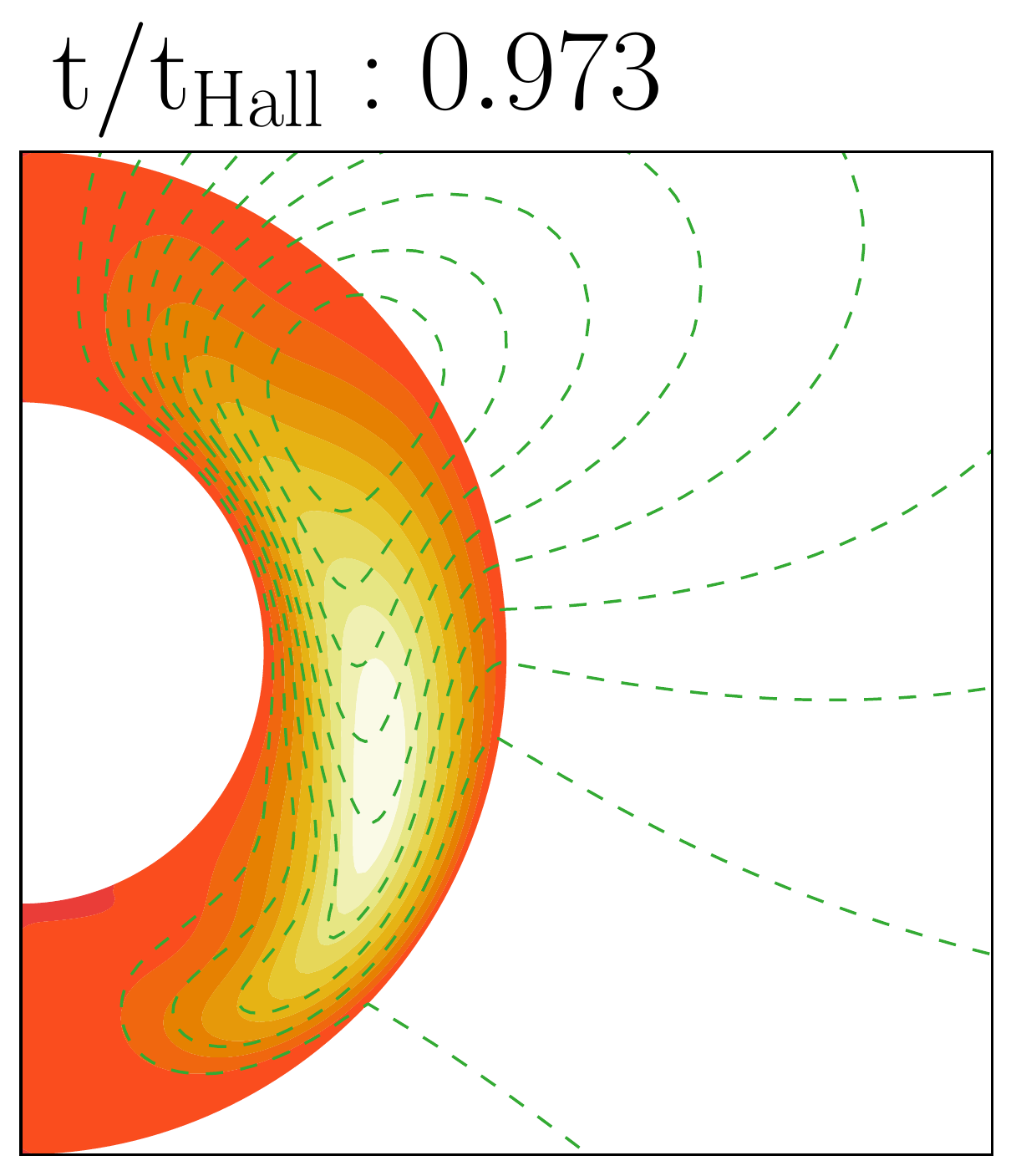}
\includegraphics[scale=0.22]{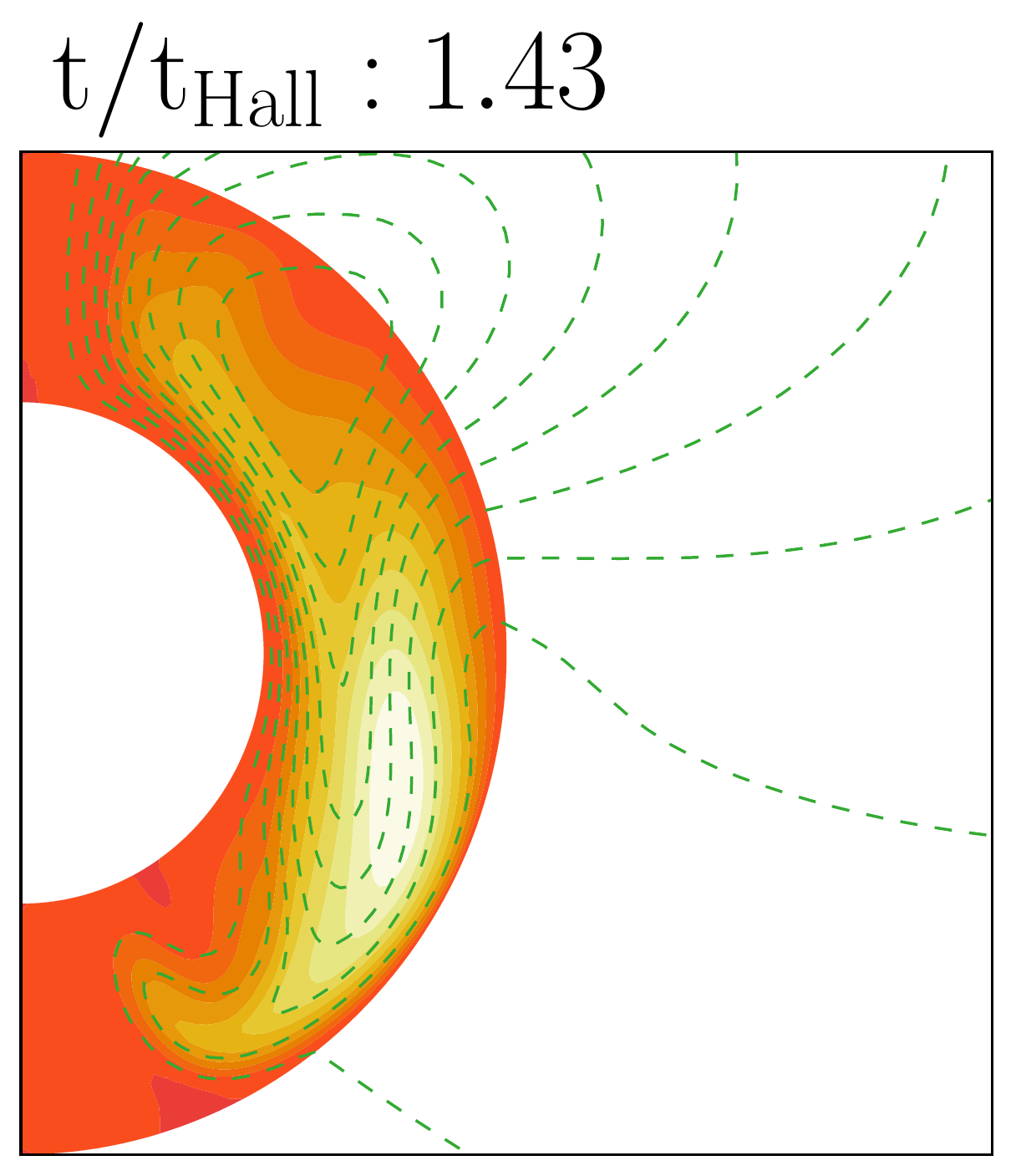}
\includegraphics[scale=0.22]{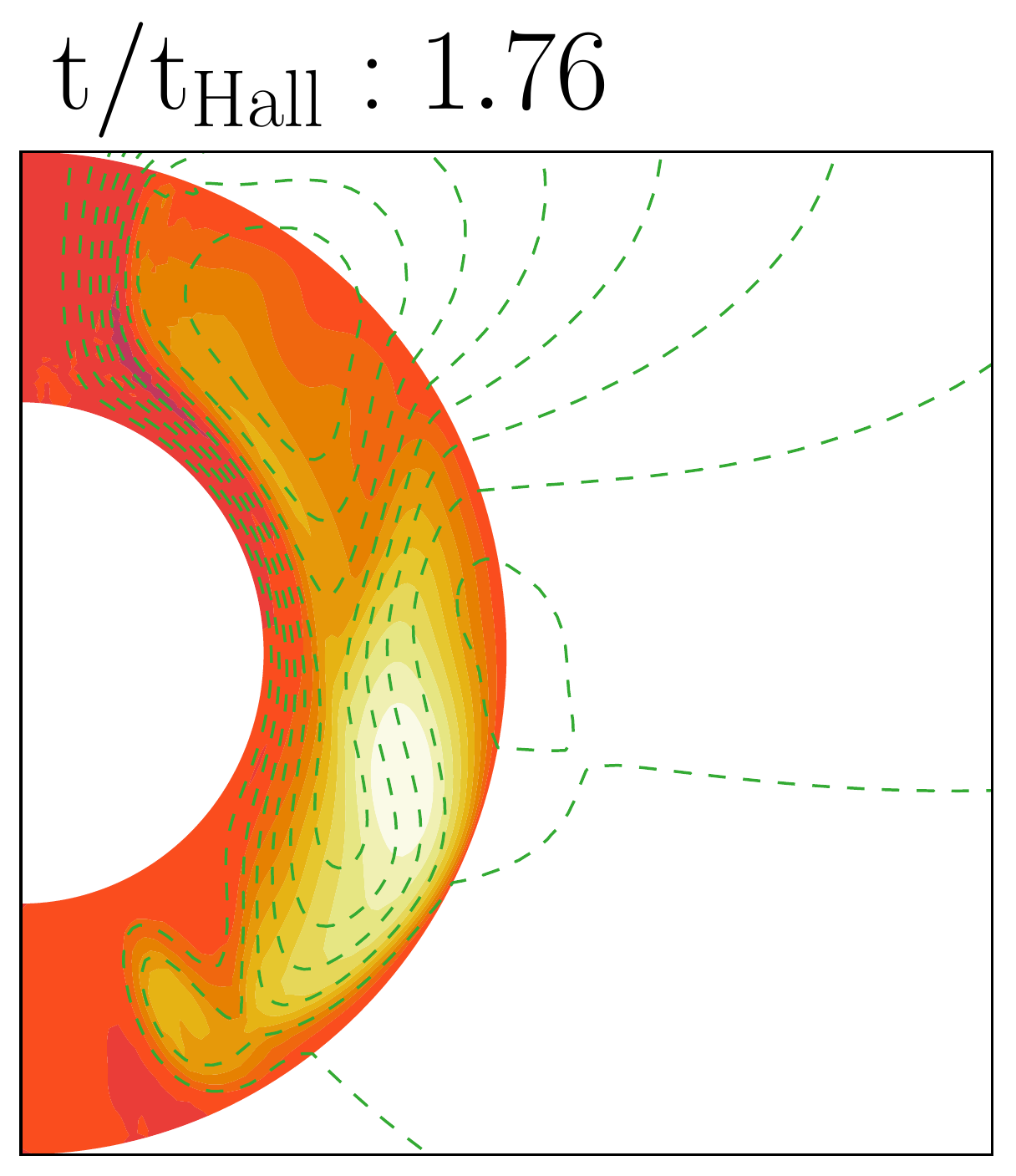}
\includegraphics[scale=0.22]{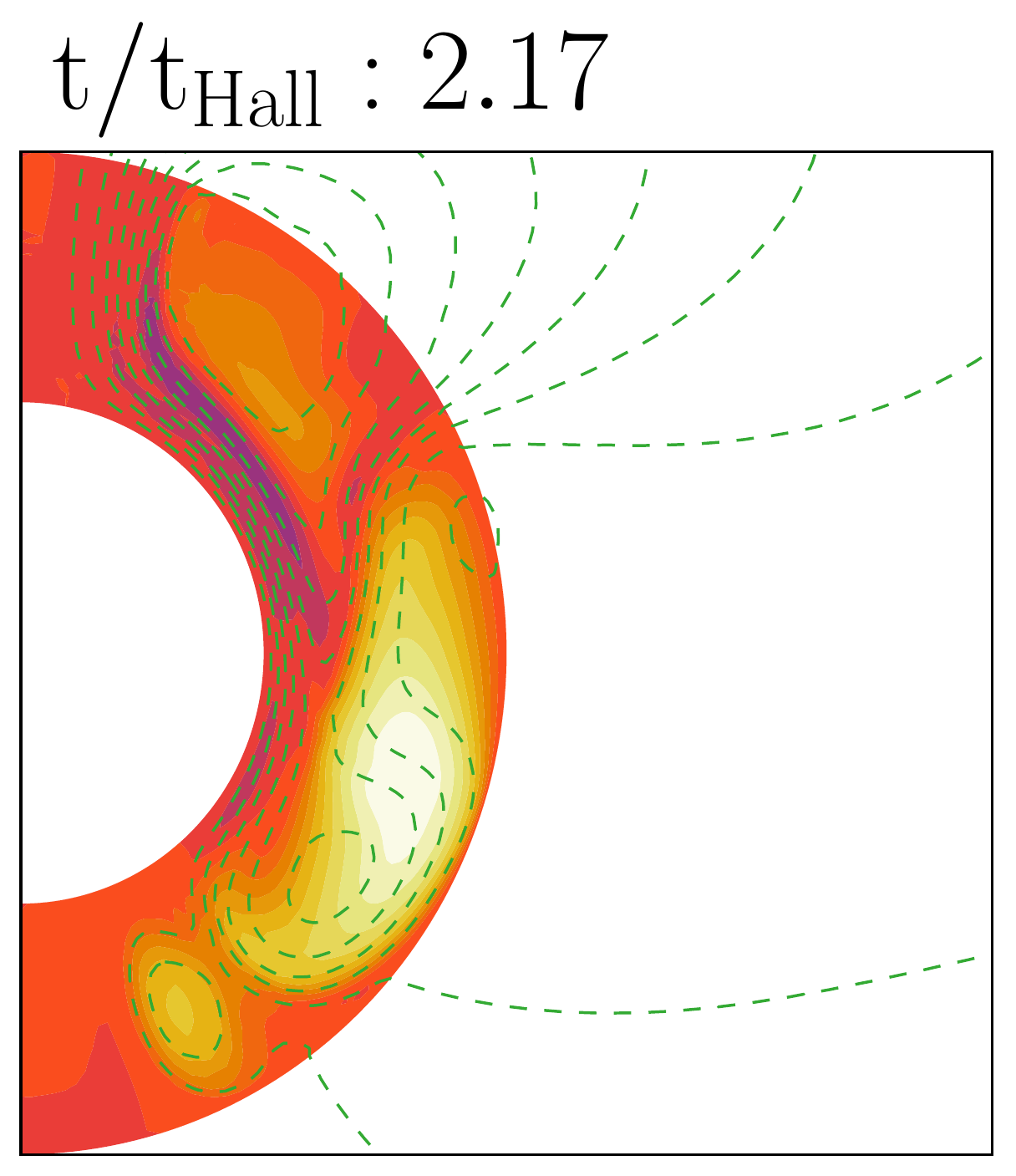}
\includegraphics[scale=0.22]{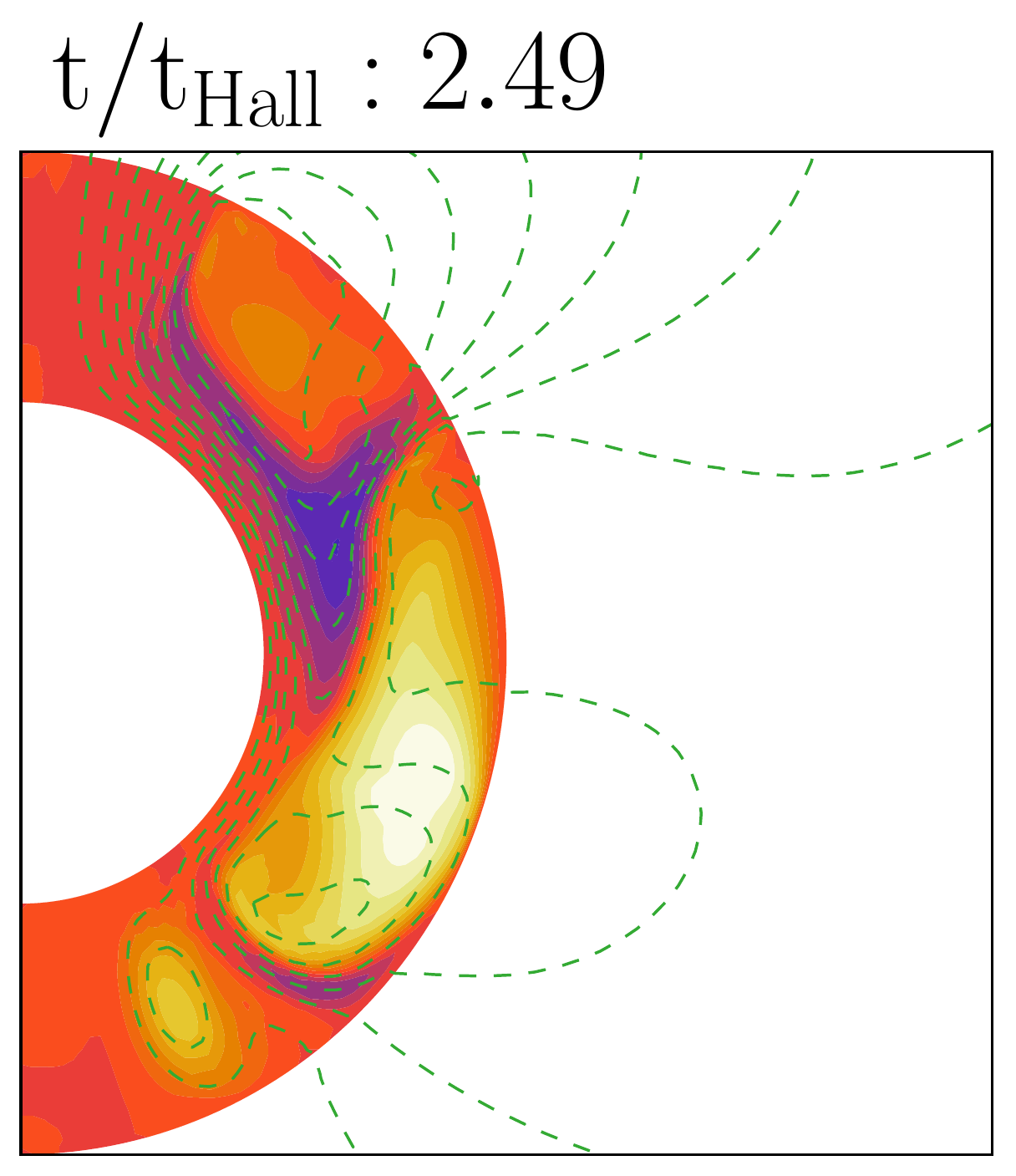}
\includegraphics[scale=0.22]{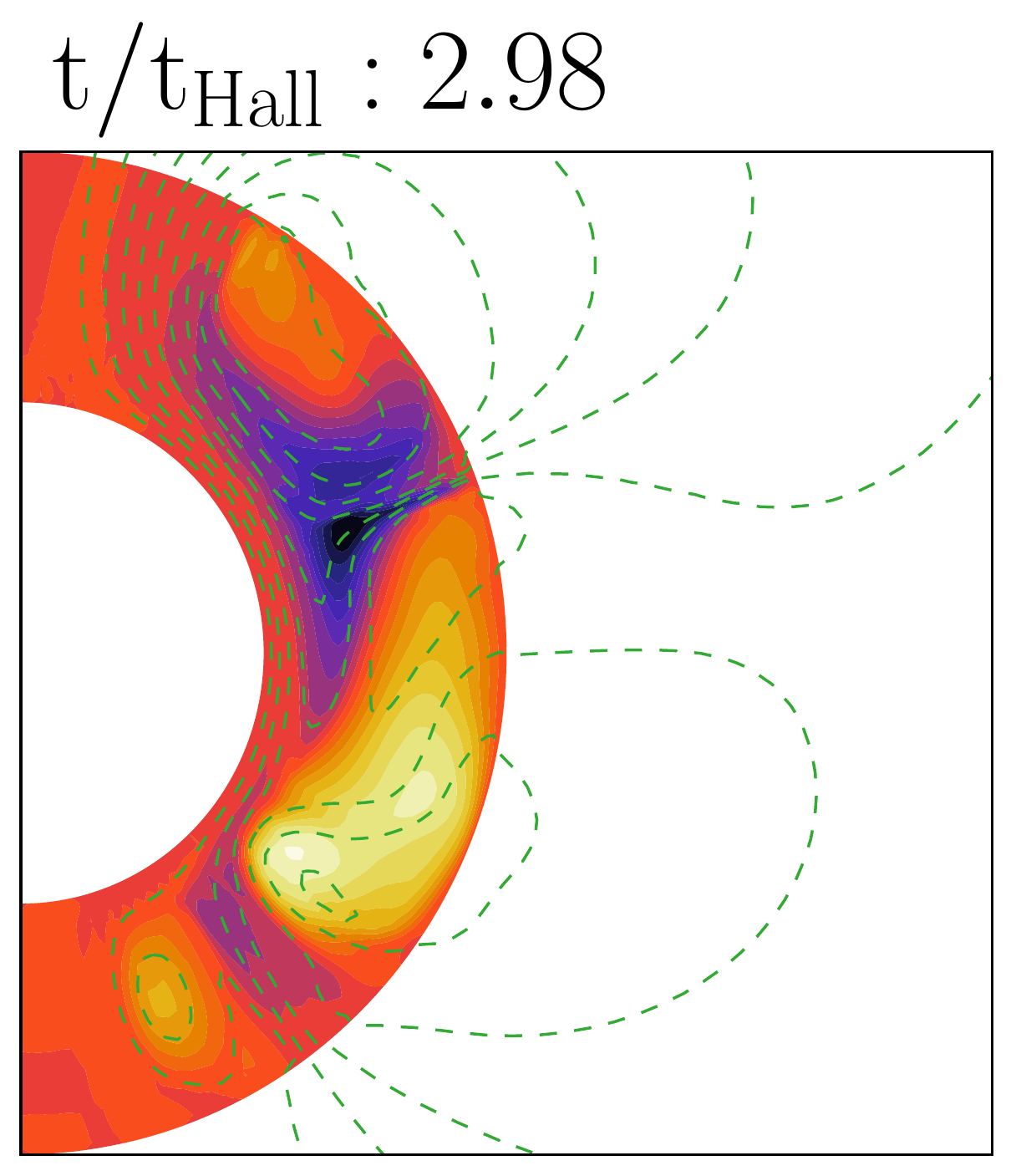}
\includegraphics[scale=0.22]{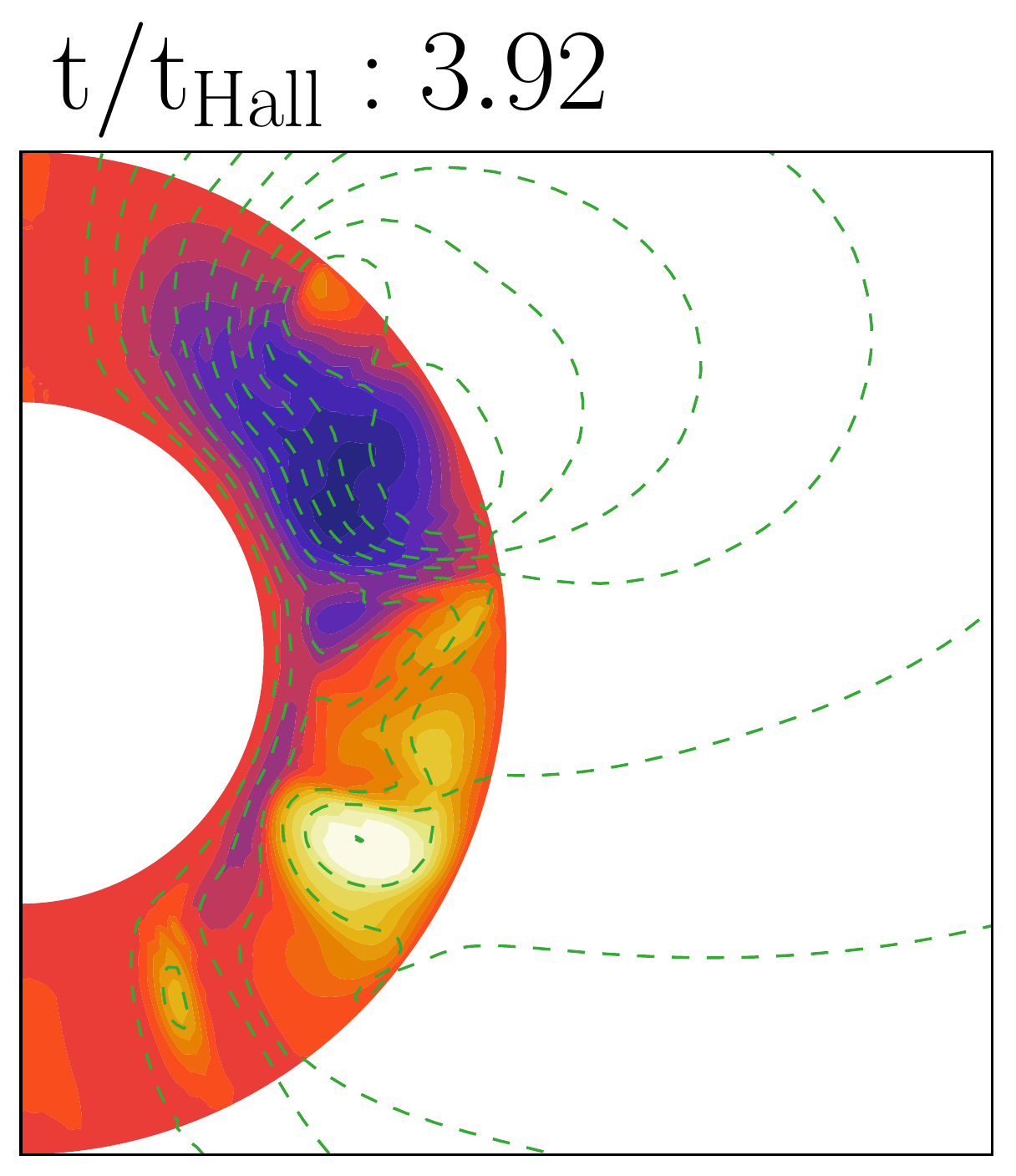}
\includegraphics[scale=0.22]{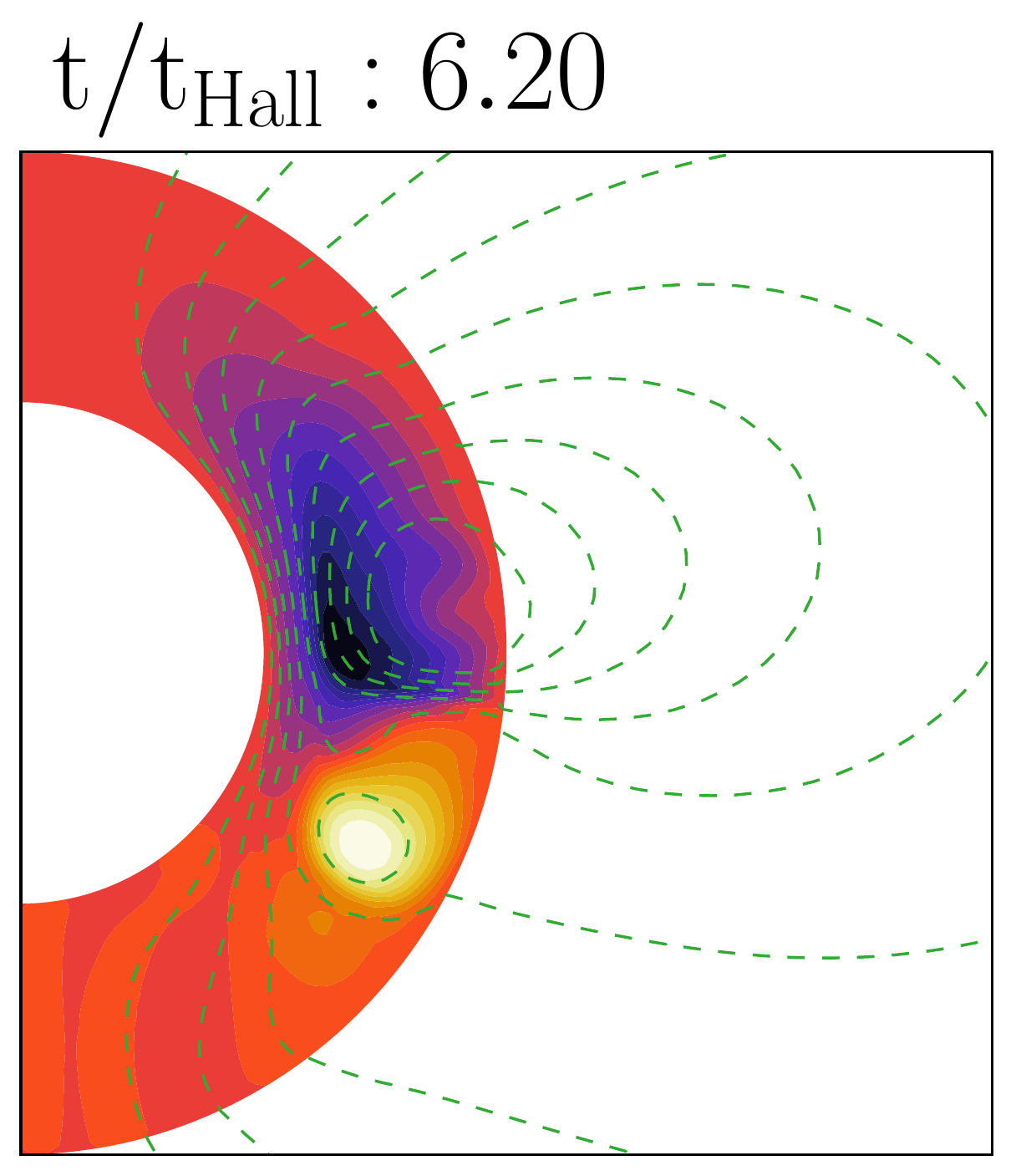}
\includegraphics[scale=0.22]{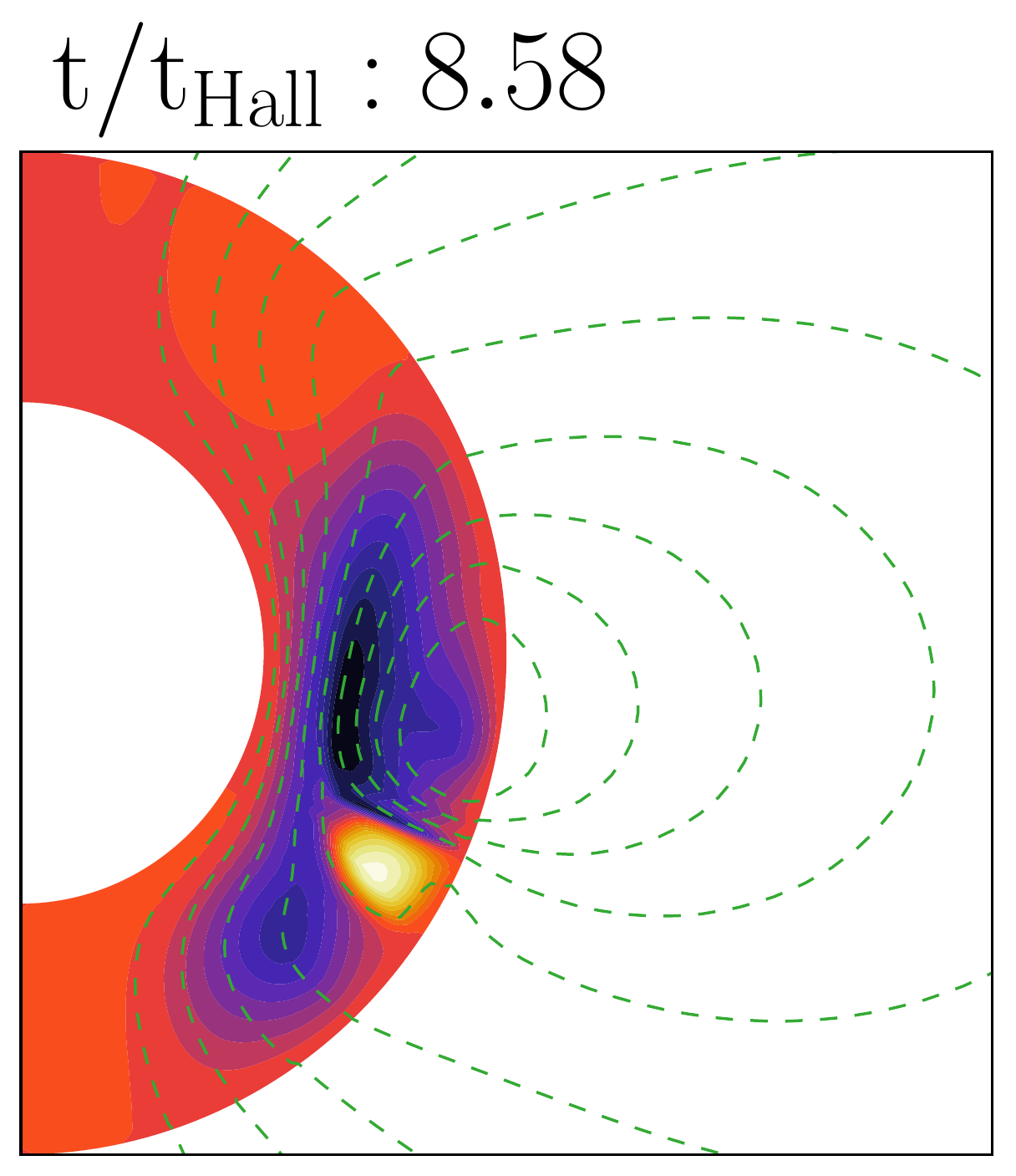}
\includegraphics[scale=0.22]{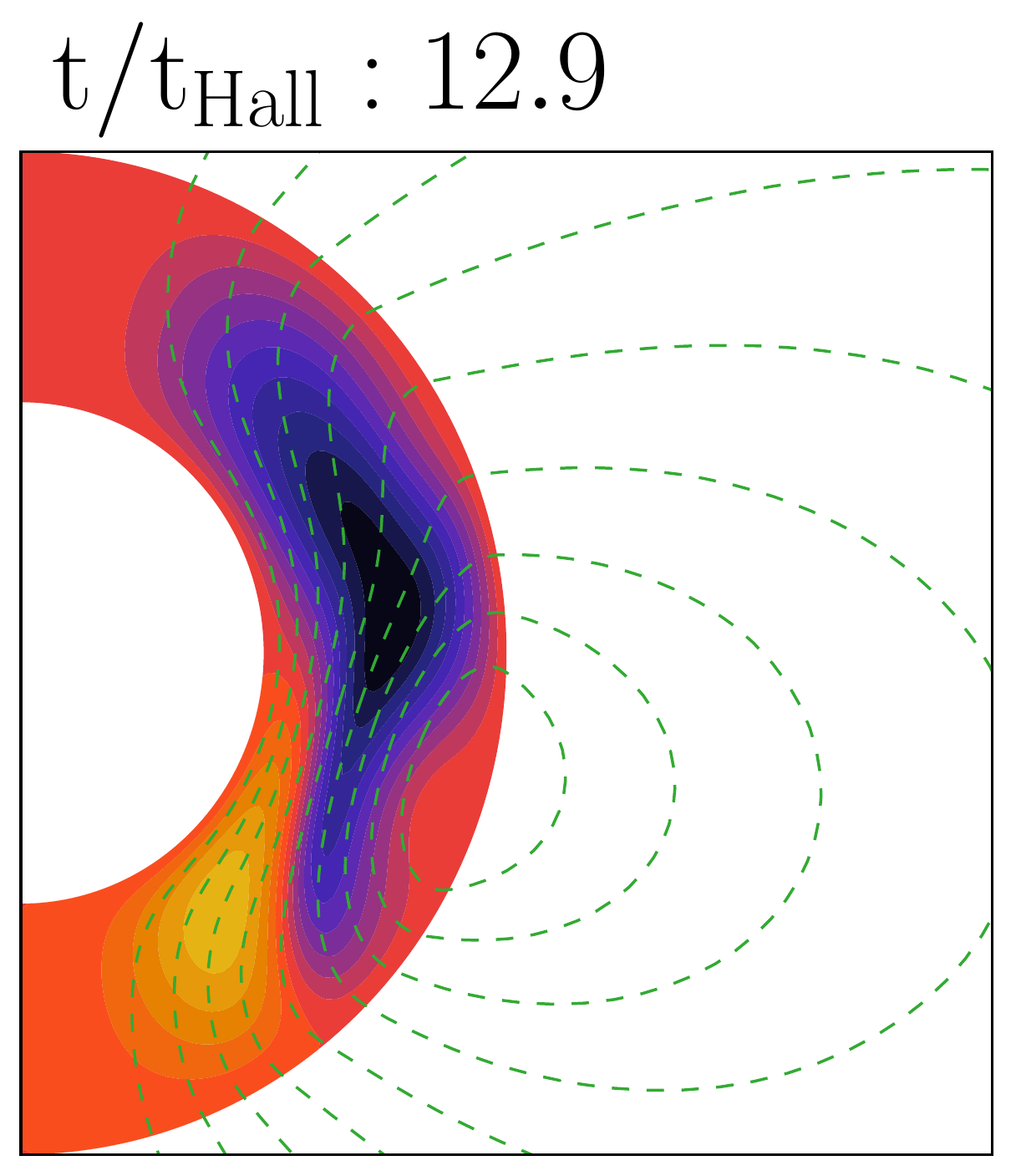}
\includegraphics[scale=0.22]{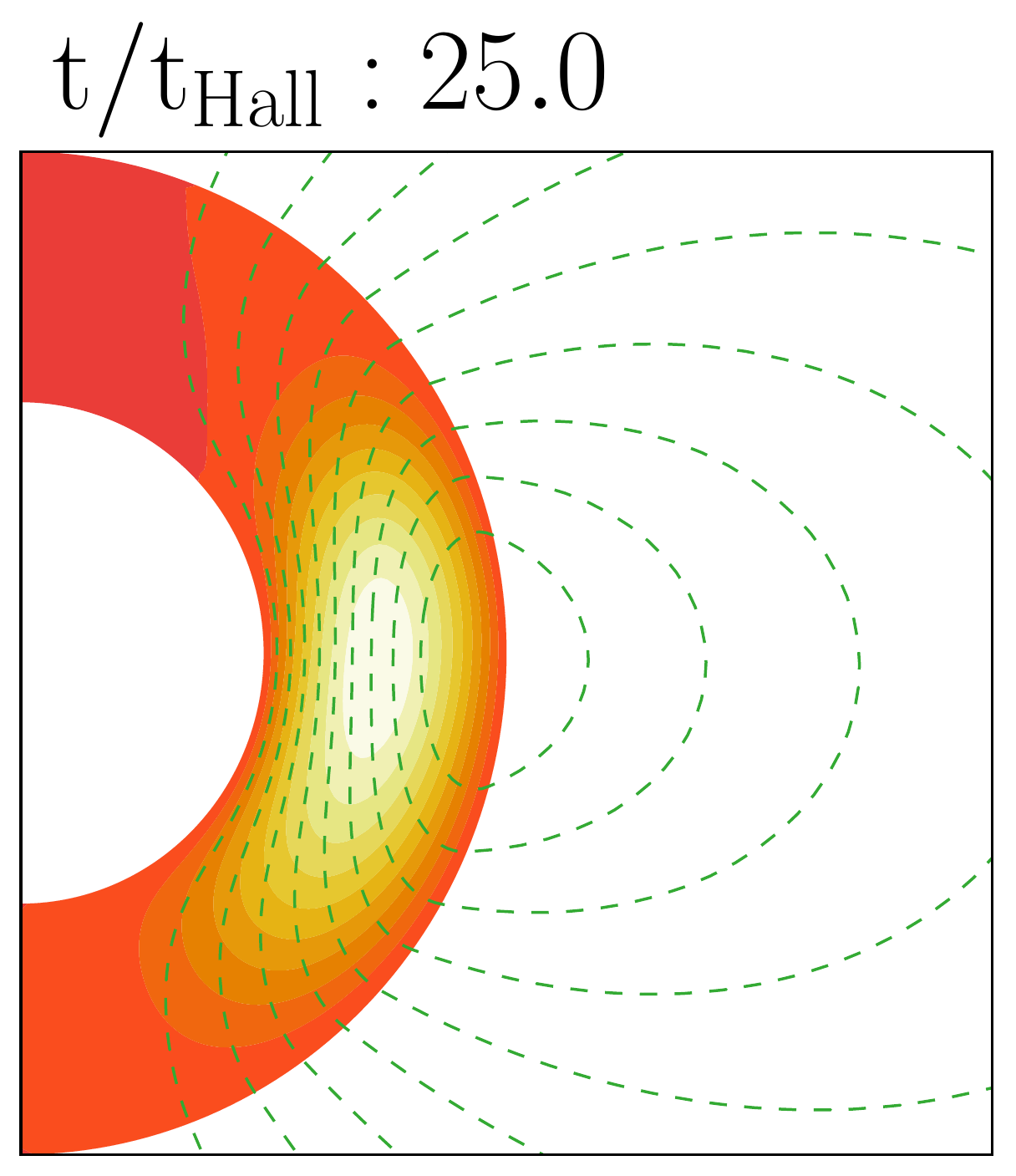}
\includegraphics[scale=0.22]{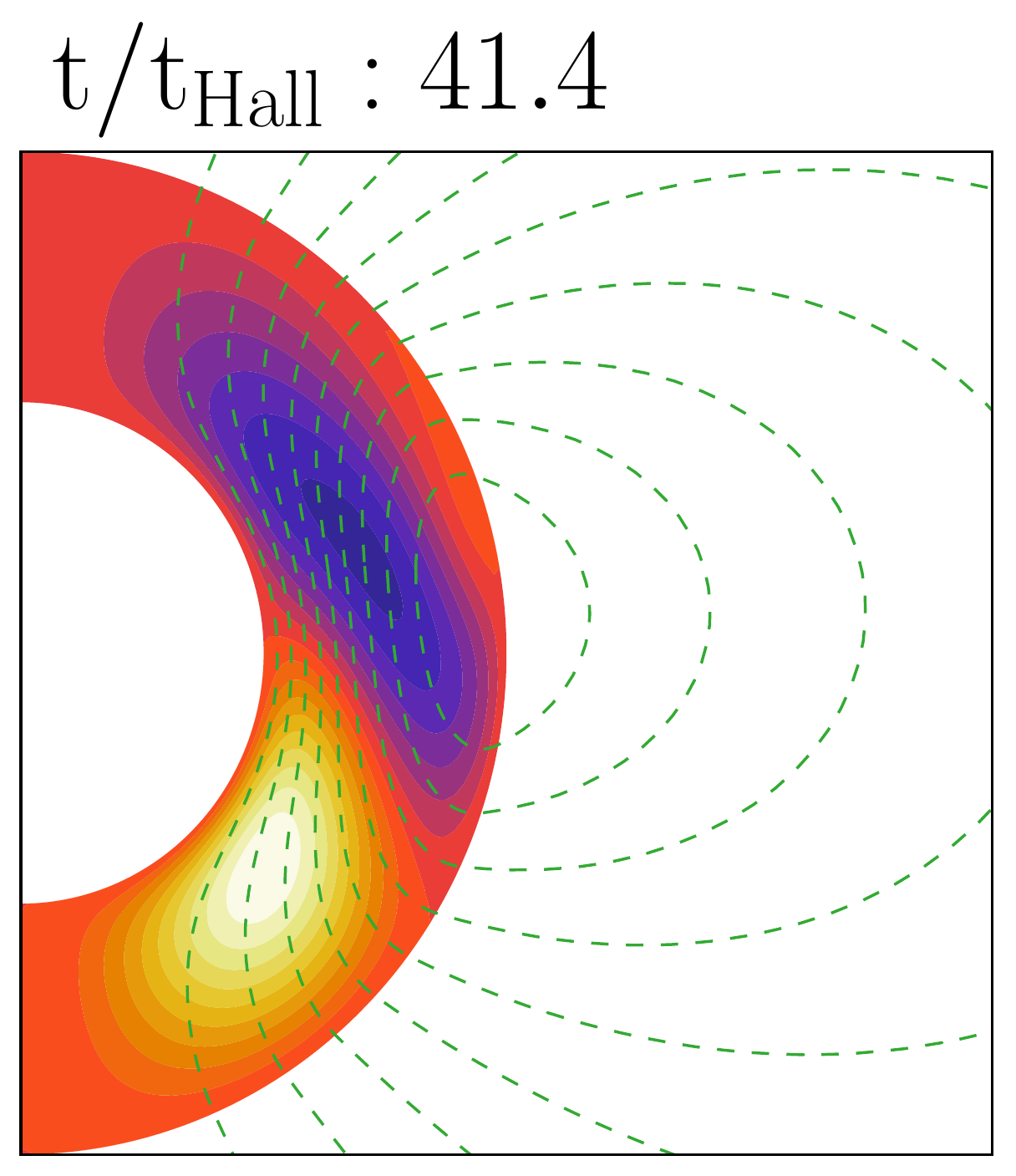}
\includegraphics[scale=0.22]{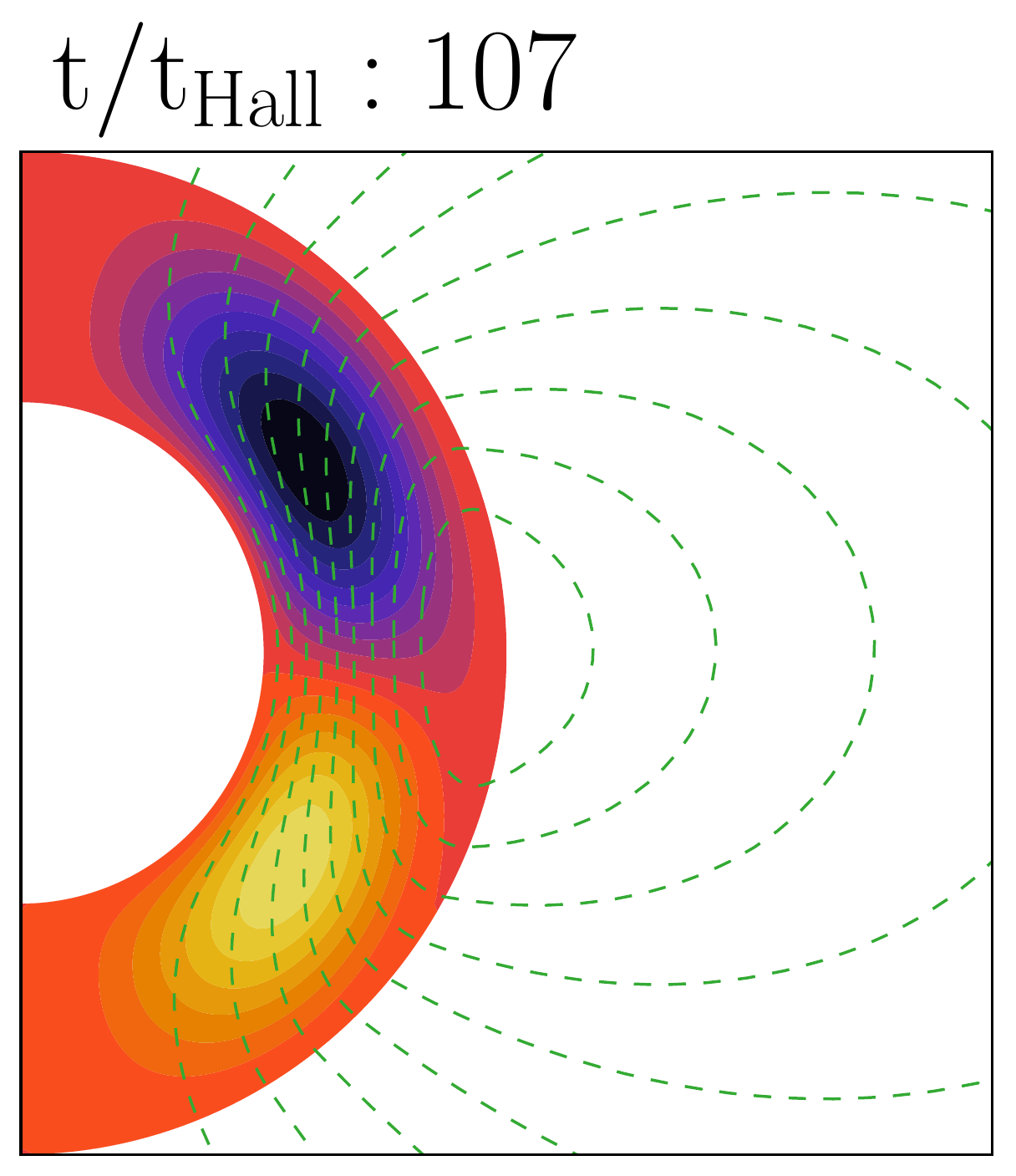}
\includegraphics[scale=0.22]{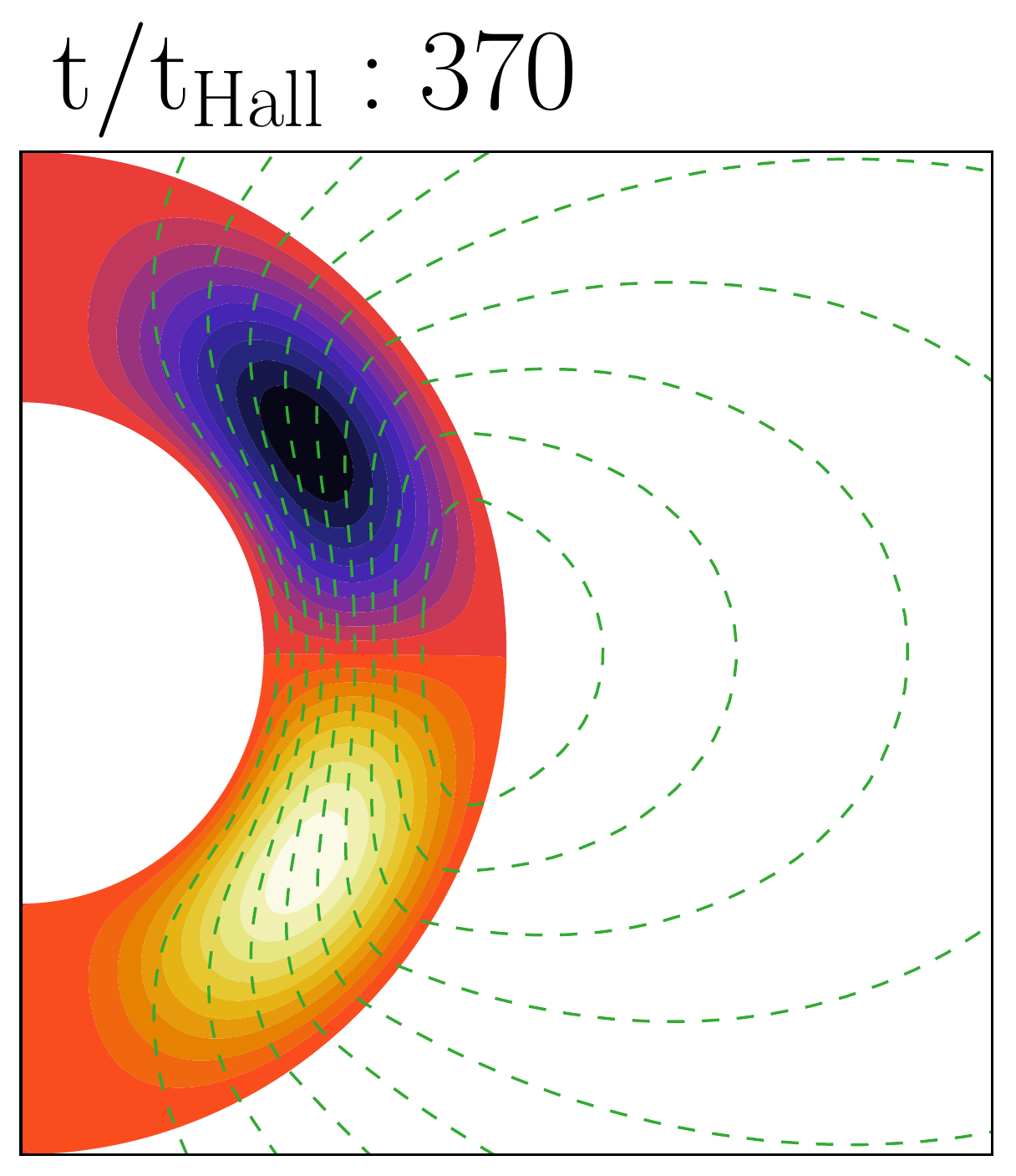}
\includegraphics[scale=0.8]{colorbar}
\end{center}
\caption{Evolution of a toroidally dominated field given by Eq. (\ref{anal::bcomb}) with $E_P/E=0.1$. Refer to the caption of Fig. \ref{anal::poldomsim} for details.}\label{anal::tordomsim}
\end{figure}

\begin{figure}
\begin{center}
\includegraphics[width=\columnwidth]{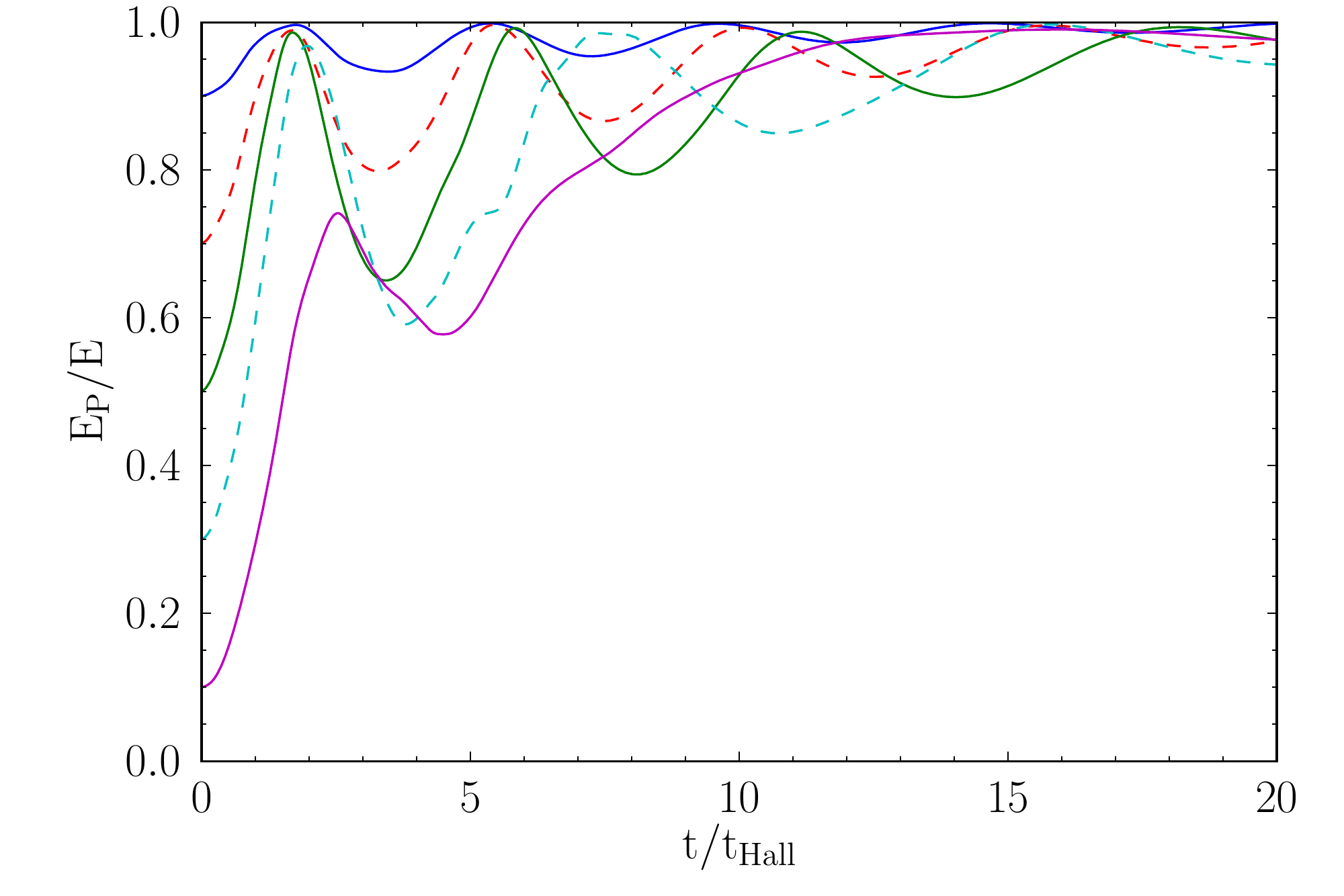}
\end{center}
\caption{Evolution of the ratio $E_P/E$ for simulations with different initial ratios, with initial conditions given by Eq. (\ref{anal::bcomb}) with $R_B=100$.}\label{anal::multipoles2}
\end{figure}

\begin{figure}
\begin{center}
\includegraphics[width=\columnwidth]{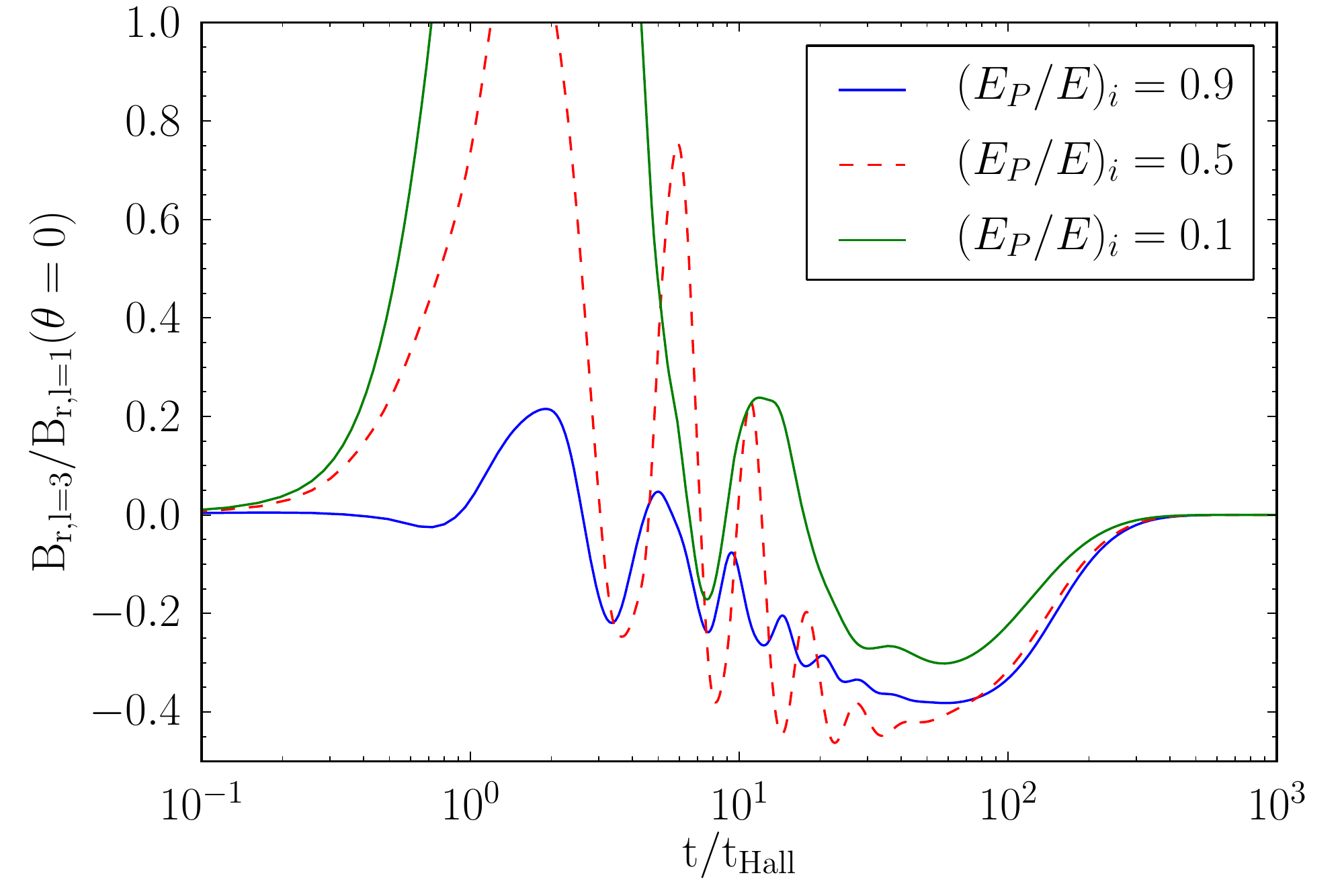}
\end{center}
\caption{Ratio between the radial component of the field corresponding to $l=1$ and $l=3$ at the pole for different initial values of $E_P/E$. Initial conditions are given by Eq. (\ref{anal::bcomb}) with $R_B=100$.}\label{anal::multipoles}
\end{figure}

The opposite case is shown in Fig. \ref{anal::tordomsim}, where the evolution of a toroidally dominated field with $E_P/E=0.1$ at the beginning is shown. In this case, the currents associated to the toroidal field drag poloidal field lines to one of the poles, just as was the case with the poloidally dominated field, but afterwards the field forms structures on much smaller scales, with strong associated currents. This produces fast Ohmic dissipation, together with energy transfer to the poloidal component. The field then becomes predominantly poloidal, and the evolution continues as in the initially poloidally dominated case.

The rapid evolution to a poloidally dominant field can be easily visualized by plotting the evolution of the ratio of poloidal to total magnetic energy, as is done in Fig. \ref{anal::multipoles2}. 
For simulations with progressively stronger toroidal fields, more energy will be lost in this stage of rapid Ohmic dissipation, producing oscillations with longer periods once they reach the poloidally dominant regime. In particular, Figs. \ref{anal::tordomsim} and \ref{anal::multipoles2} show that for the case that starts with $E_P/E=0.1$ no clear oscillatory behavior occurs, as Ohmic dissipation quickly becomes dominant.

This common evolution towards fields with a low fraction of the energy contained in the toroidal component is consistent with what is found in other studies that include an electron density gradient and also test initial configurations other than coupled fundamental Ohm modes \citep{gou+13b, geppon+07}, so it appears to be a general feature.

It is interesting that the final configuration dominated by Ohmic dissipation has a quadrupole toroidal field rather than a dipole, which has a slower decay rate. This is because the dominant poloidal dipole field acting on itself due to Hall drift produces a quadrupolar toroidal field.

Fig. \ref{anal::multipoles} displays for some of the simulations how a counter-aligned octupole forms after several Hall times. All the cases studied, except the one with an initial $E_P/E=0.1$, evolve with damped oscillations towards a point where the ratio between the $l=3$ and $l=1$ components of the radial field at the poles have a typical value $\sim -0.4$, after which Ohmic dissipation takes over and the octupole component completely decays. This point is then similar to the attractor described by \citet{gou+13c}, including also a much weaker toroidal quadrupole as seen in the last frame of Fig. \ref{anal::poldomsim}. The simulation with an initial $E_P/E=0.1$ decays very strongly at the beginning, which causes Ohmic dissipation to dominate the evolution at an earlier time, before the octupole component grows too much.

\subsection{Comparison with Meissner boundary conditions}
\label{anal::meissner}
To check whether the inclusion of Meissner boundary conditions modifies significantly the evolution, we perform an additional simulation with $R_B=20$ for the case of $E_P/E=0.9$, with and without Meissner boundary conditions. However, since combinations of the form given by Eq. (\ref{anal::bcomb}) will not satisfy the boundary conditions initially, we choose a modified toroidal field given by
\begin{eqnarray}
\boldsymbol{B}_t=(r-r_{min})\boldsymbol{B}_{11t},\label{anal::meissner_field}
\end{eqnarray}
With this choice the radial derivative of $\beta$ is zero at the crust-core interface, and thus, will satisfy the Meissner boundary conditions (see Eq. (\ref{intro::scbound})). For the poloidal component we keep the choice of using the fundamental poloidal Ohm mode.

\begin{figure}
\begin{center}
\includegraphics[width=\columnwidth]{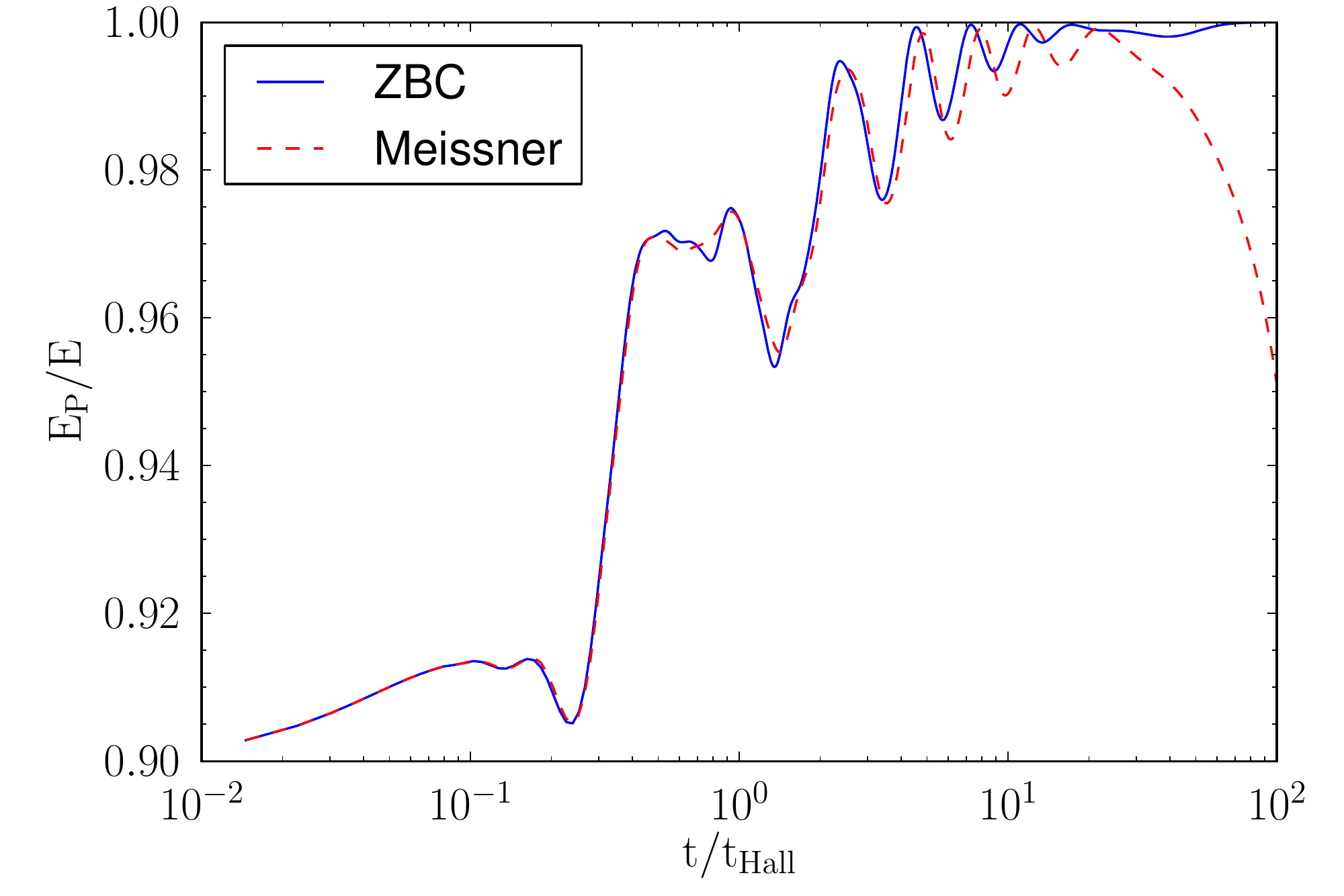}
\end{center}
\caption{Evolution of the ratio $E_P/E$ for the field described in \S\ref{anal::meissner} with $R_B=20$, $(E_P/E)_i=0.9$ and both Meissner and zero boundary conditions.}\label{anal::meissner_comp}
\end{figure}

\begin{figure}
\begin{center}
\includegraphics[scale=0.22]{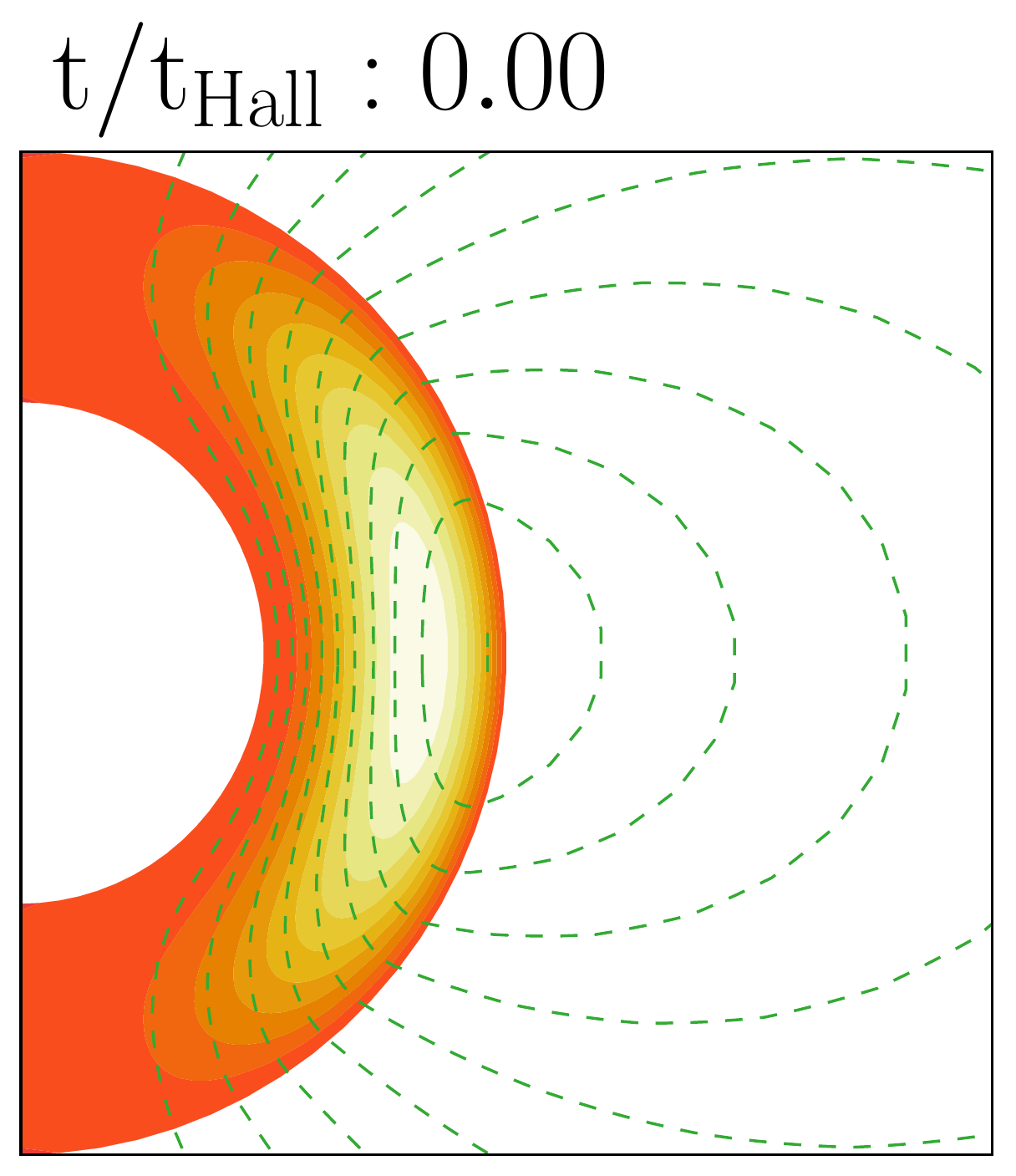}
\includegraphics[scale=0.22]{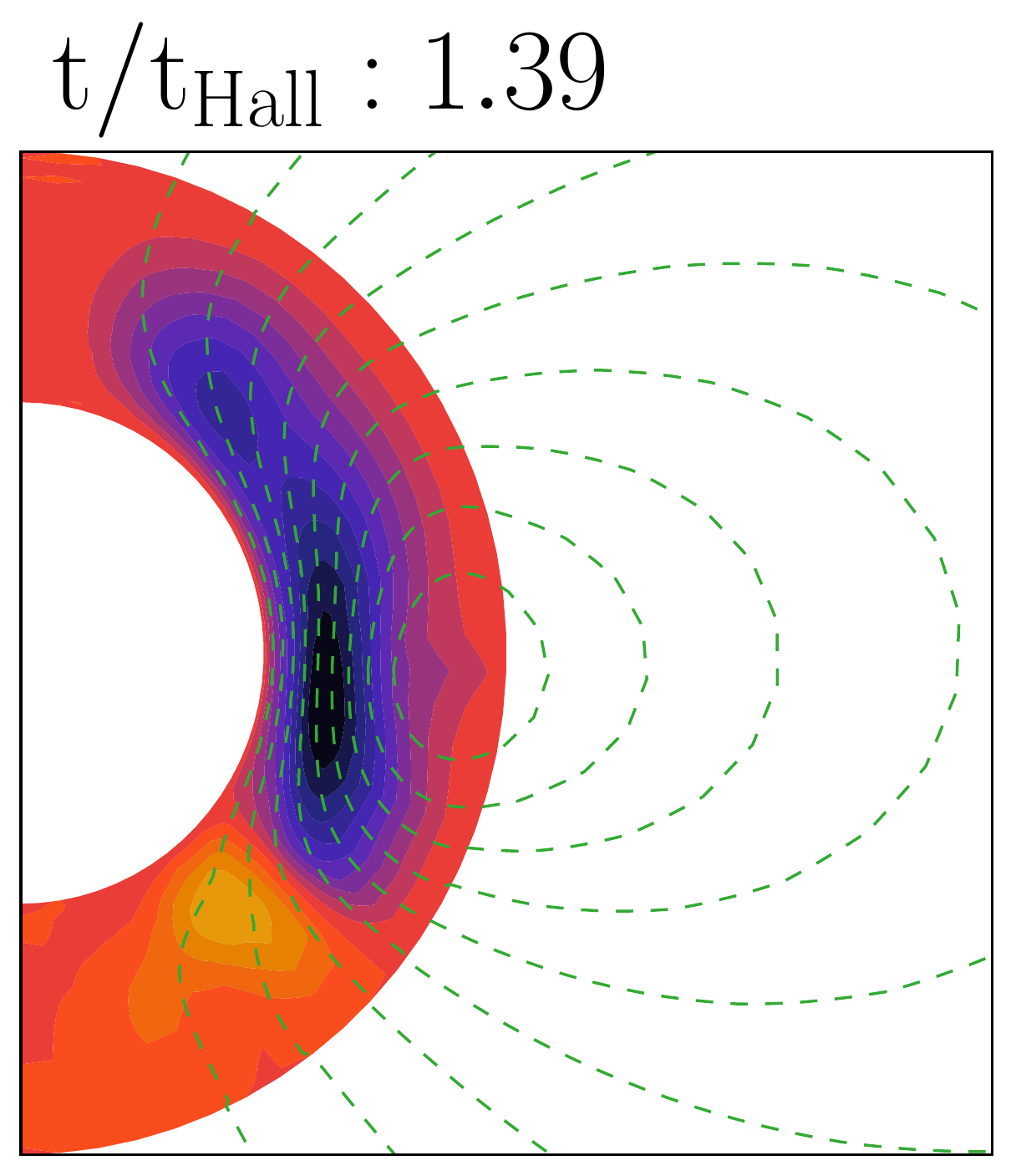}
\includegraphics[scale=0.22]{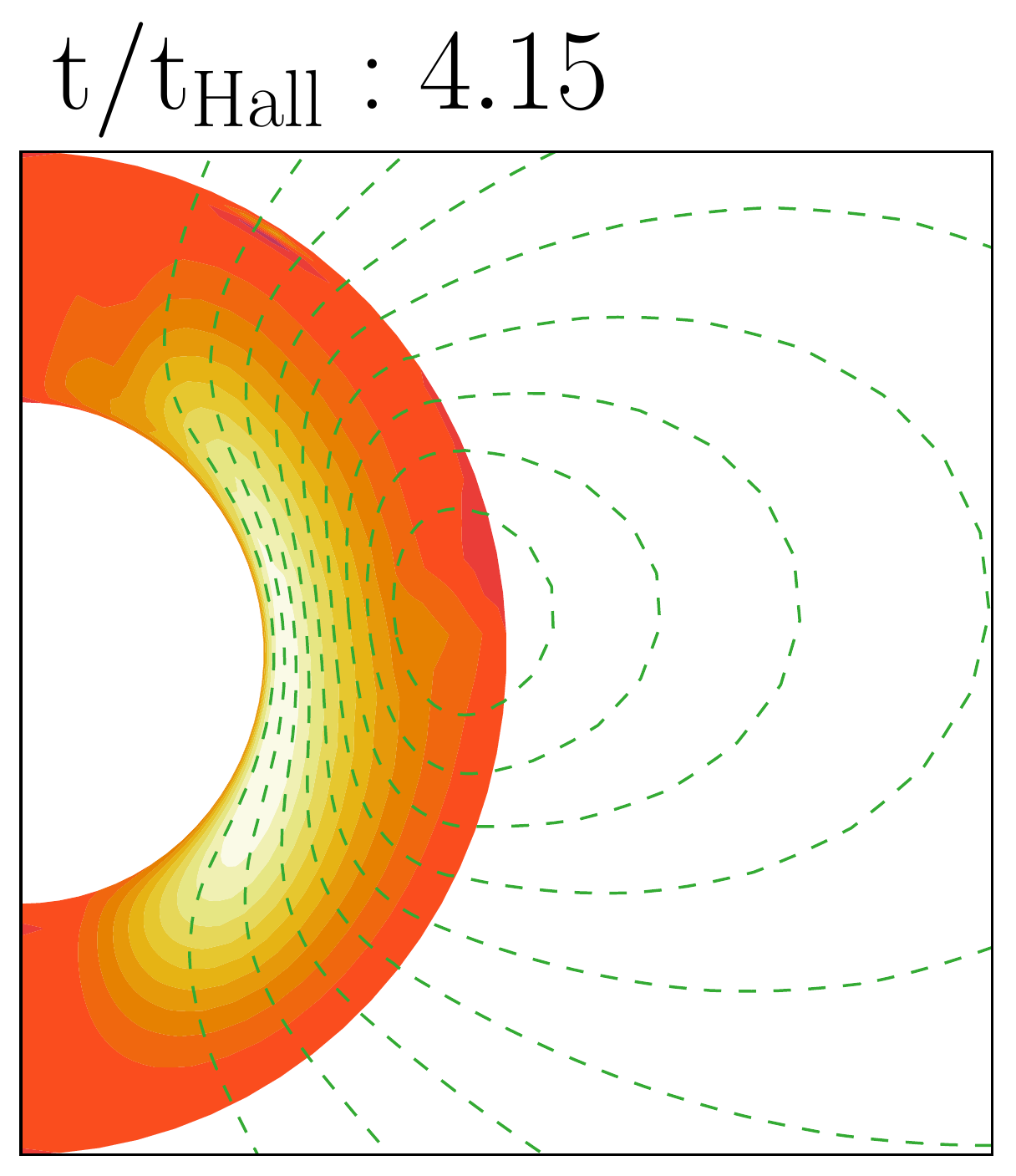}
\includegraphics[scale=0.22]{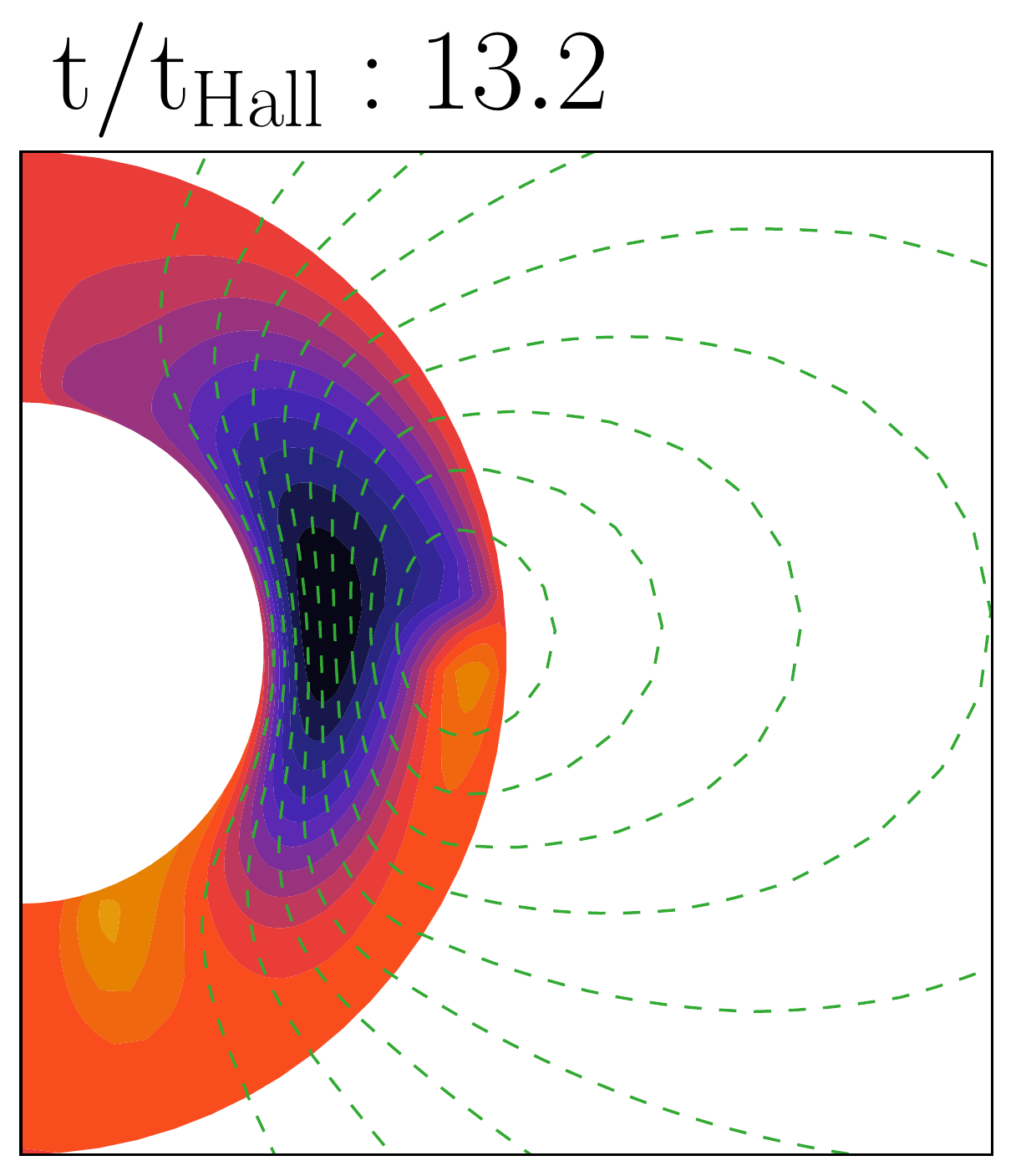}
\includegraphics[scale=0.22]{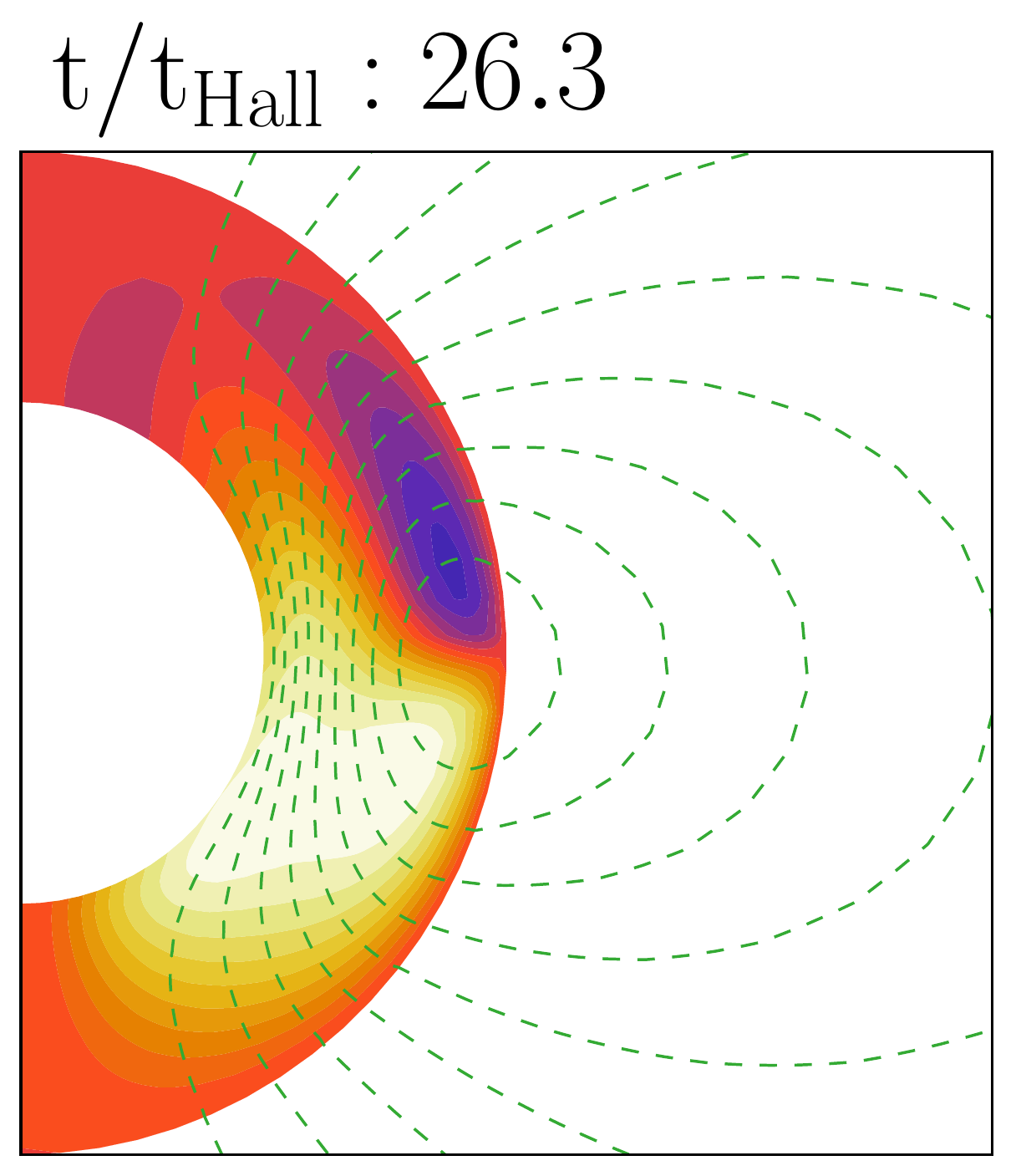}
\includegraphics[scale=0.22]{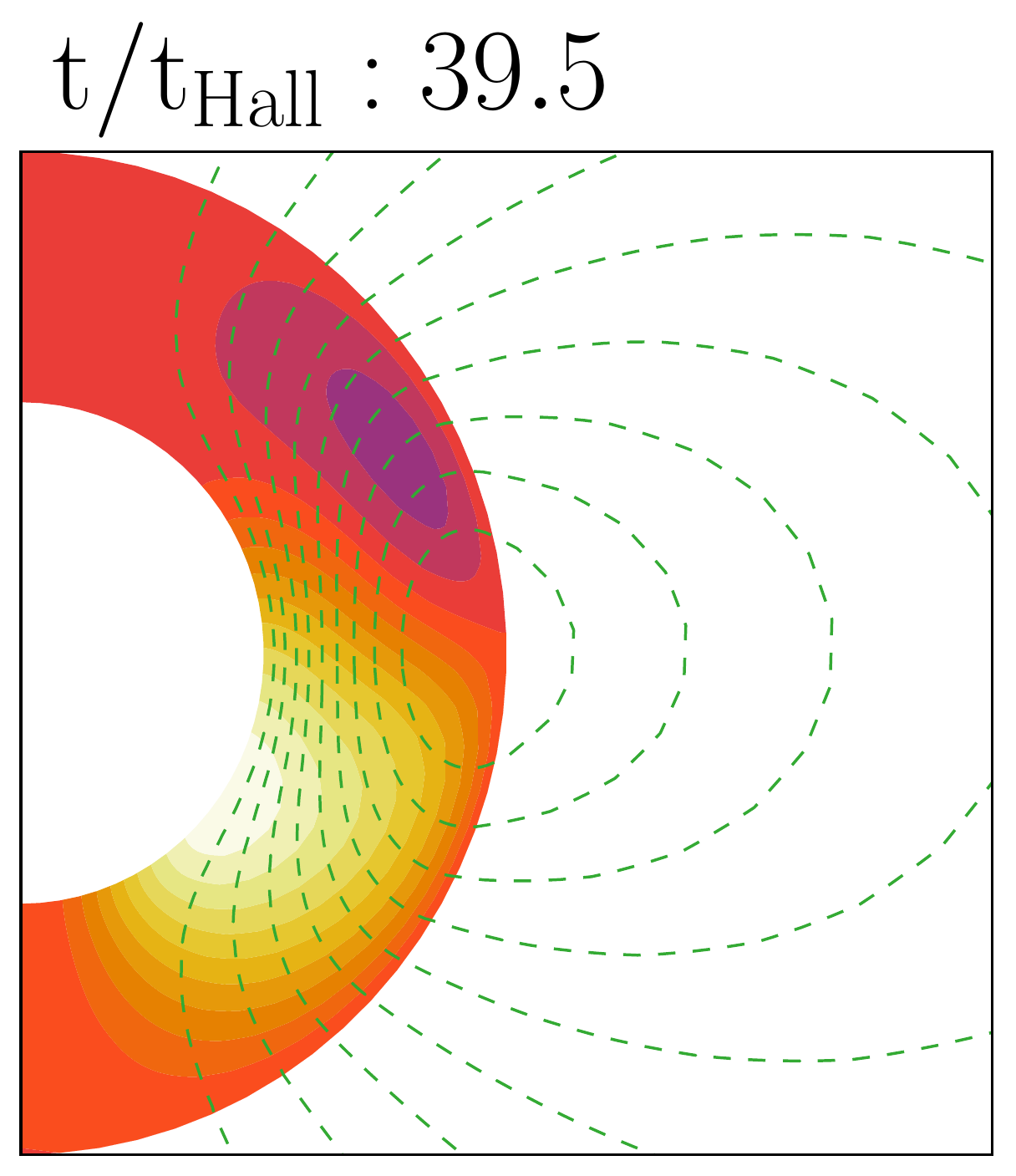}
\includegraphics[scale=0.22]{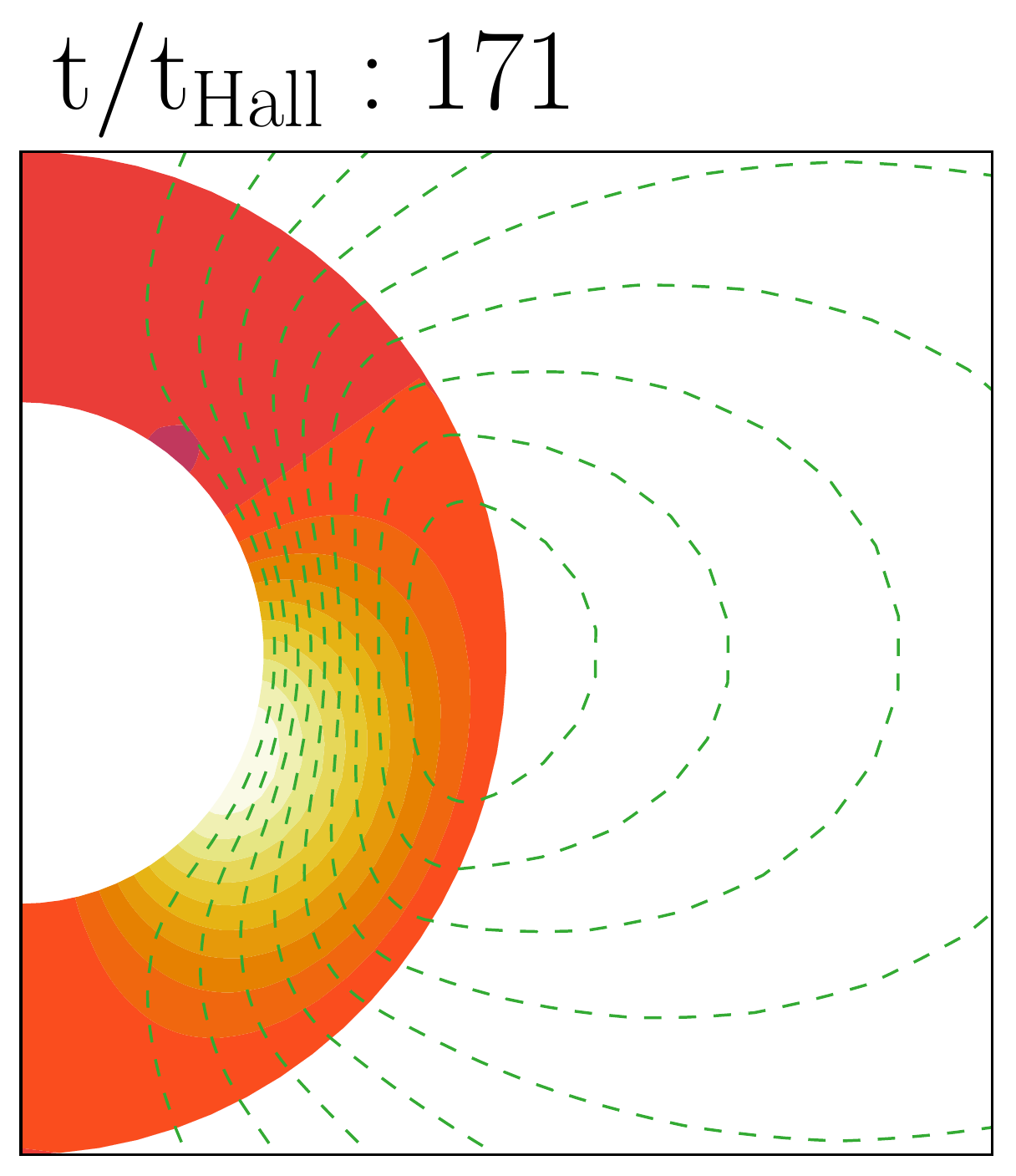}
\includegraphics[scale=0.22]{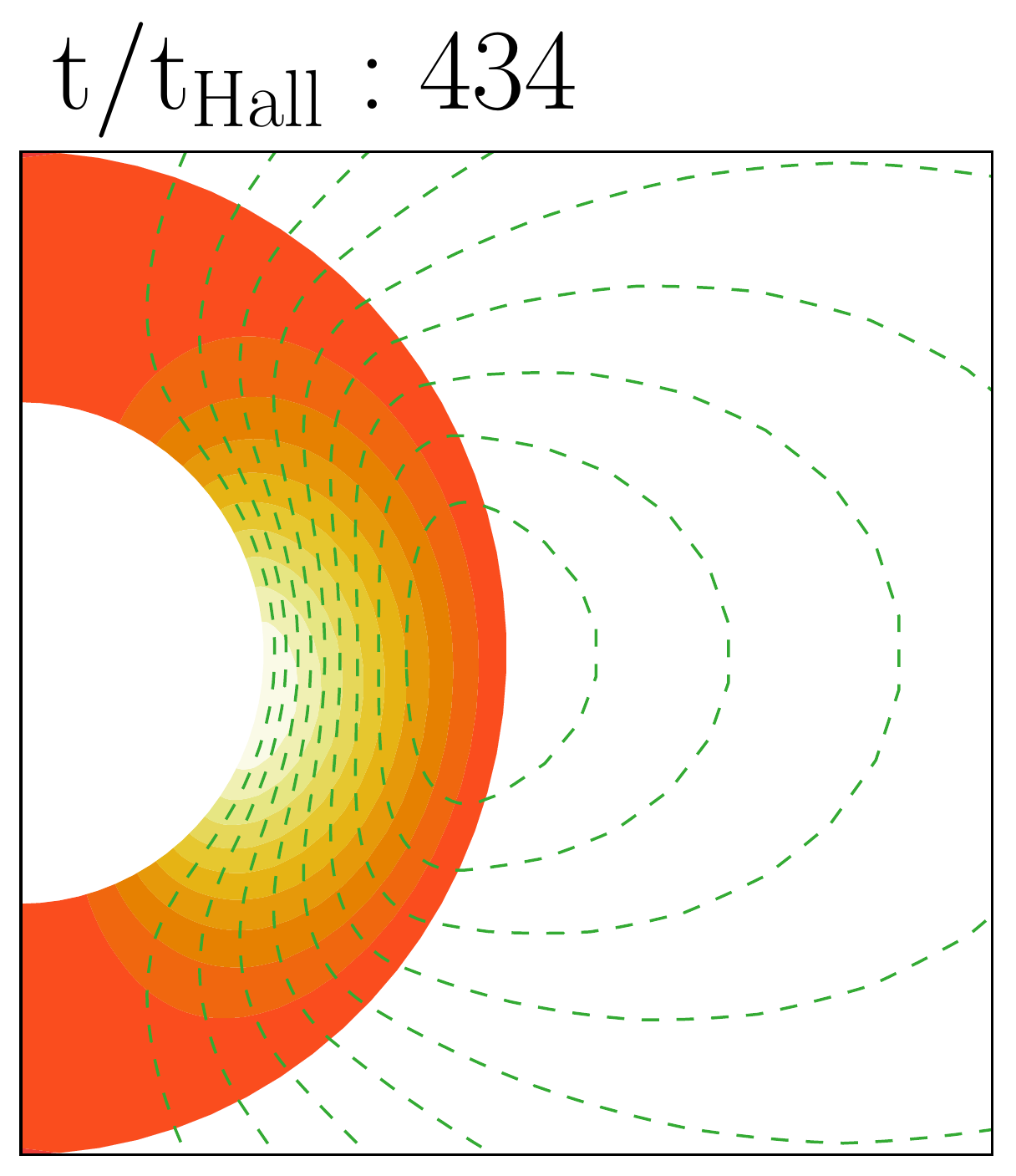}
\includegraphics[scale=0.22]{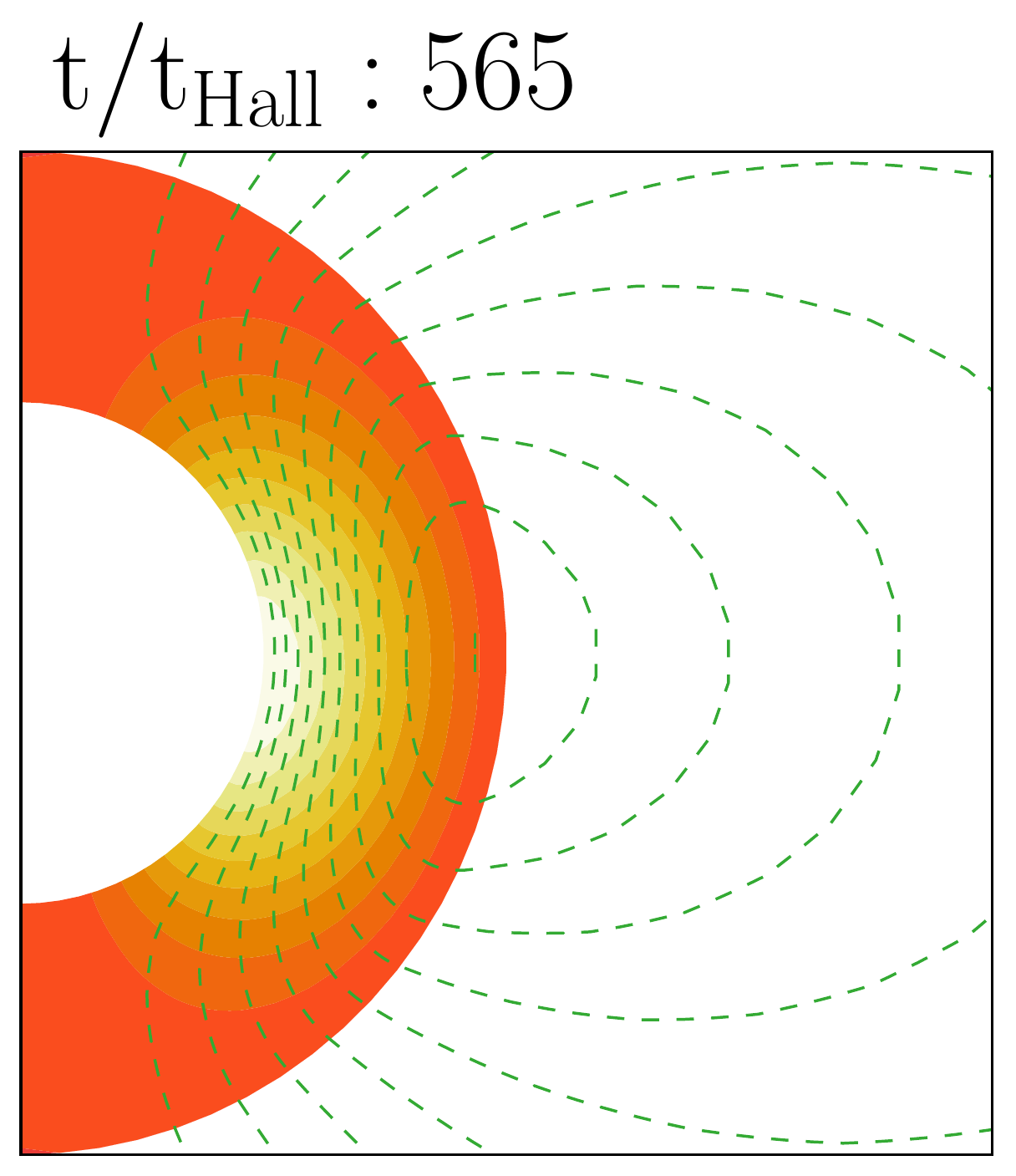}
\includegraphics[scale=0.8]{colorbar}
\end{center}
\caption{Evolution of the poloidally dominated field of \S \ref{anal::meissner} with Meissner boundary conditions. Refer to the caption of Fig. \ref{anal::poldomsim} for details.}\label{anal::meissnersim}
\end{figure}

Fig. \ref{anal::meissner_comp} shows that the evolution on short timescales is nearly identical for both kinds of boundary conditions. However, for longer timescales, where Ohmic dissipation becomes the dominant effect, simulations with Meissner boundary conditions evolve to predominantly toroidal configurations, which consist of a combination of the fundamental poloidal and toroidal Ohm mode as shown in Fig. \ref{anal::meissnersim}. This is completely different to the case with zero boundary conditions, where Ohmic dissipation drives the field to a poloidally dominated configuration. This is due to the decay of the fundamental poloidal mode being faster than that of the fundamental toroidal mode under Meissner boundary conditions (see table \ref{appendix::Ohmtable} in Appendix \ref{appendix::ohmmodes}). The toroidal field here takes a significantly longer time to settle into a final configuration than the simulations with zero boundary conditions, because this phase depends on the small difference between the decay rates of the fundamental toroidal mode and the toroidal quadrupole.

Unlike the simulations with zero boundary conditions, where the poloidal field acting on itself caused the toroidal quadrupole to remain in the later phases, the dominant toroidal mode will not produce a higher poloidal multipole, since the Hall term in the equation for $\pp\alpha/\pp t$ depends on the intensities of both the poloidal and the toroidal field and thus it will necessarily become smaller than the Ohmic term given enough time.

\section{Stability of Hall equilibria}\label{ch::equil}
A Hall equilibrium is defined as a field configuration for which the non-linear Hall drift term in Eq. (\ref{intro::timeeq}) is exactly equal to zero. As shown by \citet{gou+13a}, the functions $\alpha$ and $\beta$ describe a Hall equilibrium field if and only if $\beta$ is a function of $\alpha$ and
\begin{eqnarray}
\Delta^*\alpha+\beta\beta'=F(\alpha)nr^2\sin^2\theta,\label{intro:equilanal}
\end{eqnarray}
where $\beta'=\dd\beta/\dd\alpha$, for an arbitrary function $F(\alpha)$. In this work we will only analyse two equilibrium fields, which are solutions to this equation with the choice $F(\alpha)=F_0=$ constant.
\subsection{Stability of purely poloidal equilibrium}\label{ch::equilpol}
For $\beta=0$, \citet{gou+13a} obtained a general analytic solution of Eq. (\ref{intro:equilanal}) for the case of a spherically symmetric electron density, and field contained in a shell. In the case of uniform electron density, with the boundary condition $\alpha(r_{min},\theta)=0$ together with the continuity of the field across the surface, their equilibrium field is purely poloidal and given by $\alpha=f(r)\sin^2\theta$, where
\begin{eqnarray}
f(r)=\frac{F_0nR^4}{30}\left([5x_{min}^3-3x_{min}^5]/x+3x^4-5x^2\right),\label{intro::polequil}
\end{eqnarray}
with $x=r/R$ and $x_{min}=r_{min}/R$. The choice of $F_0$ is arbitrary and sets the intensity and direction of the field. We perform simulations for this field with $x_{min}=0.75$ for $R_B=100,200,400$.

As this equilibrium field satisfies the zero boundary conditions but not the Meissner ones at the core-crust interface, we use zero boundary condition when evolving this field. As mentioned in \S \ref{anal::meissner}, the use of Meissner boundary conditions does not seem to change the early evolution significantly, and only becomes evident at later stages when Ohmic decay becomes dominant. Thus, the choice of boundary conditions is not expected to play an important role in the stability of equilibria.

As the field is affected by Ohmic decay its structure will be modified, driving it out of equilibrium, and thus, acting as a perturbation. The simplest test that can be done to see if Hall drift plays an important role in modifying the structure of the field is comparing its evolution with and without Hall drift, and checking whether this enhances or not the decay of the field. To properly measure this enhancement, we consider the instantaneous decay timescale of the magnetic energy defined as
\begin{eqnarray}
\tau\equiv\left(\frac{1}{E}\frac{\dd E}{\dd t}\right)^{-1},\; \frac{\dd E}{\dd t}=-\frac{(\eta R_B)^{-1}}{4\pi}\int_V\boldsymbol{j}^2\dd V,\label{anal::deceq}
\end{eqnarray}
where the last expression is from \citet{holrud+02}.
\begin{figure}
\begin{center}
\includegraphics[width=\columnwidth]{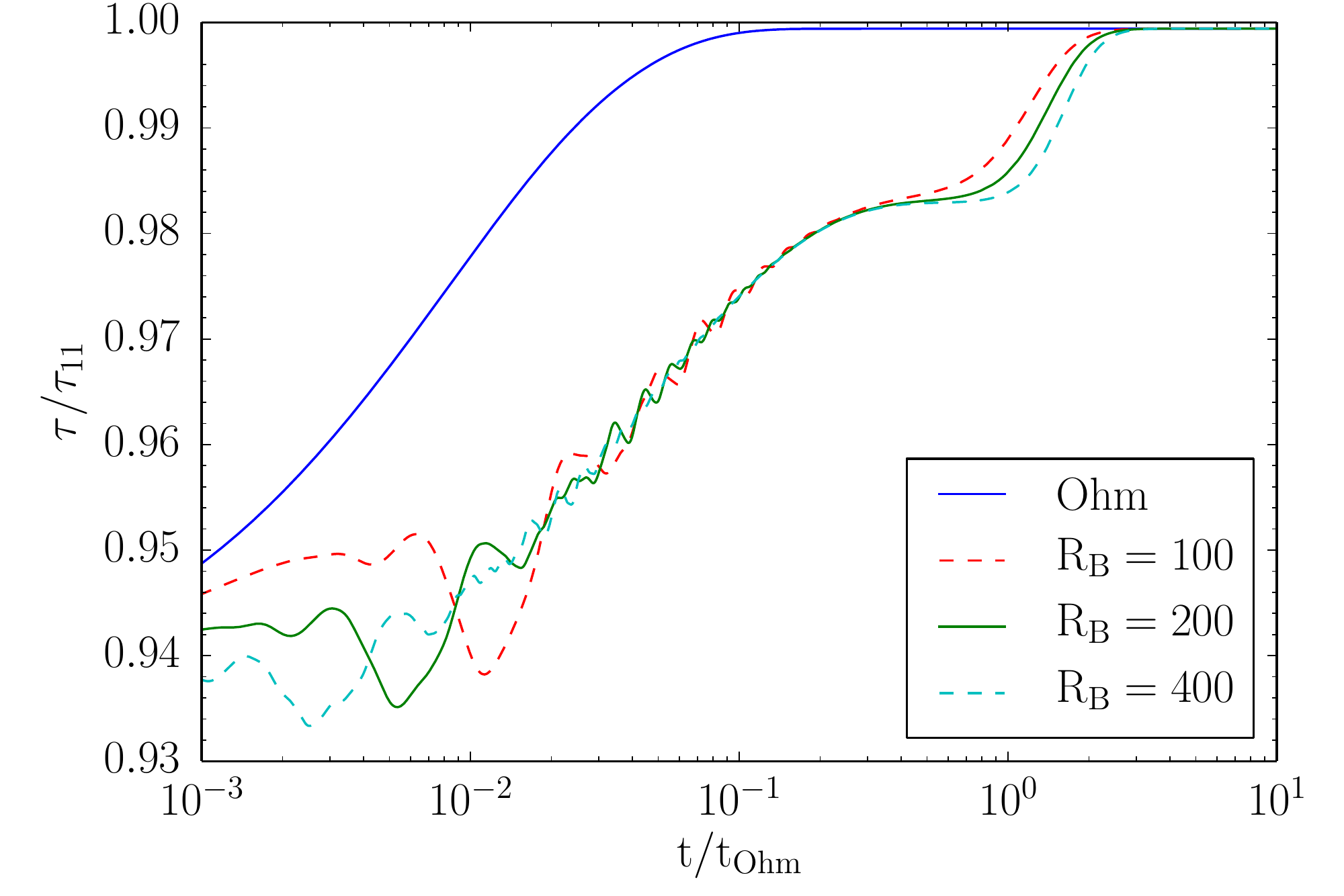}
\end{center}
\caption{Evolution of the instantaneous decay time scale $\tau$ as defined by Eq. (\ref{anal::deceq}), and measured in terms of the decay timescale $\tau_{11}$ of the energy of the fundamental poloidal Ohm mode. Results are shown for $R_B=100,200$ and for evolution through pure Ohmic decay}\label{anal::decaytime}
\end{figure}
This is shown in Fig. \ref{anal::decaytime}, where it is seen that Hall drift provides only a very slight enhancement, which seems to be almost independent of field strength.

In order to quantify the overall change of the equilibrium as it evolves, we consider the time-dependent field $\boldsymbol{B}(\boldsymbol{r},t)$, the initial equilibrium field $\boldsymbol{B}_{eq}(\boldsymbol{r})$, and the fundamental poloidal Ohm mode $\boldsymbol{B}_{11p}(\boldsymbol{r})$ normalized in terms of their characteristic fields $B_0$ as
\begin{eqnarray}
\begin{aligned}
\bhat{B}(\boldsymbol{r},t)&\equiv&\D\boldsymbol{B}(\boldsymbol{r},t)\left(\frac{V_{crust}}{\int_V\left(\boldsymbol{B}(\boldsymbol{r},t)\right)^2\dd V}\right)^{1/2},\\
\bhat{B}_{eq}(\boldsymbol{r})&\equiv&\boldsymbol{B}_{eq}(\boldsymbol{r})\left(\D\frac{V_{crust}}{\int_V\left(\boldsymbol{B}_{eq}(\boldsymbol{r})\right)^2\dd V}\right)^{1/2},\\
\bhat{B}_{11p}(\boldsymbol{r})&\equiv&\boldsymbol{B}_{11p}(\boldsymbol{r})\left(\D\frac{V_{crust}}{\int_V\left(\boldsymbol{B}_{11p}(\boldsymbol{r})\right)^2\dd V}\right)^{1/2},
\end{aligned}\label{anal:eqnorm}
\end{eqnarray}
where $V$ is just as before the volume of all space. The direction of the Ohm field is chosen in such a way that it is equal to the direction of the equilibrium field, i.e., so that $\bhat{B}_{11p}$ and $\bhat{B}_{eq}$ share the same magnetic north pole.

In terms of these normalized fields, we define the quantities
\begin{eqnarray}
\begin{aligned}
\delta_{eq}\equiv&\D\left(\frac{1}{V_{crust}}\int_V(\bhat{B}(\boldsymbol{r},t)-\bhat{B}_{eq}(\boldsymbol{r}))^2\dd V\right)^{1/2},\\
\delta_{ohm}\equiv&\D\left(\frac{1}{V_{crust}}\int_V(\bhat{B}(\boldsymbol{r},t)-\bhat{B}_{11p}(\boldsymbol{r}))^2\dd V\right)^{1/2},
\end{aligned}
\end{eqnarray}
where $V$ is the volume outside the core. These quantities are representative of the difference in shape of the time-evolved field with respect to the initial equilibrium field and the fundamental poloidal Ohm mode, to which the system will eventually decay.

\begin{figure}
\begin{center}
\includegraphics[width=\columnwidth]{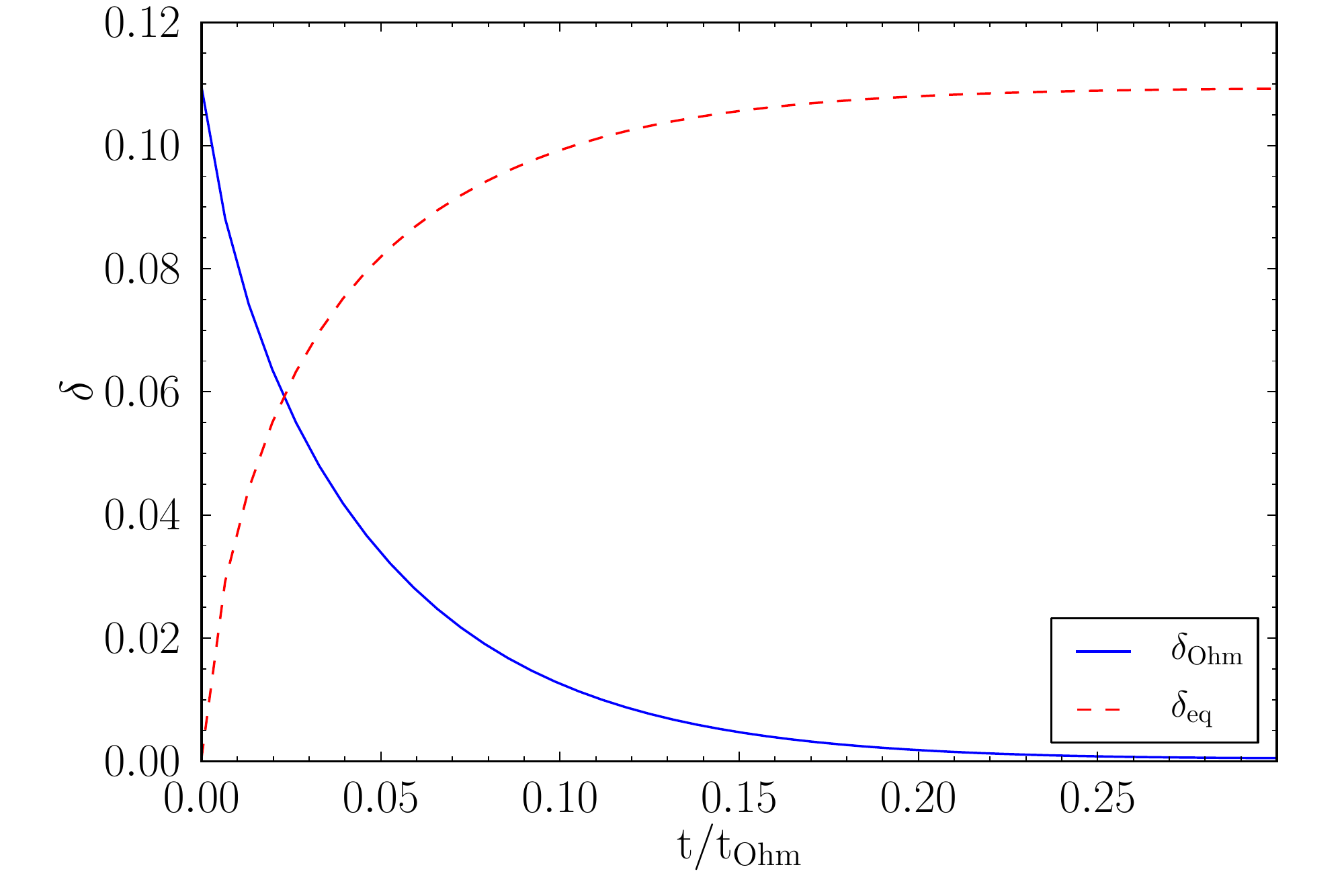}
\end{center}
\caption{Evolution of $\delta_{Ohm}$ and $\delta_{eq}$ in the case with no Hall drift.}\label{anal::deltas_ohm}
\end{figure}

The first thing to do is to check the evolution of $\delta_{eq}$ and $\delta_{Ohm}$ for the equilibrium field subject only to Ohmic dissipation. This is shown in Fig. \ref{anal::deltas_ohm}, where it can be seen that around $t/t_{Ohm}\sim 0.3$ the equilibrium field has essentially decayed to the fundamental Ohm mode. In this case, the higher modes that compose the field rapidly decay, leaving only the fundamental Ohm mode. Thus, $\delta_{Ohm}$ starts with an initial non-zero value, while $\delta_{eq}$ is equal to zero, and with time $\delta_{Ohm}$ asymptotically goes to zero, while $\delta_{eq}$ asymptotically goes to the value $\delta_{Ohm}$ had initially.

\begin{figure}
\begin{center}
\includegraphics[width=\columnwidth]{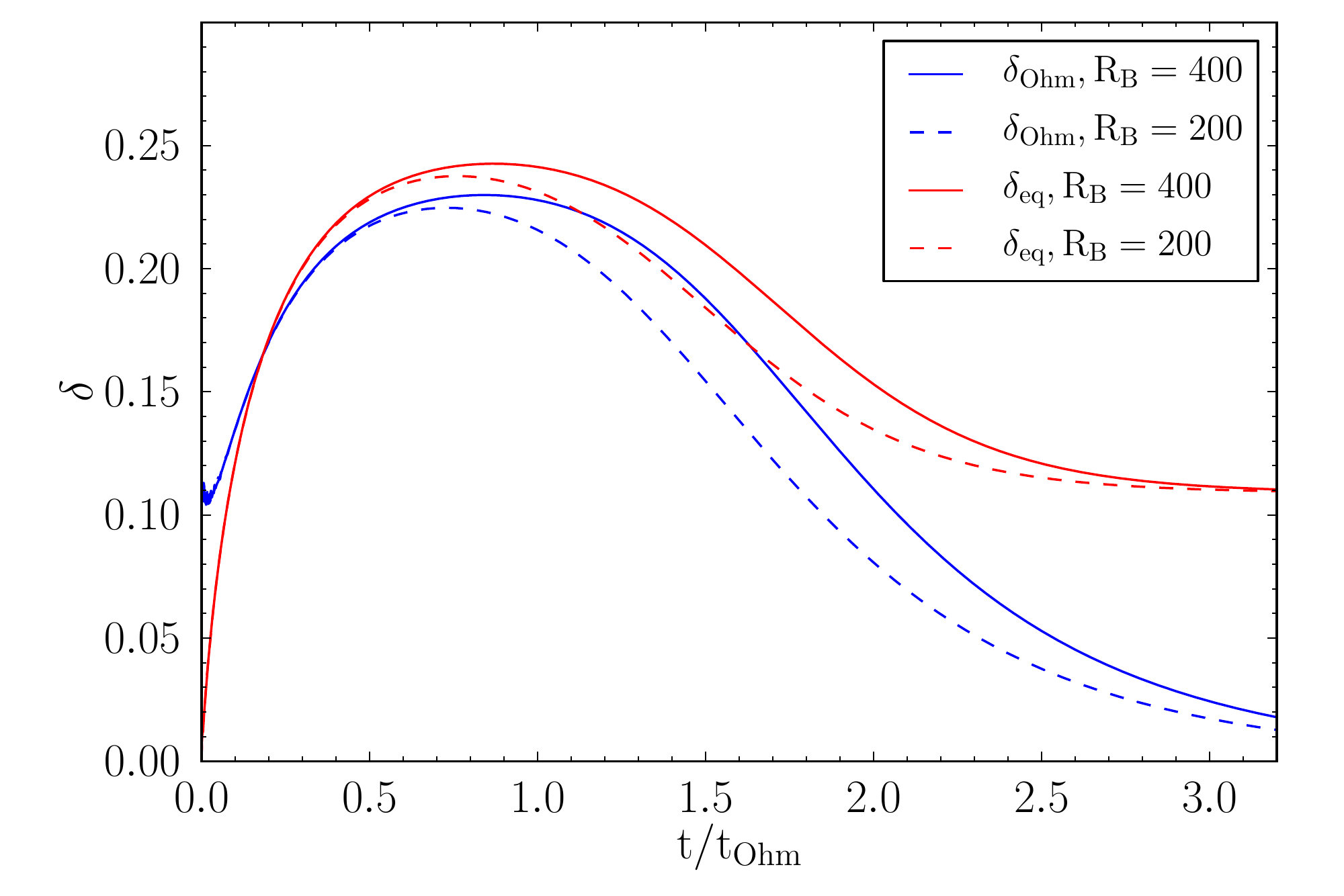}
\includegraphics[width=\columnwidth]{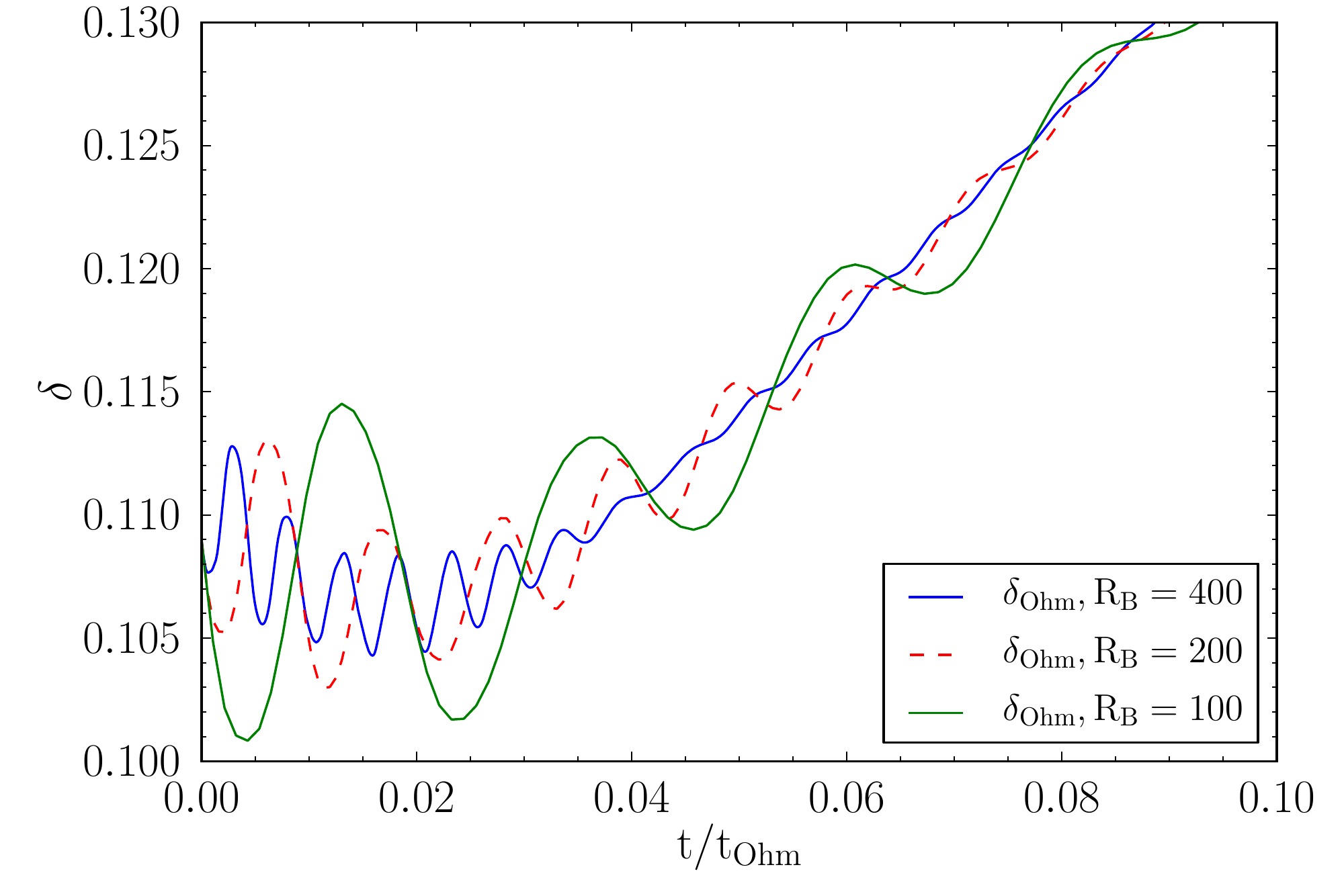}
\end{center}
\caption{Evolution of $\delta_{Ohm}$ and $\delta_{eq}$ in the case with Hall drift, for different values of $R_B=200,400$. The bottom plot shows a close-up to the beginning of the evolution, showing $\delta_{Ohm}$ for $R_B=100,200,400$. The rise scales with $t_{Ohm}$ while the small oscillations at the beginning of the simulation scale with $t_{Hall}$.}\label{anal::deltas}
\end{figure}

\begin{figure}
\begin{center}
\includegraphics[width=\columnwidth]{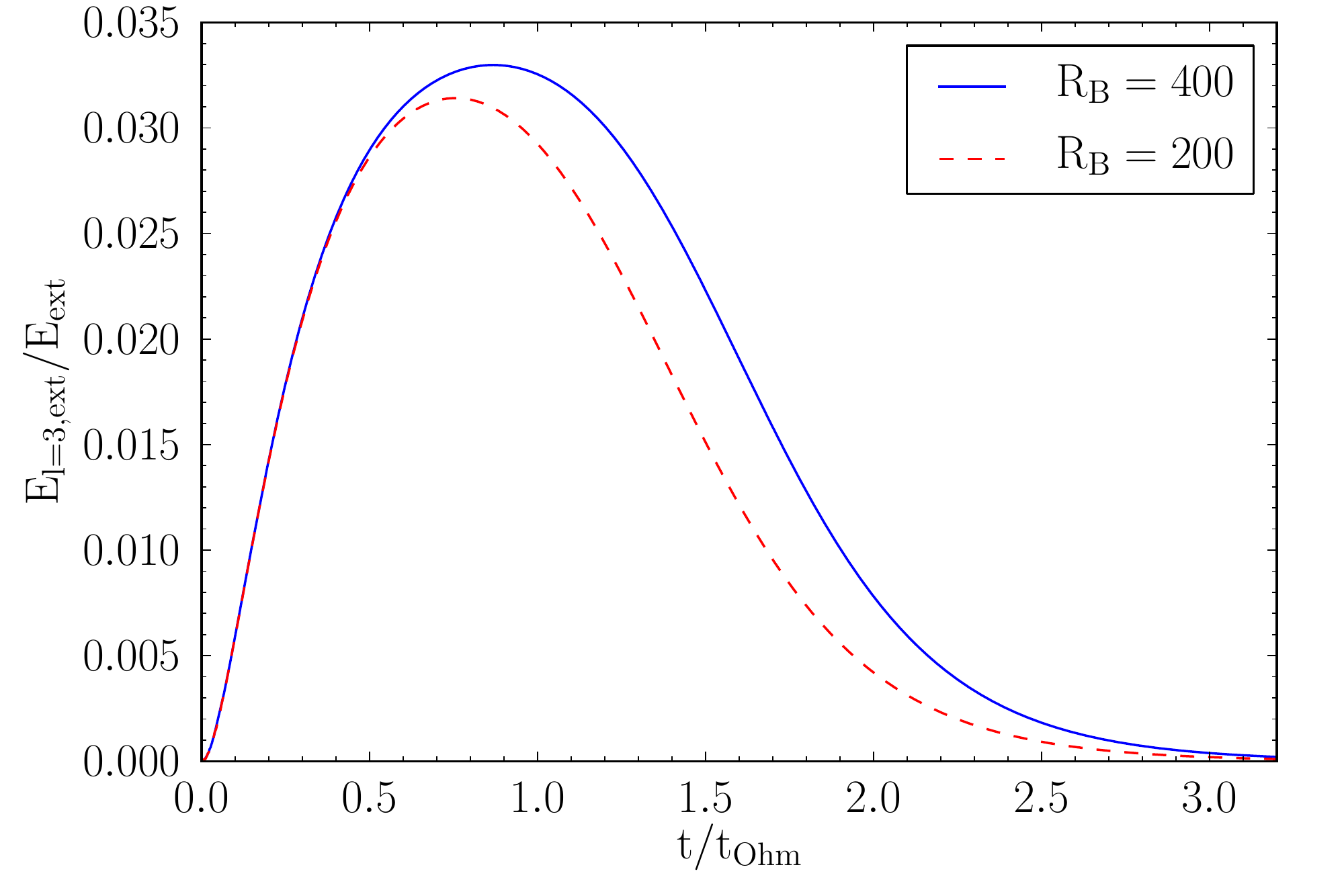}
\end{center}
\caption{Evolution of the ratio of external energy contained in the octupole to total external energy in the simulations with $R_B=100,200$.}\label{anal::octupolarsh}
\end{figure}

Now, adding Hall drift to the picture, Fig. \ref{anal::deltas} shows the evolution of $\delta_{eq}$ and $\delta_{Ohm}$ for different values of $R_B$. The most obvious changes with respect to the previous results are that the asymptotic evolution to the fundamental Ohm mode takes much longer, and that a departure from both fields used as reference happens during the initial stages of evolution. The timescale for this departure scales with $t_{Ohm}$, which means that, as we change $R_B$, this part of the evolution remains nearly unchanged when plotted as a function of $t/t_{Ohm}$, so it is not likely to be an instability driven by Hall drift. Small oscillations can be seen on top of this curve, which gradually decrease their intensity and, most importantly, have periods that scale with $t_{Hall}$. The departure from the equilibrium is small, as significant perturbations to the structure would produce values much closer to unity. By checking the digression of the external field from a dipole, the change from the purely Ohmic evolution can be understood in terms of a transfer of energy from the initial dipole to $l=3$ modes, as seen in Fig. \ref{anal::octupolarsh}.

In order to better understand what this deviation both from the initial equilibrium field and the final fundamental Ohm mode means, we construct a simplified spectral model of the system in terms of the first few Ohmic modes. To properly describe the initial equilibrium field we require at least two $l=1$ poloidal modes, and to take into account the energy transfer to higher modes we use one $l=3$ mode. In addition, a toroidal mode is required in order for Hall drift to transfer energy between different multipoles. Taking all of this into account, we choose to describe the field as
\begin{eqnarray}
\begin{aligned}
\frac{\vec{B}(t,\vec{r})}{B_0}\simeq & a_{11}(t)\bhat{B}_{11p}(\vec{r})+a_{21}(t)\bhat{B}_{21p}(\vec{r})\\
&\qquad\qquad+a_{13}(t)\bhat{B}_{13p}(\vec{r})+b_{12}(t)\bhat{B}_{12t}(\vec{r}),
\end{aligned}
\end{eqnarray}
where $\bhat{B}_{nlp}$ and $\bhat{B}_{nlt}$ are the Ohmic modes for zero boundary conditions at the core-crust interface, with the $p$ and $t$ subscripts denoting poloidal and toroidal modes respectively, $n$ and $ l$ denoting the radial and latitudinal indexes (see Appendix \ref{appendix::ohmmodes}), and normalized in the same way as in Eq. (\ref{anal:eqnorm}). These modes are orthogonal, and the energy of each component is simply its coefficient $a_{nl}$ or $b_{nl}$ squared and multiplied by $B_0^2V_{crust}/(8\pi)$. As a field that initially has $\alpha$ symmetric and $\beta$ antisymmetric will preserve those symmetries \citep{holrud+02}, we excluded in this model toroidal modes with $l=1$ and poloidal modes with $l=2$. These symmetries are explicitly seen in the simulations, so if any symmetry-breaking instability exists, which could be induced because the numerical initial conditions will not be perfectly symmetric (or antisymmetric), it is not observed and thus we ignore it in this analysis.

Just as before, the direction of the fields is relevant, and it should be defined in an unambiguous way. We will consider all the poloidal modes chosen in such a way that they are aligned with the equilibrium field (which means $B_r(r=R,\theta=0)>0$) and the toroidal mode in such a way that right below the surface of the Northern hemisphere ($0<\theta<\pi/2$) the azimuthal component $B_\phi$ is positive. Decomposing the equilibrium field in terms of the Ohm modes so defined, we get
\begin{eqnarray}
\frac{\boldsymbol{B}_{eq}(\boldsymbol{r})}{B_0}\simeq 0.9975 \bhat{B}_{11p}-0.0581\bhat{B}_{21p}+0.0348\bhat{B}_{31p},
\end{eqnarray}
where also the third radial dipole mode was included for comparison. Note that this field is very close to the fundamental poloidal Ohmic mode, so the simulations of \S \ref{ch::poltor} can be thought as the evolution of this equilibrium with different perturbations. From the total energy of this equilibrium $\sim 99.5\%$ is in the fundamental mode, while the other two components shown contain $0.34\%$ and $0.12\%$ of the total energy respectively. It seems reasonable to assume that the remaining components will not affect the evolution significantly. For simplicity we also ignore the $a_{31}$ coefficient in our model.

We now decompose Eq. (\ref{intro::timeeq}) in terms of our Ohmic eigenmodes. The Ohmic term is trivial, but the Hall term requires numerical integrations to obtain all the relevant terms. Since $a_{11}$ is initially much larger than all other terms, and the simulations show that the structure of the field does not change substantially, we ignore all non-linear terms that do not contain this coefficient.
For the initial values $a_{11,i}=0.9975$ and $a_{21,i}=-0.0581$ the Hall term is approximately zero under this approximation, which means that
\begin{eqnarray}
\begin{aligned}
0\simeq&-a_{11,i}^2\nabla\times([\nabla\times\bhat{B}_{11p}]\times\bhat{B}_{11p})\\
&\qquad\qquad-a_{11,i}a_{21,i}\nabla\times([\nabla\times\bhat{B}_{11p}]\times\bhat{B}_{21p}\\
&\qquad\qquad\qquad\qquad\qquad\qquad+[\nabla\times\bhat{B}_{21p}]\times\bhat{B}_{11p}).
\end{aligned}\label{anal::eqmodes2}
\end{eqnarray}
Using this, and defining $\delta$ as
\begin{eqnarray}
\delta\equiv\left(a_{21}-a_{11}\frac{a_{21,i}}{a_{11,i}}\right),
\end{eqnarray}
the Hall term can be written as
\begin{eqnarray}
\begin{aligned}
-\nabla\times\left([\nabla\times\boldsymbol{B}]\times\boldsymbol{B}\right)\simeq\qquad\qquad\qquad\qquad\qquad\quad\qquad\\
-a_{11}\delta\nabla\times([\nabla\times\bhat{B}_{11p}]\times\bhat{B}_{21p}\qquad\qquad\qquad\\
+[\nabla\times\bhat{B}_{21p}]\times\bhat{B}_{11p})\quad\qquad\\
-a_{11}a_{13}\nabla\times([\nabla\times\bhat{B}_{11p}]\times\bhat{B}_{13p}\qquad\\
+[\nabla\times\bhat{B}_{13p}]\times\bhat{B}_{11p})\quad\\
-a_{11}b_{12}\nabla\times([\nabla\times\bhat{B}_{12t}]\times\bhat{B}_{11p}).
\end{aligned}
\end{eqnarray}
We decompose this expression in terms of the Ohmic modes, after which Eq. (\ref{intro::timeeq}) turns into four ordinary differential equations for the coefficients $a_{nl}$ and $b_{nl}$, namely,
\begin{eqnarray}
\begin{aligned}
\dot{a}_{11}=&0.187 a_{11}b_{12}-3.10R_B^{-1}a_{11}\\
\dot{a}_{21}=&2.82 a_{11}b_{12}-22.9R_B^{-1}a_{21}\\
\dot{a}_{13}=&1.36 a_{11}b_{12}-4.66R_B^{-1}a_{13}\\
\dot{b}_{12}=&-2.70 a_{11}\delta-1.10 a_{11}a_{13}-10.4R_B^{-1}b_{12},
\end{aligned}\label{ecu::spec_eq}
\end{eqnarray}
where the thickness of the crust is used as the unit of length. We combine the first two equations in order to produce an equation for $\dot{\delta}$ to use instead of the equation for $\dot{a}_{21}$,
\begin{eqnarray}
\dot{\delta}=\D 2.83a_{11}b_{12}-22.9 R_B^{-1}\delta + 1.51R_B^{-1}a_{11}.
\end{eqnarray}
The final approximation we make is to assume $a_{11}$ constant, which is justified since simulations show that the field never digresses significantly from the initial configuration where most of the energy is contained in this term, and Eq. (\ref{ecu::spec_eq}) shows that the timescale associated with the evolution of $a_{11}$ is the longest one in both the Hall and Ohm dominated regimes. This leaves us with a $3\times3$ inhomogeneous system of differential equations, namely
\begin{eqnarray}
\begin{pmatrix}
\dot{\delta}\\
\dot{a}_{13}\\
\dot{b}_{12}
\end{pmatrix}=
A
\begin{pmatrix}
\delta\\
a_{13}\\
b_{12}
\end{pmatrix}
+
b,
\end{eqnarray}
where
\begin{eqnarray}
\begin{aligned}
A=&\begin{pmatrix}
-22.9 R_B^{-1} & 0 & 2.83a_{11}\\
0 & -4.66 R_B^{-1} & 1.36a_{11}\\
-2.70a_{11} & -1.10a_{11} & -10.37 R_B^{-1}
\end{pmatrix},\\
b=&
\begin{pmatrix}
1.51R_B^{-1}a_{11}\\
0\\
0
\end{pmatrix},
\end{aligned}
\end{eqnarray}
and we neglect the time dependence of $a_{11}$. Note that, without the Ohmic terms (the ones $\propto R_B^{-1}$), the system becomes homogeneous, in which case, with the time-dependence written as $e^{\lambda t}$, it has a continuous family of solutions with eigenvalue $\lambda=0$ describing Hall equilibria,
\begin{eqnarray}
b_{12}=0,\quad \delta = -0.407 a_{13},\label{hall_equils}
\end{eqnarray}
one of which is the initial equilibrium ($a_{13}=\delta=b_{12}=0)$. The other two eigenvalues are imaginary, representing oscillations around these equilibria.

When including Ohmic diffusion, but keeping only the lowest orders in the small parameter $R_B^{-1}$, a particular solution for the inhomogeneous system can be readily obtained by setting $\dot{\delta}=\dot{a}_{13}=\dot{b}_{12}=0$, for which we get a simple linear algebraic equation resulting in
\begin{eqnarray}
\begin{aligned}
\delta=&0.0323 a_{11}+O(R_B^{-1})\\
a_{13}=&-0.0791 a_{11}+O(R_B^{-1})\\
b_{12}=&-0.274 R_B^{-1}+O(R_B^{-2}),
\end{aligned}\label{part_sol}
\end{eqnarray}
which for $R_B\rightarrow\infty$ approaches one particular equilibrium of the family described by Eq. (\ref{hall_equils}). For the homogeneous part, approximate eigenvalues can be obtained to order $R_B^{-1}$,
\begin{eqnarray}
\begin{aligned}
\lambda_1 =& -7.64R_B^{-1}+O(R_B^{-2}),\\
\lambda_\pm=&\pm 3.02 a_{11}i-15.1R_B^{-1}+O(R_B^{-2}),
\end{aligned}
\end{eqnarray}
where the first eigenvalue corresponds to a decay towards the particular solution of Eq. (\ref{part_sol}), while the complex ones correspond to damped oscillations around the different equilibria described by Eq. (\ref{hall_equils}). Using the initial value of $a_{11}$, the imaginary part results in a period of $2.09t_{Hall}$, which is very close to the actual period of the oscillations in the simulation. The coupled effect of Hall drift and Ohmic decay can then be understood as a drift through a continuum of Hall equilibria, driven by Ohmic decay, towards the attractor configuration given by Eq. (\ref{part_sol}), which allows us to explain the part of the evolution that scales with $t_{Ohm}$ in Fig. \ref{anal::deltas}. This attractor consists of a dominant dipolar poloidal field, coupled to a counter-aligned octupole through a weak toroidal quadrupole (with amplitude $\propto R_B^{-1}$). These properties are exactly the same as those described for the attractor of \citet{gou+13c}.

\subsection{Stability of poloidal+toroidal confined equilibrium}\label{ch::equilpoltor}
\citet{gou+13a} also provide a solution for Eq. \ref{intro:equilanal} with non-zero toroidal field, which assumes $\alpha=f(r)\sin^2\theta$, a functional dependence $\beta=s\alpha$, with $s$ constant, and field lines contained inside the shell, which is equivalent to the boundary condition $\alpha(R,\theta)=0$. Applying also the boundary conditions $\alpha(r_{min},\theta)=0$ and $\pp\alpha/\pp r=0$ at the surface ($r=R$) to avoid surface currents, gives the following solutions for uniform electron density
\begin{eqnarray}
\begin{aligned}
f(r)=&F_0nR^4\left(a\left[\frac{\sin(sr)}{sr}-\cos(sr)\right]\right.\\
&\left.\qquad\qquad+b\left[\frac{\cos(sr)}{sr}+\sin(sr)\right]-\frac{(sr)^2}{(sR)^4}\right),
\end{aligned}
\end{eqnarray}
with $s=20.9R^{-1}=5.23L^{-1}$, $a=0.00135$, and $b=0.00185$ being the solution with the smallest value of $s$ for the choice $r_{min}=0.75R$. Just as in the purely poloidal case, the choice of $F_0$ sets the strength and orientation of the field.

We evolve this field keeping it confined to the crust by applying zero boundary conditions both at the crust-core interface and the surface of the star. In principle, we could allow the field to escape the star due to Ohmic dissipation, but this turns out to be numerically unstable. Note that the condition $(\pp\alpha/\pp r)_{r=R}=0$ is not enforced throughout the evolution, so surface currents can (and do) develop.

Simulations for this initial condition turn out to decay significantly during the first few $t_{Hall}$, quickly reducing the value of $R_B$ down to a point where Ohmic decay becomes significant compared to Hall drift. This suggests the initial field is unstable, but to further test this, we perform different simulations, in which we rescale the field at each timestep to keep its energy at its initial value and thus yielding a constant value of $R_B$ (which otherwise would decay $\propto B_0\propto\sqrt{E}$).

\begin{figure}
\begin{center}
\includegraphics[scale=0.22]{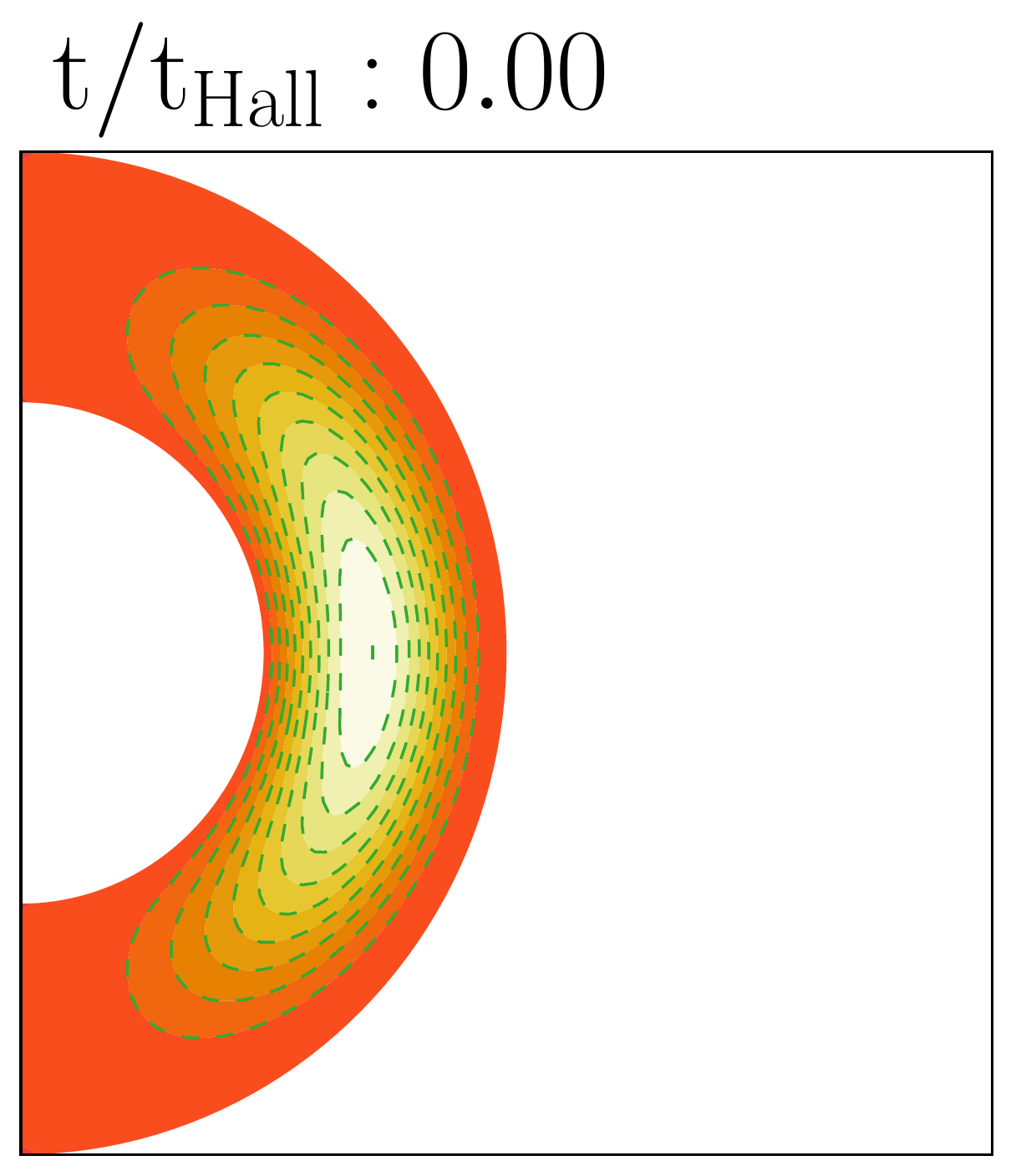}
\includegraphics[scale=0.22]{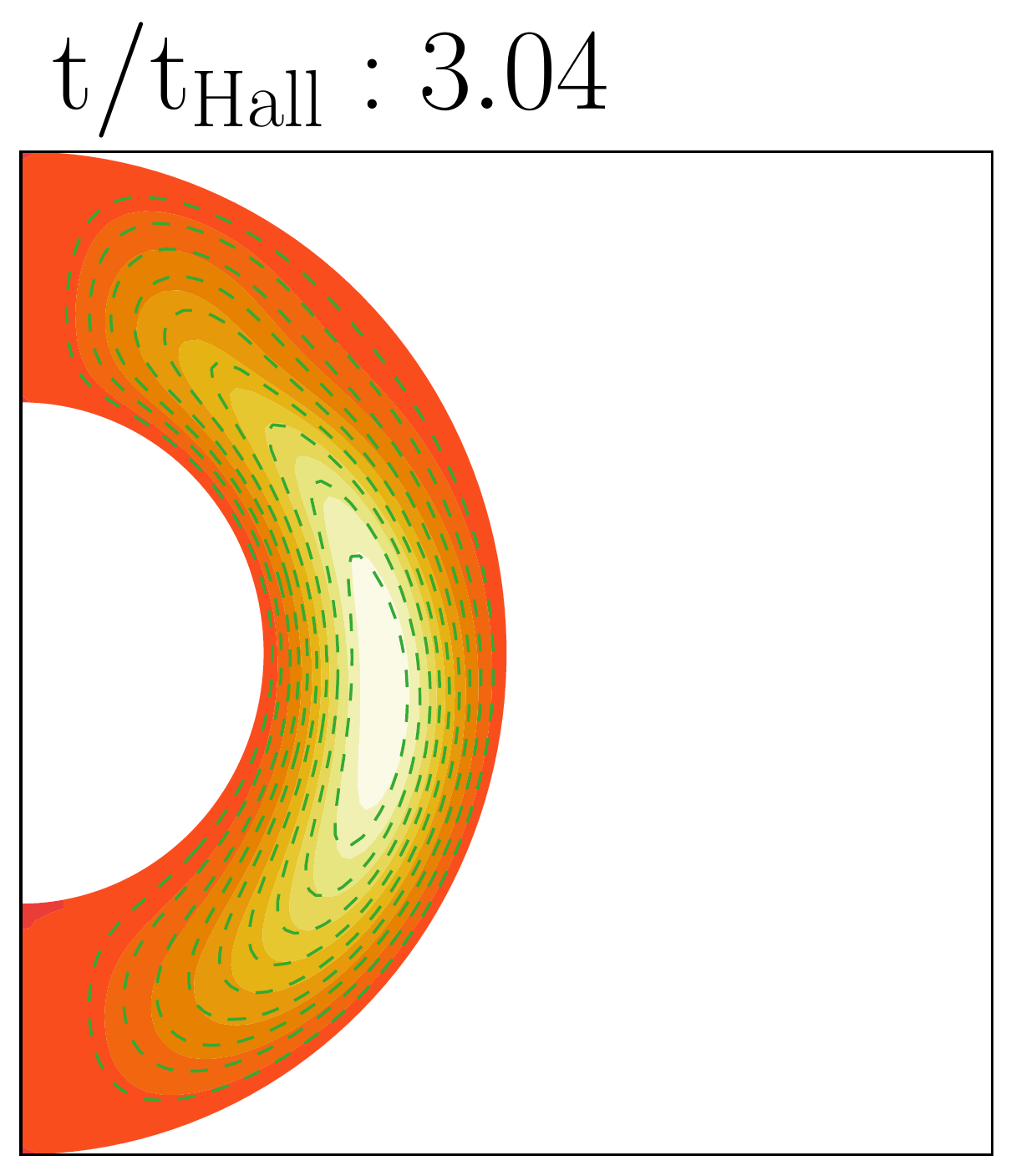}
\includegraphics[scale=0.22]{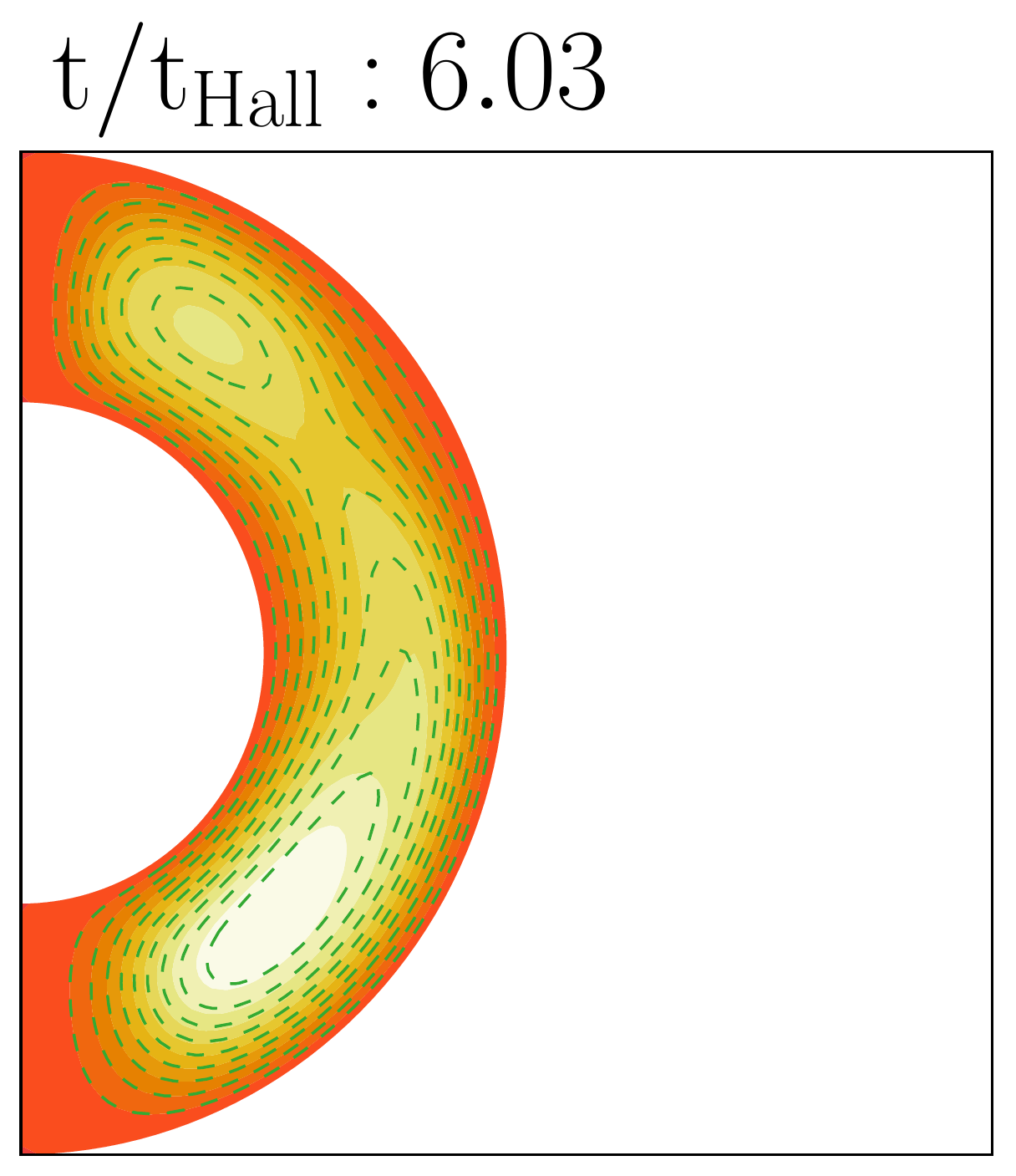}
\includegraphics[scale=0.22]{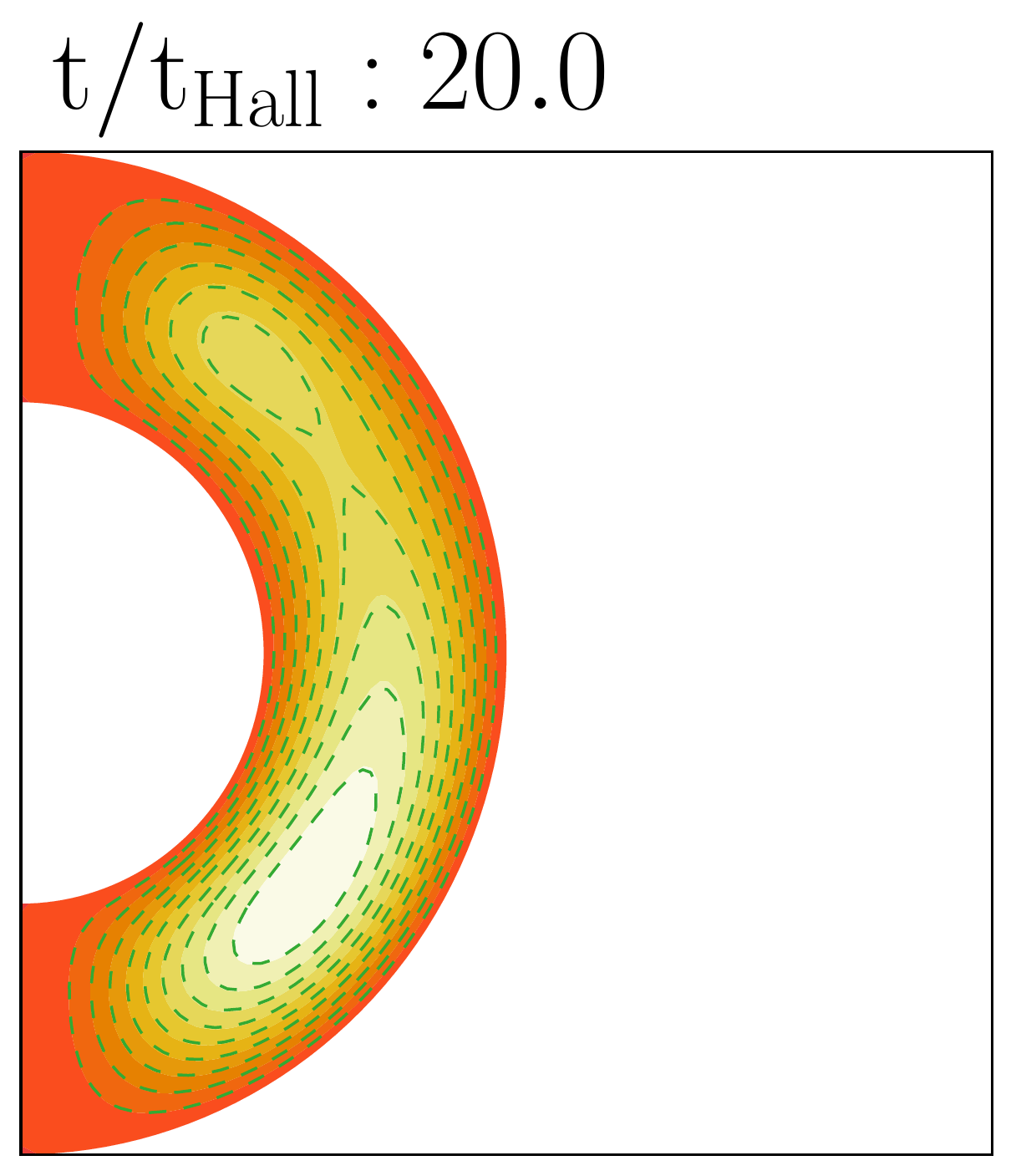}
\includegraphics[scale=0.22]{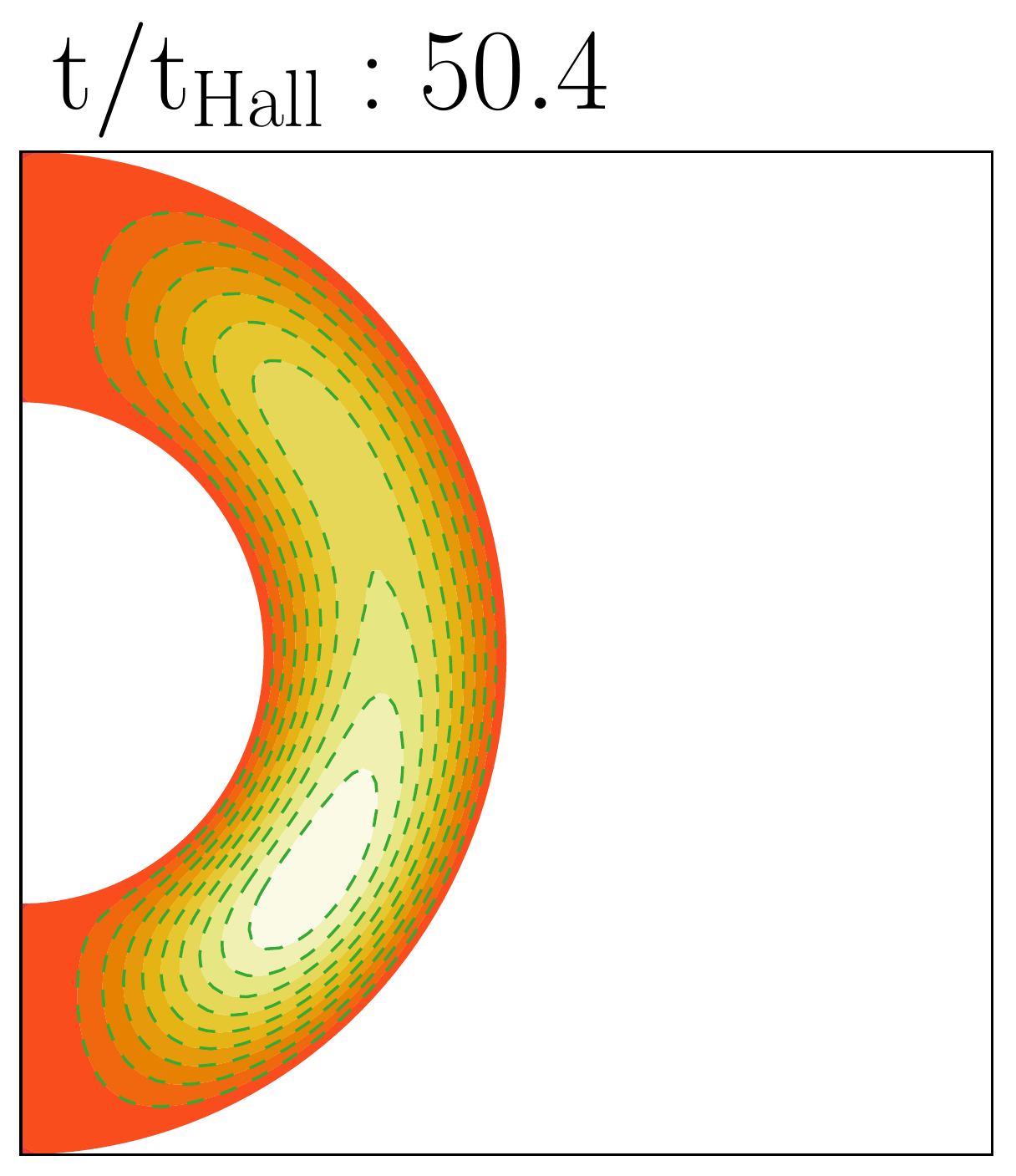}
\includegraphics[scale=0.22]{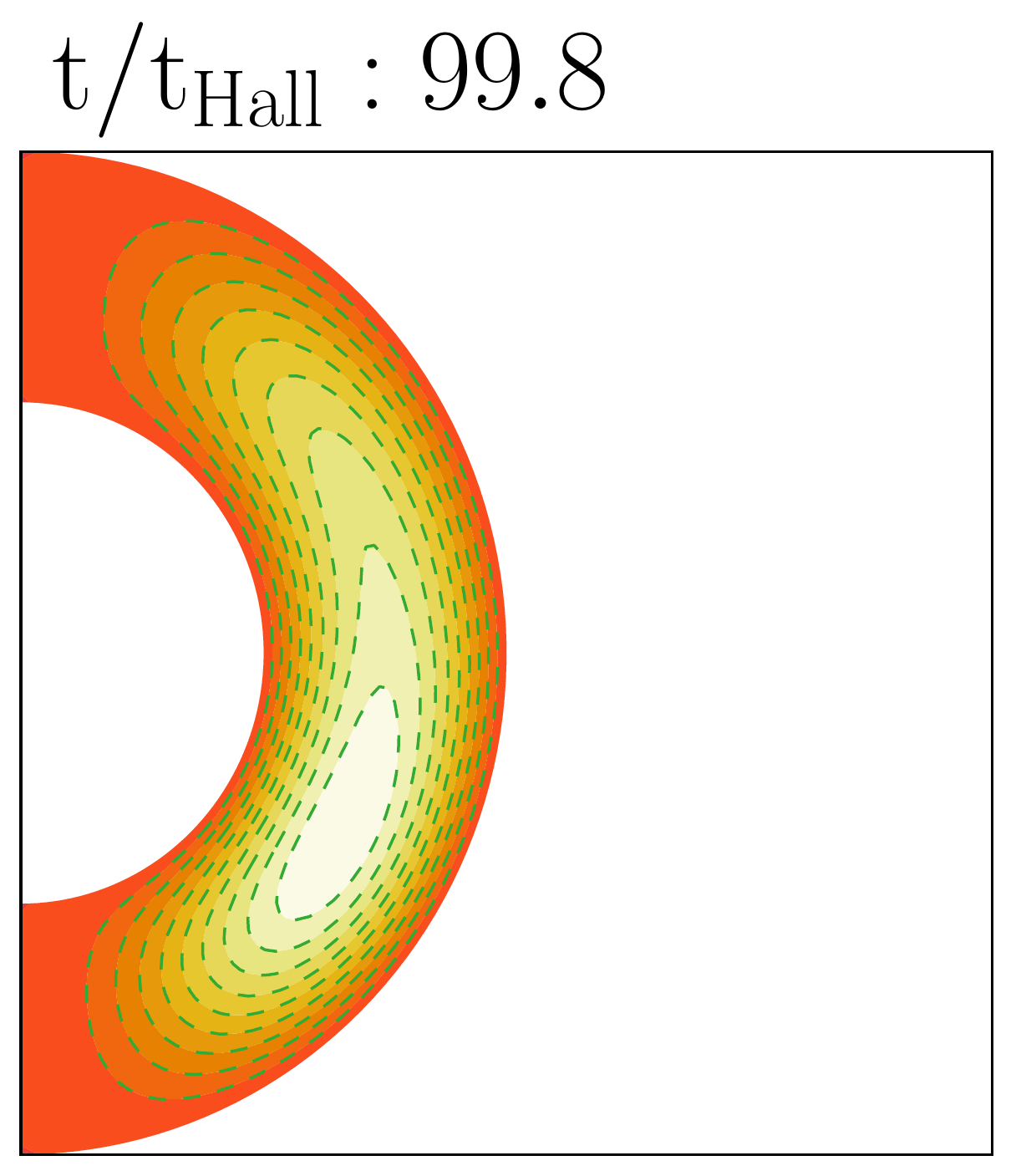}
\includegraphics[scale=0.22]{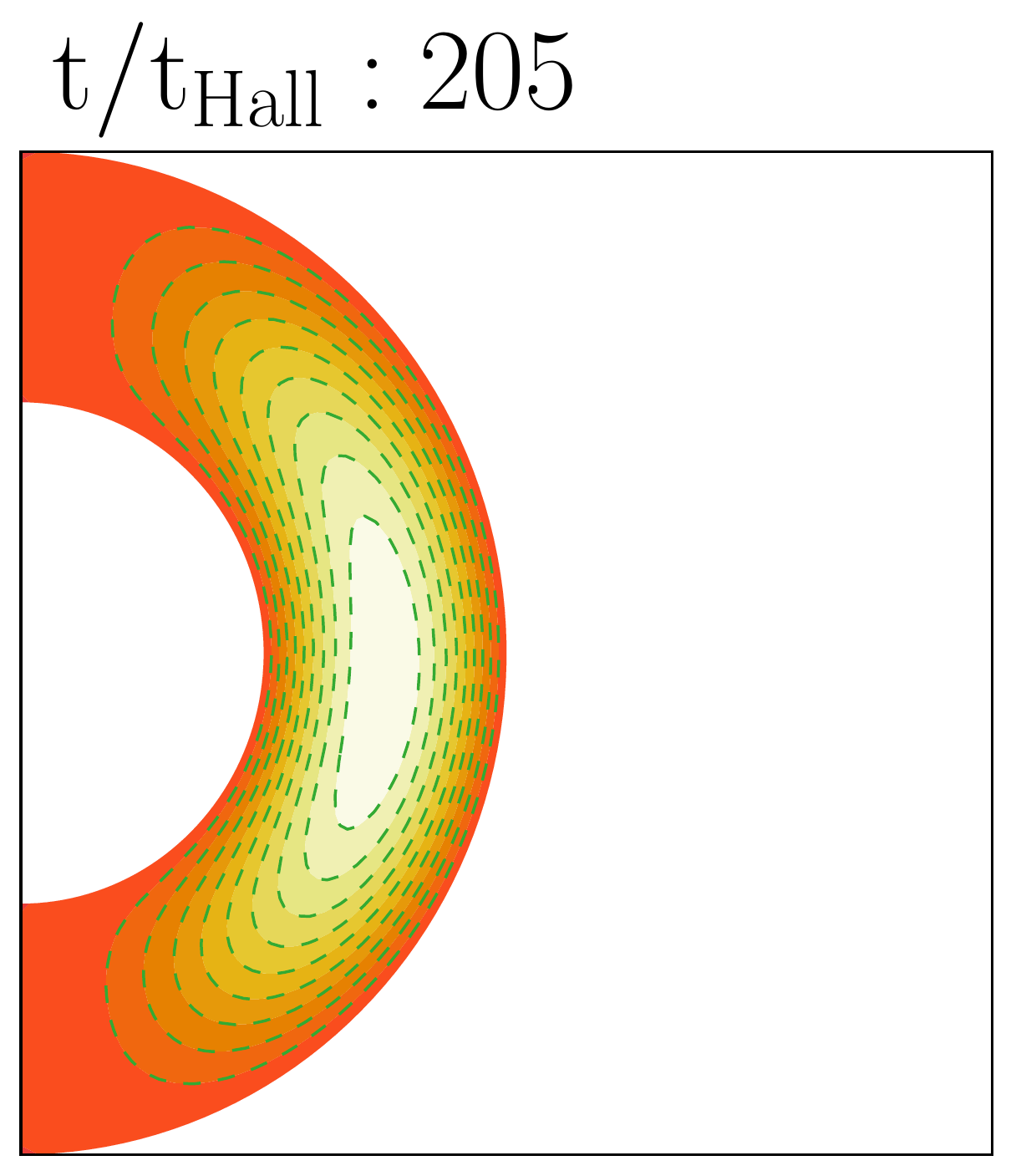}
\includegraphics[scale=0.22]{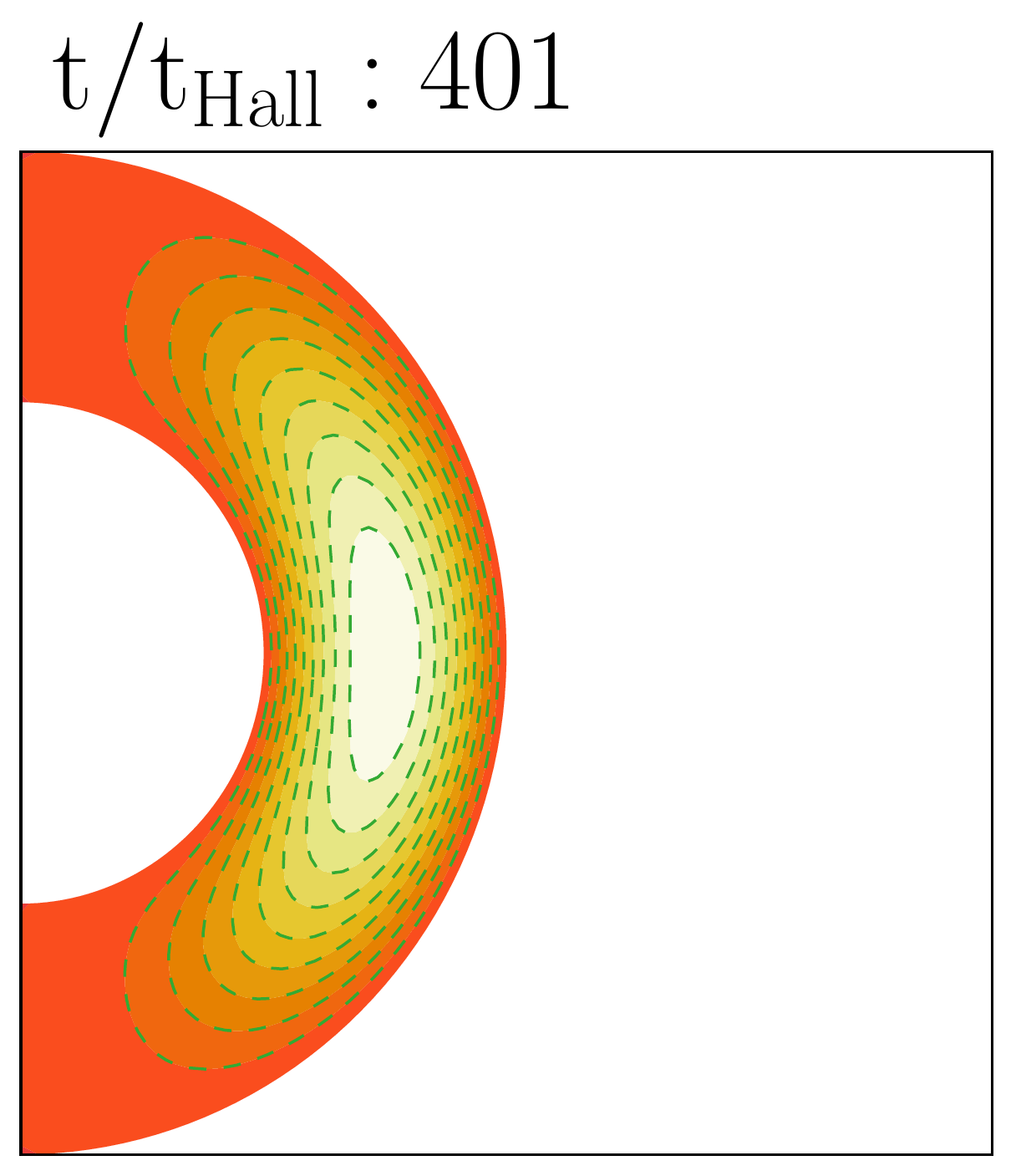}
\includegraphics[scale=0.22]{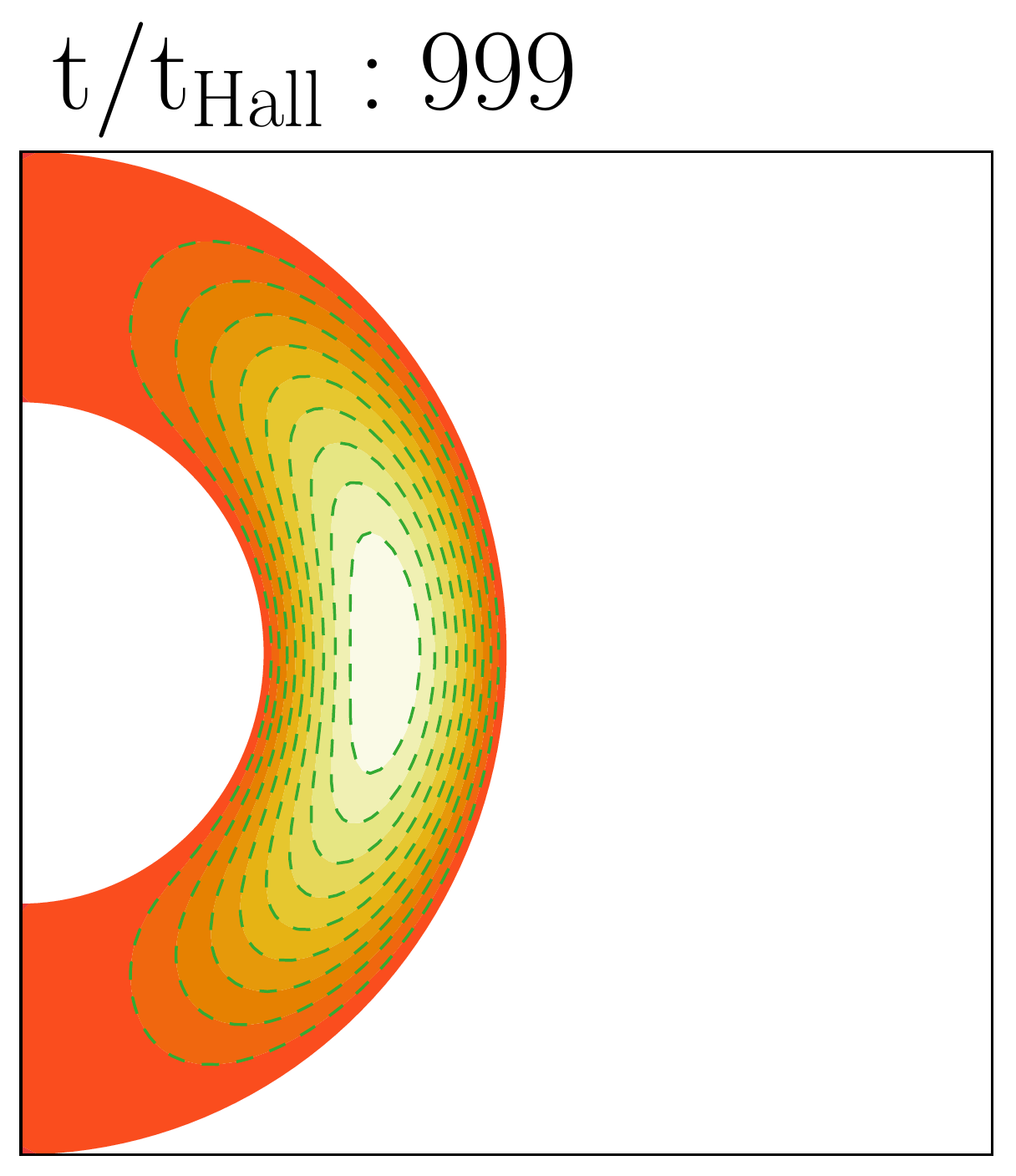}
\includegraphics[scale=0.8]{colorbar}
\end{center}
\caption{Evolution of the poloidal+toroidal equilibrium field for $R_B=100$, applying zero boundary conditions both at the core-crust interface and the surface. In this simulation, the field is rescaled at each timestep in order to keep it at a constant energy. Refer to the caption of Fig. \ref{anal::poldomsim} for details.}\label{anal::ptequilevol}
\end{figure}

\begin{figure}
\begin{center}
\includegraphics[width=\columnwidth]{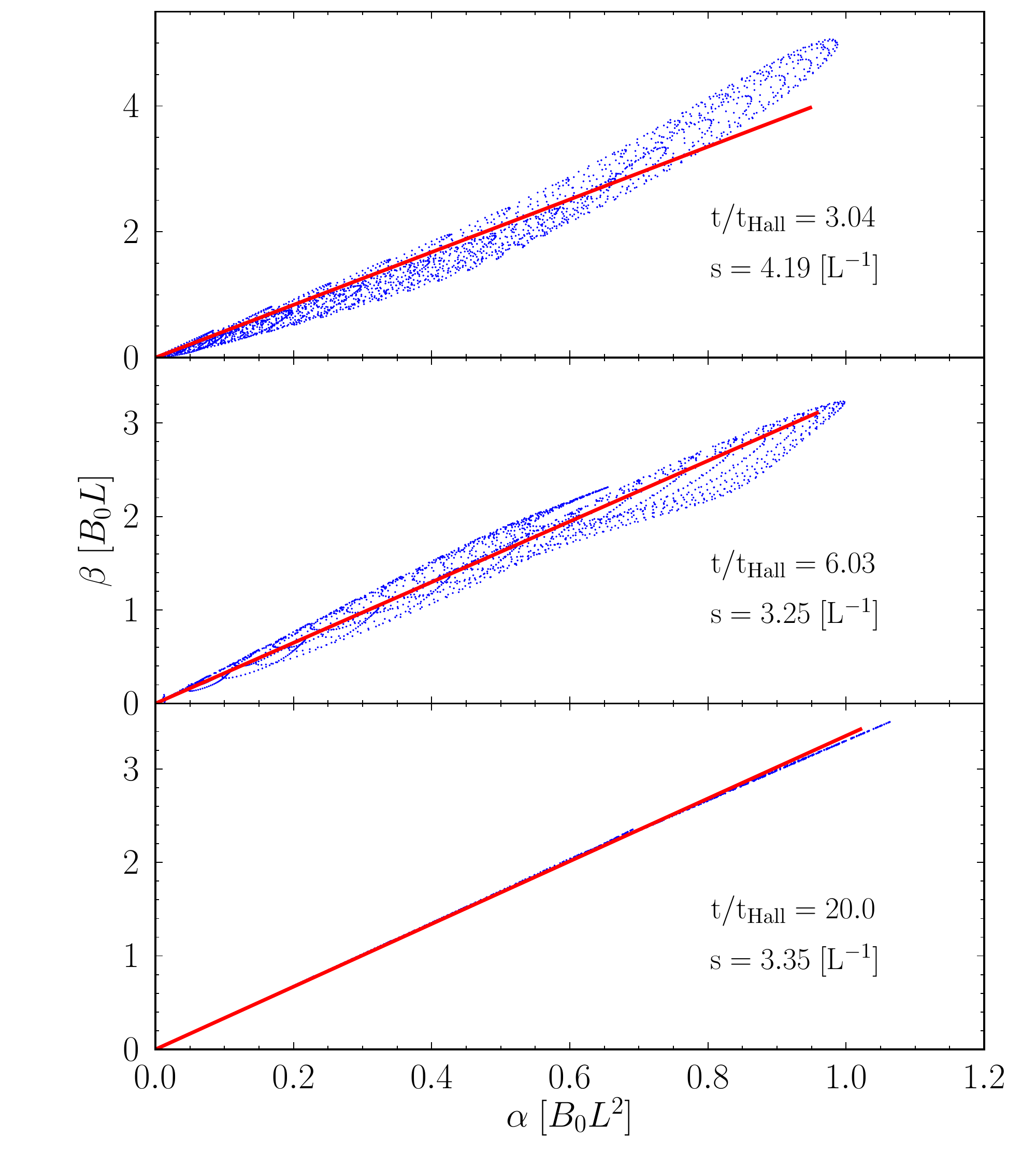}
\end{center}
\caption{Values of $\beta$ plotted against values of $\alpha$ marked with blue dots for each grid point for a snapshot in the simulation of the poloidal+toroidal equilibrium with $R_B=100$. In this simulation, the field is rescaled at each timestep in order to keep it at a constant energy. The red line represents $s\alpha$, where $s=\sum \beta/\sum \alpha$.}\label{anal::balpha_ptequil}
\end{figure}

The evolution of this equilibrium field, including the rescaling of the energy, is shown in Fig. \ref{anal::ptequilevol} for $R_B=100$, where it is seen that it evolves to a configuration that is asymmetric with respect to the equator, but then eventually becomes symmetric again. In this final configuration both the toroidal and poloidal fields are in their fundamental modes (see Appendix \ref{appendix::ohmmodes}), even though $R_B$ is kept high so Hall drift remains the dominant process. As illustrated in Fig. \ref{anal::balpha_ptequil}, the field first evolves away from a simple functional relationship $\beta=\beta(\alpha)$, which is a necessary condition for equilibrium, and afterwards goes into a very tight linear relationship with $\beta\propto\alpha$. This suggests that the field is initially driven away from the equilibrium configuration, and settles into a different one afterwards.

\begin{figure}
\begin{center}
\includegraphics[width=\columnwidth]{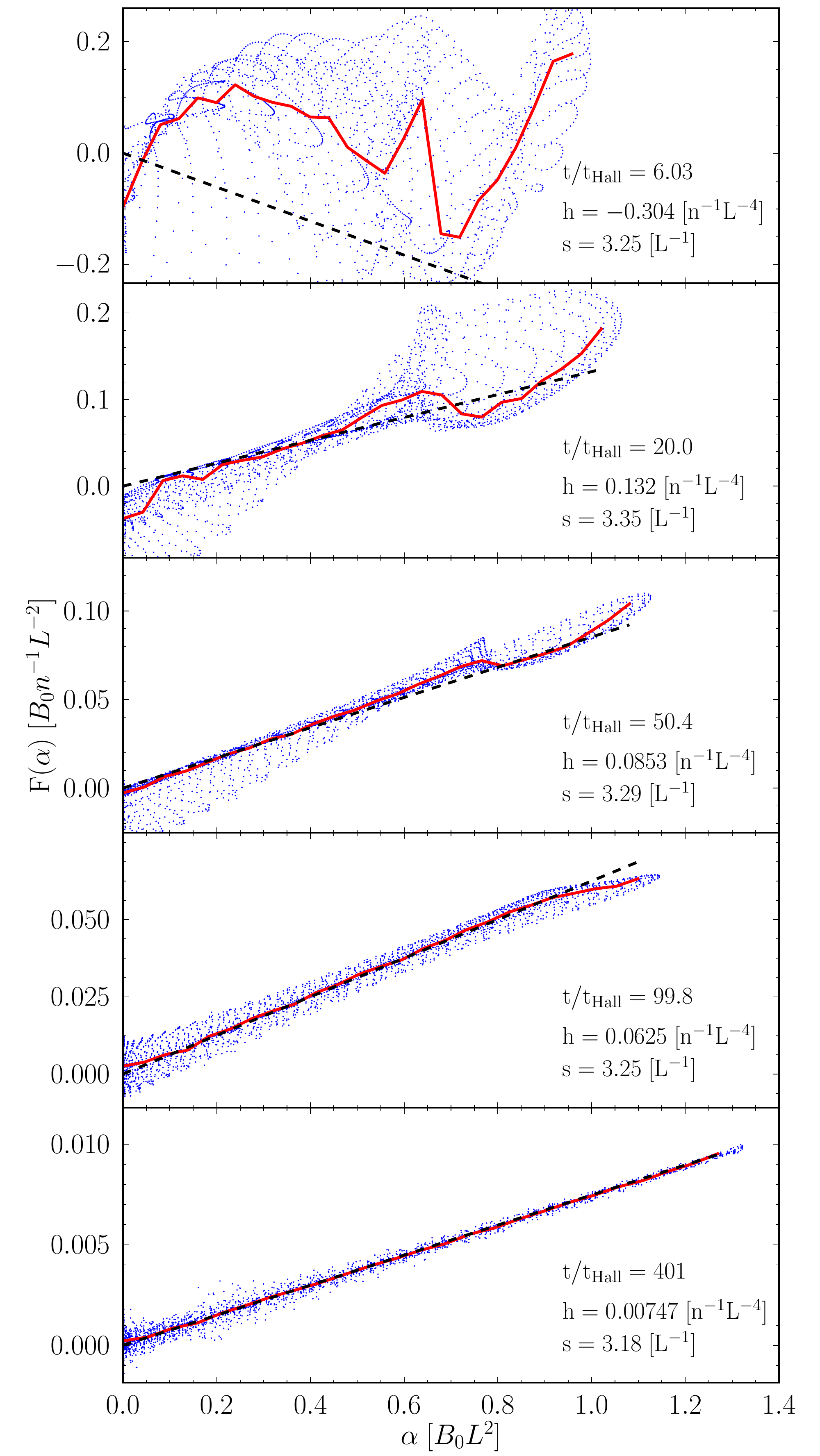}
\end{center}
\caption{Value of $F(\alpha)=(\Delta^*\alpha+s^2\alpha)/(nr^2\sin^2\theta)$ marked with blue dots for each grid point at different times in a simulation of the poloidal+toroidal equilibrium for $R_B=100$. In this simulation, the field is rescaled at each timestep in order to keep it at a constant energy. The red solid line is given by the median values of $F(\alpha)$ in bins of size $\Delta\alpha=\max(|\alpha|)/25$, and the black dashed line is given by $h\alpha$, where $h=\sum F(\alpha)/\sum \alpha$}.\label{anal::func_ptequil}
\end{figure}

In order to check this, we consider that an equilibrium should be a solution of the Grad-Shafranov equation (\ref{intro:equilanal}) for an arbitrary function $F(\alpha)$. As the field evolves into a configuration for which $\beta\propto\alpha$, we assume $\beta=s\alpha$, and $s$ is computed in the simulations as $\sum \beta/\sum \alpha$, where the summation is done over all points in the grid. Using this, the Grad-Shafranov equation can be rewritten as
\begin{eqnarray}
\frac{\Delta^*\alpha+s^2\alpha}{nr^2\sin^2\theta}=F(\alpha).
\end{eqnarray}
In Fig. \ref{anal::func_ptequil}, the left-hand side of this equation is plotted against its value of $\alpha$ for each grid point, and for different times.
This, together with Figs. \ref{anal::ptequilevol} and \ref{anal::balpha_ptequil}, allows us to get a clearer picture:
\begin{enumerate}
\item The initial Hall equilibrium evolves in $t\sim 6 t_{Hall}$ into a non-equilibrium configuration, meaning that it is unstable to Hall drift. This can be seen from the lack of functional dependences $\beta=\beta(\alpha)$ and $F=F(\alpha)$.
\item At $t\sim 20 t_{Hall}$ the field has evolved towards a stable equilibrium that is asymmetric with respect to the equator. The toroidal field then satisfies a very tight linear relationship with $\alpha$, and $F$ also approaches a linear relationship $F(\alpha)=h\alpha$. 
\item Up to $t\sim400 t_{Hall} = 4 t_{Ohm}$ the field evolves driven by Ohmic dissipation through different stable Hall equilibria characterized by different values of $s,h$.
\item In the end, a final configuration is reached, symmetric with respect to the equator and with $h=0$ and $s\simeq 3.17L^{-1}$. This acts as an attractor under the coupled effects of Hall drift and Ohmic dissipation.
\end{enumerate}
This final attractor state can easily be obtained from Eq. (\ref{intro:equilanal}) with $\beta=s\alpha$ and $F(\alpha)=0$, as that is the same eigenvalue equation for the Ohmic eigenmodes, for which solutions are readily available (see Appendix \ref{appendix::ohmmodes}). This means that in this final state, the poloidal and toroidal fields are in their fundamental Ohmic modes, so Ohmic dissipation cannot provide a perturbation to this equilibrium, and the configuration remains steady, thus acting as an attractor. Note, however, that this final attractor state has $E_P/E=0.50$, so half of the energy is contained in the toroidal field, regardless of the value of $R_B$, which is an important difference with respect to the attractor of the previous section for which the amplitude of the toroidal field scaled as $R_B^{-1}$.

\begin{figure}
\begin{center}
\includegraphics[width=\columnwidth]{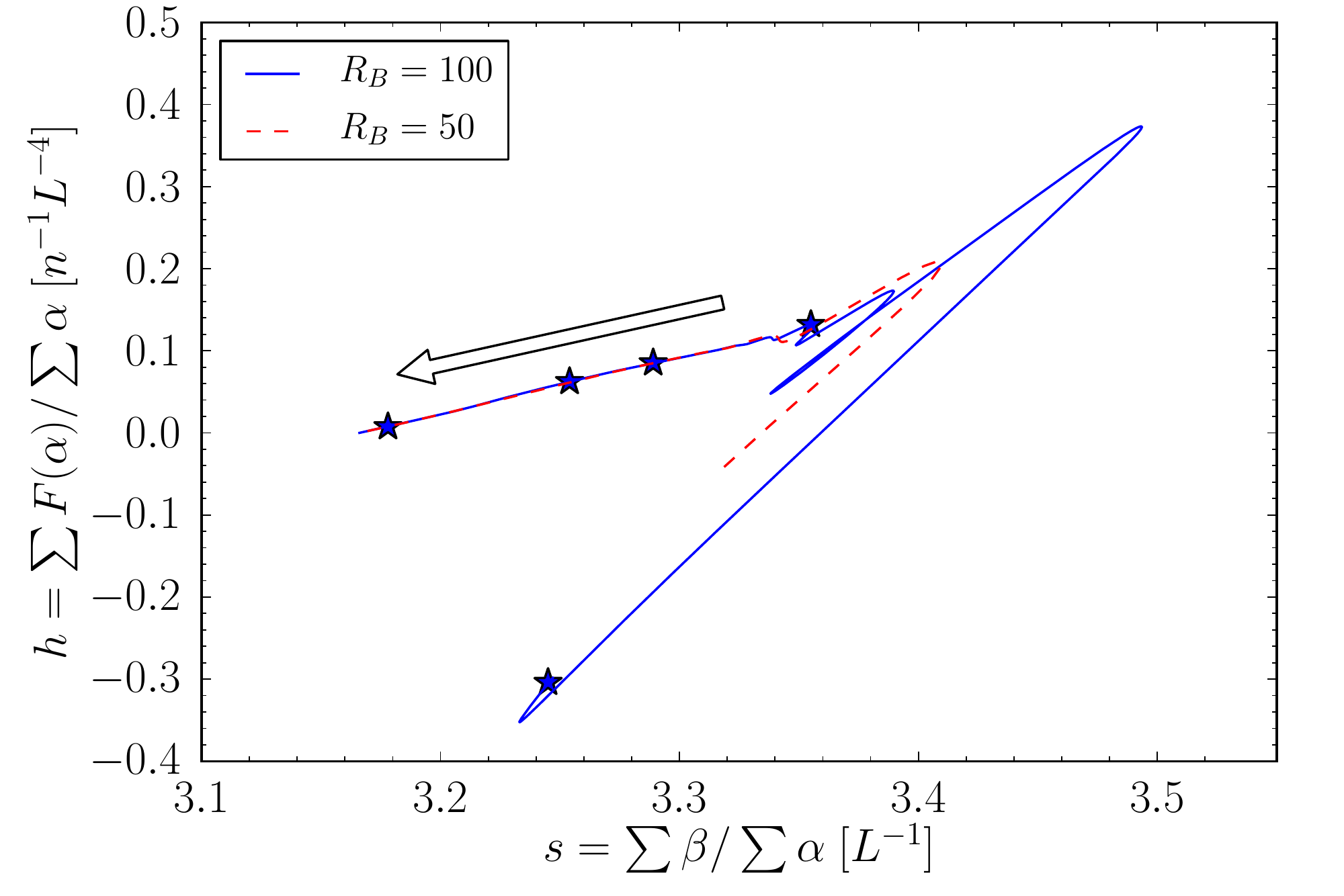}
\end{center}
\caption{Evolution of $s$ and $h$ for the poloidal+toroidal equilibrium normalized at each step to the same energy. The arrow indicates the direction of time evolution, and only data for $t>6 t_{Hall}$ are shown. Stars indicate the snapshots shown in Fig. \ref{anal::func_ptequil} for the simulation with $R_B=100$}\label{anal::ptequil_sk}
\end{figure}

The coupled evolution of $s$ and $h$, is shown in Fig. \ref{anal::ptequil_sk} for two different values of $R_B$. At the beginning of the simulation, the field is driven out of equilibrium, so the values of $s$ and $h$ do not really have a meaning, but once the field approaches an equilibrium, they start following very closely the same pairs of values, even though the conditions from which the equilibrium is reached are different due to the different choice of $R_B$.

A possible explanation of the one-to-one relation between $s$ and $h$ can be given in terms of the Grad-Shafranov equation, which, with $\beta=s\alpha$ and $F(\alpha)=h\alpha$, takes the form
\begin{eqnarray}
\Delta^*\alpha+s^2\alpha-h\alpha nr^2\sin^2\theta=0 \label{anal::pteqequfin}
\end{eqnarray}
becoming a homogenous linear partial differential equation for $\alpha$. It is to be expected that, given the appropriate boundary conditions and a fixed value of $h$, there is a discrete set of values of $s$ for which non-trivial solutions are possible (of which the lowest-order one would be chosen by the evolution of the system). As $h$ is modified, this set of allowed values of $s$ should change continuously, yielding the behaviour seen in Fig. \ref{anal::ptequil_sk}.

\section{Conclusions}\label{ch:conclusions}
Although Hall drift was proposed as a relevant mechanism for magnetic field evolution in neutron stars more than two decades ago, purely analytical studies of its effects have not yet been very conclusive. The non-linear nature of this process has made it very complicated to do so, and numerical simulations have become essential to acquire a better understanding of it.

In this work we have studied the stability of Hall equilibria in axial symmetry (2D), finding that there are both stable equilibria (such as the purely poloidal equilibrium of \S\ref{ch::equilpol}) and unstable ones (the confined equilibrium of \S\ref{ch::equilpoltor}). Even stable equilibria show an evolution, driven by Ohmic dissipation, towards an attractor state \citep[equivalent to the one identified by][]{gou+13b,gou+13c}, which is a particular case of a Hall equilibrium that retains its structure under Ohmic decay. For the unstable case, the field ends up evolving first towards a stable equilibrium, and then towards an attractor. In both cases, this attractor state is approached through a slow drift along nearby Hall equilibria, around which damped oscillations are present.

For the particular case of a field completely contained in the crust, although the evolution towards an attractor follows the same pattern, the final state is qualitatively different from the case of a field that crosses the surface of the star \citep[which is also the one described by][]{gou+13b,gou+13c}. The attractor for the confined field consists of a combination of poloidal and toroidal field components with a fixed fraction of the energy contained in each, in stark contrast with the $R_B^{-1}$ scaling the toroidal field amplitude has in the case where field lines can penetrate the surface. Whether similar attractors with non-negligible toroidal field components exist when we do not confine the field to the crust is not clear, though the simulations of \citet{gou+13b,gou+13c} seem to indicate that they do not (or at least the field is unlikely to evolve towards them).

If stable equilibrium configurations (the same or others) still exist when the assumption of axial simmetry is removed, it could have very important consequences for the observational properties of magnetars, as it was shown for the case of the purely poloidal equilibrium that Hall drift could produce only a slight enhancement in the dissipation rate of magnetic energy, which was independent of the intensity of the magnetic field. In this way, objects with similar field strengths could be very different, behaving as very active SGRs if the initial field is far away from an equilibrium configuration, or as relatively quiescent AXPs in the opposite case.

 It is however difficult to establish whether the field could be close to a Hall equilibrium when the neutron star crust solidifies, since at this time the field is expected to be an MHD equilibrium, and in general these will not be Hall equilibria \citep{gou+13a}. In particular, the formation of a purely poloidal field (akin to the equilibrium field of \S\ref{ch::equilpol}) is unlikely, as it has been extensively shown that these are MHD unstable \citep{martay+73,florud+77,braspru+06,mar+11}, and, although we found that it is possible to have stable Hall equilibria with an important toroidal component, these were completely confined to the interior of the crust. Attempts to construct Hall equilibria with field present in the exterior seem to suggest that it is only possible to find solutions with only a few percent of the total energy contained in the toroidal field \citep{lanjon+12,gou+13a}, though \citet{ciorez+13} claim otherwise. Moreover, a proper treatment of the problem should also include evolution of the field inside the core of the neutron star, which might end up being the dominat process.

\acknowledgments
This project was supported by the FONDECYT Regular Grants 1110213, 1110135, 1130273; the CONICYT International Collaboration Grant DFG-06; the research project \#655 of the university of Medell\'in; a CONICYT Scholarship for Master Studies; and the Basal Center for Astronomy and Associated Technologies (CATA) PFB-06/2007. We also thank G. R\"udiger for providing us with the code used by \citet{holrud+02,holrud+04}, and the anonymous referee for helpful comments that substantially improved the presentation of this paper.

\appendix

\section{Boundary conditions} \label{appendix::boundary}
\subsection{Conditions at the axis ($\theta=0,\pi$)}
The only requirement at the axis is that the magnetic field is a single-valued vector field. In axial symmetry, this implies that $B_{\phi}=B_{\theta}=0$, yielding
\begin{eqnarray}
\beta(\theta=0,\pi)=0,\quad \frac{\pp\alpha}{\pp r}(\theta=0,\pi)=0,
\end{eqnarray}
which means $\alpha$ is constant along the axis. Since $\alpha$ has an arbitrary ``zero-point'', we choose $\alpha=0$ at the axis, which also makes $\alpha\to 0$ as $r\to\infty$.

\subsection{Meissner boundary conditions $(r=r_{min})$}

At the interface  between the solid crust and the fluid core, we require the normal component of the magnetic field and the tangential component of the electric field to be continuous, that is,
\begin{eqnarray}
B_r|_{in}=B_r|_{out},\quad E_{\theta}|_{in}=E_{\theta}|_{out},\quad E_{\phi}|_{in}=E_{\phi}|_{out}.
\end{eqnarray}
Due to the (assumed) Meissner effect, the magnetic and electric fields in the core are zero, and thus these conditions are simply
\begin{eqnarray}
B_r(r=r_{min})=0,\quad E_{\theta}(r=r_{min})=E_{\phi}(r=r_{min})=0.
\end{eqnarray}
These conditions allow for surface charges and currents, which can make $E_r$, $B_\theta$, and $B_\phi$ non-zero at the boundary. The easiest of these three to apply is the first one. Since $B_r\propto \pp\alpha/\pp\theta$, it implies that $\alpha$ is constant along the boundary, and since we already fixed $\alpha=0$ on the symmetry axis we must have $\alpha=0$ at  $r=r_{min}$ also.

The condition on the electric field produces a much more complex boundary condition. In terms of the magnetic field, the electric field in the crust is given by
\begin{eqnarray}
-c\boldsymbol{E}=\frac{c}{4\pi n e}(\nabla\times\boldsymbol{B})\times\boldsymbol{B}+\eta\nabla\times\boldsymbol{B}
\end{eqnarray}
so requiring the tangential component of the electric field to be zero at the boundary is equivalent to
\begin{eqnarray}
0=\left[\varpi^2\chi(\nabla\times\boldsymbol{B})\times\boldsymbol{B}+\eta R_B^{-1}\nabla\times\boldsymbol{B}\right]_\parallel
\end{eqnarray}
at $r=r_{min}$, where $\parallel$ denotes the tangential component. In terms of $\alpha$ and $\beta$, this produces the following two (non-linear) boundary conditions in spherical coordinates
\begin{eqnarray}
\begin{aligned}
0=\left(\chi\beta\frac{\pp\beta}{\pp\theta}+R_B^{-1}\frac{\eta}{\sin\theta}\frac{\pp\beta}{\pp r}\right)_{r=r_{min}},\qquad
0=\left(\sin\theta\chi\frac{\pp\beta}{\pp\theta}\frac{\pp\alpha}{\pp r}+R_B^{-1}\eta\Delta^*\alpha\right)_{r=r_{min}},
\end{aligned}\label{intro::scbound}
\end{eqnarray}
where it has been explicitly used that $\alpha(r_{min},\theta)=0$. These ``Meissner boundary conditions'' have been written in a form that closely resembles the terms in Eqs. (\ref{intro::eqssph2}), allowing us to see that the time derivative of $\alpha$ is zero at the inner boundary, and that no toroidal magnetic flux is lost to the core of the star.

\subsection{Zero boundary conditions}
These conditions correspond to taking the limit $R_B\rightarrow\infty$ for Eq. (\ref{intro::scbound}), namely
\begin{eqnarray}
\begin{aligned}
0=\left(\chi\beta\frac{\pp\beta}{\pp\theta}\right)_{r=r_{min}},\qquad
0=\left(\sin\theta\chi\frac{\pp\beta}{\pp\theta}\frac{\pp\alpha}{\pp r}\right)_{r=r_{min}}.
\end{aligned}
\end{eqnarray}
In order for these to be satisfied, together with the condition at the axis, we must have $\beta(r_{min},\theta)=0$, which together with $\alpha(r_{min},\theta)=0$ form the boundary conditions. Note that these conditions are also used at the surface of the star in \S\ref{ch::equilpoltor}.

\subsection{Matching an external vacuum field $(r=R)$}\label{intro::condoutside}
If outside the star we consider a perfect vacuum, then the magnetic field there is completely determined by its radial component at the surface of the star, which must be continuous. Furthermore, we expect surface currents to dissipate on timescales much smaller than those of interest to us, so not only the radial component of the magnetic field must be continuous, but also the tangential one.

The condition imposed on $\beta$ because of this is trivial. Since there are no currents outside the star, we must have $\beta=0$ there, and the continuity of the azimuthal component of the field immediately gives $\beta=0$ as a boundary condition at $r=R$.

The condition on $\alpha$ is much more complex, as it is non-local. As shown in \citet{mar+11}, the continuity of the radial component of the magnetic field implies that the field outside the star is given by
\begin{eqnarray}
\begin{aligned}
\boldsymbol{B}=\nabla\Psi,\qquad
\Psi(r,\theta,\phi)=\sum_{l=1}^{\infty}\sum_{m=-l}^{l}\frac{a_{lm}}{r^{l+1}}Y_{lm}(\theta,\phi),\qquad a_{lm}=-\frac{R^{l+2}}{l+1}\int_{4\pi}(B_r)_{r=R}Y_{lm}^*\dd\Omega,\label{intro::Boutexp}
\end{aligned}\label{intro::al}
\end{eqnarray}
where $(B_r)_{r=R}$ is the radial field $\boldsymbol{B}\cdot\bhat{r}$ at the surface of the star, which can be expressed in terms of $\alpha$ as $([r^2\sin\theta]^{-1}\pp\alpha/\pp\theta)_{r=R}$. For axially symmetric fields these coefficients can be written in terms of $\alpha$ as (omitting the unnecessary $m=0$ subscript)
\begin{eqnarray}
a_l=\frac{R^l}{l+1}\sqrt{\pi(2l+1)}\int_0^{\pi}P_l^1(\cos\theta)\alpha(R,\theta)\dd\theta. \label{intro::alint}
\end{eqnarray}
This selection of the $a_l$ coefficients will give a continuous radial component of the magnetic field, but so far we have not imposed continuity on its $\theta$ component. Equating the value of $B_{\theta}$ just inside the star, as given by $\alpha$, and just outside the star, as given by the combination of the $a_l$, we have
\begin{eqnarray}
\begin{aligned}
\left(B_\theta\right)_{r=R}=\sum_{l=1}^\infty\frac{a_l}{R^{l+2}}\sqrt{\frac{2l+1}{4\pi}}\frac{\pp}{\pp\theta}P_l(\cos\theta)=-\frac{1}{R\sin\theta}\left[\frac{\pp\alpha}{\pp r}\right]_{r=R}.
\end{aligned}
\end{eqnarray}
Solving for $\pp\alpha/\pp r$ at the surface, we have the boundary condition required for the continuity of the $\theta$ component of the field, namely,
\begin{eqnarray}
\left[\frac{\pp\alpha}{\pp r}\right]_{r=R}=-\sum_{l=1}^\infty\frac{a_l}{R^{l+1}}\sqrt{\frac{2l+1}{4\pi}}\sin\theta P_l^1(\cos\theta),\label{intro::bcsurface}
\end{eqnarray}
with $a_l$ given by Eq. (\ref{intro::alint}). We cannot evaluate these infinite summations numerically, so we cut them at a maximum multipole $L$ defined at the beginning of each simulation.

\section{Numerical implementation}\label{appendix::num}
We compute Eqs. (\ref{intro::eqs}) in spherical coordinates, i.e.,
\begin{eqnarray}
\frac{\pp\alpha}{\pp t}&=&\sin\theta\chi\left(\frac{\pp\beta}{\pp\theta}\frac{\pp \alpha}{\pp r}-\frac{\pp\beta}{\pp r}\frac{\pp \alpha}{\pp \theta}\right)+R_B^{-1}\eta\Delta^*\alpha,\label{intro::eqssph_alpha}\\
\frac{\pp}{\pp t}\left(\frac{\beta}{\sin\theta}\right)&=&-\frac{\pp}{\pp r} F_r-\frac{\pp}{\pp\theta}F_\theta,\label{intro::eqssph_beta}
\end{eqnarray}
where
\begin{eqnarray}
\begin{aligned}
F_r=-\chi\Delta^*\alpha\frac{\pp\alpha}{\pp\theta}-\chi\beta\frac{\pp \beta}{\pp\theta}-R_B^{-1}\frac{\eta}{\sin\theta}\frac{\pp\beta}{\pp r},
\qquad
F_\theta=\chi\Delta^*\alpha\frac{\pp\alpha}{\pp r}+\chi\beta\frac{\pp\beta}{\pp r}-R_B^{-1}\frac{\eta}{r^2\sin\theta}\frac{\pp\beta}{\pp\theta}.
\end{aligned}\label{intro::eqssph2}
\end{eqnarray}

Note that, for completeness, in this section we do not assume $n$ and $\eta$ to be constants. The functions $\alpha$ and $\beta$ are discretized on a regular spherical grid with $N_{\theta}$ points in the $\theta$ direction (including the axis), and $N_{r}$ points in the radial direction (including the surface and the inner boundary), plus a point just outside the surface and one just below the inner boundary. These two additional points are required to set boundary conditions on the derivatives.

To describe the discretized values of the functions, we use the notation $\alpha_{i,j}^k$, where $i$ and $j$ denote the grid points in the radial and angular directions respectively (with $i$ going from $0$ to $N_r+1$ and $j$ going from $0$ to $N_{\theta}-1$), and $k$ denotes the timestep. The numerical method used for the temporal discretization of the system of Eqs. (\ref{intro::eqssph_alpha},\ref{intro::eqssph_beta}) is simply of the form
\begin{eqnarray}
\alpha_{i,j}^{k+1}=\alpha_{i,j}^{k}+\Delta t\left(\frac{\pp\alpha}{\pp t}\right)_{i,j}^k,\;\beta_{i,j}^{k+1}=\beta_{i,j}^{k}+\Delta t\left(\frac{\pp\beta}{\pp t}\right)_{i,j}^k
\end{eqnarray}
However, the calculation of the spatial derivatives is done in fundamentally different ways for the $\alpha$ and $\beta$ equations. For Eq. \ref{intro::eqssph_alpha} we simply use the usual three-point stencil approximation to all the spatial derivatives involved, namely,
\begin{eqnarray}
\begin{aligned}
\left(\frac{\pp\alpha}{\pp t}\right)_{i,j}^k=&\frac{\sin\theta_j\chi_{i,j}}{4\Delta r\Delta\theta}\left[(\beta_{i,j+1}^k-\beta_{i,j-1}^k)(\alpha_{i+1,j}^k-\alpha_{i-1,j}^k)\right.\\
&\qquad\left.-(\beta_{i+1,j}^k-\beta_{i-1,j}^k)(\alpha_{i,j+1}^k-\alpha_{i,j-1}^k)\right]+R_B^{-1}\eta_{i,j}\left(\Delta^*\alpha\right)_{i,j}^{k},\label{num::alphaeq}
\end{aligned}
\end{eqnarray}
where $\theta_j=j\Delta\theta$, and $\left(\Delta^*\alpha\right)_{i,j}^{k}$ is computed as
\begin{eqnarray}
\begin{aligned}
\left(\Delta^*\alpha\right)_{i,j}^{k}=&\frac{\alpha_{i+1,j}^k+\alpha_{i-1,j}^k-2\alpha_{i,j}^k}{\left(\Delta r\right)^2}+\frac{\alpha_{i,j+1}^k+\alpha_{i,j-1}^k-2\alpha_{i,j}^k}{(r_j)^2\left(\Delta \theta\right)^2}-\cot\theta_j\frac{\alpha_{i,j+1}^k-\alpha_{i,j-1}^k}{2(r_j)^2\Delta \theta},\label{num::gsa}
\end{aligned}
\end{eqnarray}
and $r_j=r_{min}+(i-1)\Delta r$. For the Eq. \ref{intro::eqssph_beta} we take advantage of its flux-conservative properties by using the discretization
\begin{eqnarray}
\begin{aligned}
\left(\frac{\pp\beta}{\pp t}\right)_{i,j}^k=&
-\frac{\sin\theta_j}{\Delta r}\left(\left[F_r\right]_{i+1/2,j}^k-\left[F_r\right]_{i-1/2,j}^k\right)-\frac{\sin\theta_j}{\Delta\theta}\left(\left[F_\theta\right]_{i,j+1/2}^k-\left[F_\theta\right]_{i,j-1/2}^k\right),
\end{aligned}\label{num::betaeq}
\end{eqnarray}
where $(i\pm1/2,j)$ and $(i,j\pm1/2)$ denote edge-centered values of the grid, obtained for example as
\begin{eqnarray}
\begin{aligned}
\left[\frac{\eta}{\sin\theta}\frac{\pp\beta}{\pp r}\right]_{i+ 1/2,j}^k=
\frac{\eta_{i+ 1/2,j}}{\sin\theta_j}\left(\frac{\beta_{i+1,j}^k-\beta_{i,j}^k}{\Delta r}\right),
\end{aligned}
\end{eqnarray}
in which the remaining edge-centered value $\eta_{i+ 1/2,j}$ can be solved exactly. The main feature of the discretization given by Eq. (\ref{num::betaeq}) is that the time derivative of the toroidal magnetic flux
\begin{eqnarray}
\Phi_{tor}=\sum_{i=1}^{N_{r}}\sum_{j=1}^{N_{\theta}-1}\frac{\beta_{i,j}^k}{\sin\theta_j}\Delta r\Delta \theta
\end{eqnarray}
will depend only on boundary values of the functions and its derivatives. This embodies the flux-conserving property of the equation for $\beta$.

For the evolution of $\beta$, we use Eq. (\ref{num::betaeq}) for the grid points with $2\le i \le N_{r}-1$ and $1\le j \le N_{\theta}-1$. For the case of $\alpha$, Eq. (\ref{num::alphaeq}) is used to solve the evolution for the grid points with $2\le i \le N_{r}$ and $1\le j \le N_{\theta}-1$ when using zero boundary conditions, and for the grid points with $1\le i \le N_{r}$ and $1\le j \le N_{\theta}-1$ when using Meissner boundary conditions.

To time-evolve $\beta_{1,j}^k$ for the Meissner boundary conditions, we use a modified form of Eq. (\ref{num::betaeq}),
\begin{eqnarray}
\begin{aligned}
\left(\frac{\pp\beta}{\pp t}\right)_{1,j}^k=&
-\frac{\sin\theta_j}{\Delta r/2}\left(\left[F_r\right]_{3/2,j}^k-\left[F_r\right]_{1,j}^k\right)
-\frac{\sin\theta_j}{\Delta\theta}\left(\left[F_\theta\right]_{1,j+1/2}^k-\left[F_\theta\right]_{1,j-1/2}^k\right).
\end{aligned}
\end{eqnarray}
Noting that $\alpha_{1,j}^k=0$ together with the first of the Meissner boundary conditions (\ref{intro::bcsurface}) implies $\left[F_r\right]_{1,j}^k=0$ , we have that
\begin{eqnarray}
\begin{aligned}
\left(\frac{\pp\beta}{\pp t}\right)_{1,j}^k=&
-\frac{\sin\theta_j}{\Delta r/2}\left[F_r\right]_{3/2,j}^k
-\frac{\sin\theta_j}{\Delta\theta}\left(\left[F_\theta\right]_{1,j+1/2}^k-\left[F_\theta\right]_{1,j-1/2}^k\right).
\end{aligned}
\end{eqnarray}

Also, when evolving the equilibrium field of \S \ref{ch::equilpol}, a slight numerical problem arose at points right below the surface, where a small error in the resolution of $\Delta^{*}\alpha$ produced numerical artifacts in the evolution of the toroidal field function. In order to overcome these problems, for this particular case we solve the toroidal field function just below the surface of the star by interpolation,
\begin{eqnarray}
\beta_{N_r-1,j}^k=\frac{1}{2}\left(\beta_{N_r-2,j}^k+\beta_{N_{r},j}^k\right)=\frac{1}{2}\beta_{N_r-2,j}^k,
\end{eqnarray}
where the zero boundary condition at the surface was used. Since in these poloidally dominated cases the toroidal field should not produce steep current sheets at the surface, this approximation should not significantly alter the evolution.

\subsection{Implementation of boundary conditions} \label{appendix::imp_boundary}

In this section, we describe how the boundary conditions shown in Appendix \ref{appendix::boundary} are implemented in this finite-difference scheme. The condition at the axis is trivial, but at other points more care is required.
\subsubsection{Matching an external vacuum field}
For the surface of the star, we have the boundary condition given by Eq. (\ref{intro::bcsurface}), which can be expressed in finite-difference form as
\begin{eqnarray}
\begin{aligned}
\frac{\alpha_{N_{r}+1,j}^k-\alpha_{N_{r}-1,j}^k}{2\Delta r}=
-\sum_{l=1}^L \frac{a_l}{R^{l+1}}\sqrt{\frac{2l+1}{4\pi}}\sin\theta_j P_l^1(\cos\theta_j).
\end{aligned}
\end{eqnarray}
This expression gives the value of $\alpha_{N_{r}+1,j}^k$ at each timestep. The sum must be limited to a finite number of multipoles $L$, and the $a_l$ are given by Eq. (\ref{intro::alint}), but with the integral computed as,
\begin{eqnarray}
\begin{aligned}
a_l=\frac{R^l}{l+1}\sqrt{\pi(2l+1)}
\sum_{j=0}^{N_{\theta}-1}\frac{(\alpha_{N_{r},j}^k+\alpha_{N_{r},j+1}^k)}{2} P_l^1(\cos\theta_{j+1/2})\Delta\theta.
\end{aligned}
\end{eqnarray}
For $\beta$, right at the surface the continuity of the azimuthal component of the field requires $\beta_{N_r,j}^k=0$. The value of $\beta$ just outside the surface, $\beta_{N_r+1,j}$, is only required to solve $\pp\beta/\pp r$ right at the surface. However, this cannot be evaluated as $\beta_{N_r+1,j}=0$ (no toroidal flux outside the star), because the derivative of $\beta$ needs not be continuous at the surface. Instead, we set $\beta_{N_r+1,j}^k$ in such a way that the radial derivative of $\beta$ computed at the surface is the same as computing it backwards, i.e.
\begin{eqnarray}
\frac{\beta_{N_r+1,j}^k-\beta_{N_r-1,j}^k}{2\Delta r}=\frac{\beta_{N_r,j}^k-\beta_{N_r-1,j}^k}{\Delta r}.
\end{eqnarray}
\subsubsection{Zero boundary conditions}
The zero boundary conditions are simple, except that an extra one is needed at the crust-core interface to set the value of $\alpha$ right below the surface. This is done by noting that, since $\pp \alpha/\pp t=0$ at the inner boundary (its value is fixed to zero), Eq. \ref{intro::eqssph_alpha} implies that $\Delta^*\alpha=0$ there, which reduces to
\begin{eqnarray}
\alpha_{0,j}=-\alpha_{2,j},
\end{eqnarray}
and similarly when it is applied at the outer boundary.
\subsubsection{Meissner boundary conditions}
To implement the Meissner boundary conditions given by Eqs. (\ref{intro::scbound}), we solve for $\alpha_{0,j}^k$ and $\beta_{0,j}^k$ from the discretized versions of these, which results in
\begin{eqnarray}
\begin{aligned}
\beta_{0,j}^{k}=&\D\beta_{2,j}^{k}+\frac{R_B}{n_{i,j}\eta_{i,j}r_{min}^2\sin\theta_j}\frac{\Delta r}{\Delta \theta}\left(\beta_{1,j+1}^k-\beta_{1,j-1}^k\right)\\
\alpha_{0,j}^{k}=&\D \frac{\D(\beta_{1,j+1}^{k}-\beta_{1,j-1}^{k})\Delta r+4 n_{i,j}\eta_{i,j}R_B^{-1}r_{min}^2\sin\theta_j\Delta\theta}{\D(\beta_{1,j+1}^{k}-\beta_{1,j-1}^{k})\Delta r-4 n_{i,j}\eta_{i,j} R_B^{-1}r_{min}^2\sin\theta_j\Delta\theta}\alpha_{2,j}^{k}.
\end{aligned}
\end{eqnarray}
Unfortunately, the denominator in this last expression can get very close to zero, causing numerical problems for simulations with large $R_B$.

\subsection{Variable time step}\label{appendix::timestep}
Usually it will be the case that some time intervals in the simulation will require a much smaller $\Delta t$ to converge without producing numerical instabilities. Since Hall drift will advect field lines with the electron velocity $\boldsymbol{v}_e=-\boldsymbol{j}/(ne)$, we have a Courant condition of the form
\begin{eqnarray}
\frac{|\boldsymbol{v}_e|\Delta t}{\Delta l}<k_c,
\end{eqnarray}
where $k_c<1$ is chosen at the beginning of each simulation and $\Delta l$ is the smallest dimension of any grid cell. Using that $\boldsymbol{j}=c(\nabla\times \boldsymbol{B})/(4\pi)$, the timestep is solved as
\begin{eqnarray}
\Delta t=k_c\frac{4\pi ne}{c}\frac{\Delta l}{|\nabla\times\boldsymbol{B}|},
\end{eqnarray}
which is exactly the same condition used by \citet{vig+12}. In the thin crust, unless the number of grid points in the radial direction is much larger than the number of grid points in the $\theta$ direction, the smallest dimension of a grid cell will always be $\Delta r$. Because of this we assume in our implementation $\Delta l = \Delta r$.

At each step of the simulation, this is evaluated at all points in the grid, and the smallest value is chosen. In order to avoid having the timestep increase indefinitely as Ohmic diffusion becomes dominant, we also define a dissipative timestep,
\begin{eqnarray}
\Delta t=k_c\frac{(\Delta r)^2}{\eta},
\end{eqnarray}
and the smallest of the two values is used at each step of the simulation.

Although in principle we only need $k_c<1$, in the simulations performed for this work numerical problems arose when $k_c$ was chosen close to unity, sometimes requiring values as small as $k_c=0.01$ for the simulations to converge. The need for such small timesteps should be related to the use of purely explicit (and first order in time) methods for the time evolution, and the explicit third order
spatial derivatives that appear in the equations.

\section{Ohmic modes}
\label{appendix::ohmmodes}
If we only consider the effect of Ohmic decay in Eqs. (\ref{intro::eqs}) the resulting differential equations are linear, namely
\begin{eqnarray}
\frac{\pp \alpha}{\pp t}=\eta\Delta^*\alpha,\quad \frac{\pp \beta}{\pp t}=\varpi^2\nabla\cdot\left(\frac{\eta\nabla\beta}{\varpi^2}\right),
\label{appendix::equohm}
\end{eqnarray}
which is a well studied problem for the case where currents are present inside a sphere \citep{lam+83,cow+45,chagab+72,cum+02}. Exponentially decaying solutions can be easily computed for the case of $\eta=\eta(r)$ and currents contained in a spherical shell \citep{cum+04}. For the simple case of constant resistivity the right hand side of Eq. \ref{appendix::equohm} for $\beta$ reduces to $\eta\Delta^*\beta$, and analytical solutions for these Ohmic modes are readily available,
\begin{eqnarray}
\alpha(r,\theta,t)=r\left[Aj_l\left(kr\right)+By_l\left(kr\right)\right]P_l^1(\cos\theta)\sin\theta e^{-t/\tau},\\
\beta(r,\theta,t)=r\left[Cj_l\left(kr\right)+Dy_l\left(kr\right)\right]P_l^1(\cos\theta)\sin\theta e^{-t/\tau},
\end{eqnarray}
where $j_l$ and $y_l$ are the spherical Bessel functions of order $l$. Also, $\tau=1/(\eta k^2)$, and the values of $k$ and the ratios $A/B$ and $C/D$ depend on the boundary conditions used for $\alpha$ and $\beta$. A few values of $\tau$ for different modes are shown in Table \ref{appendix::Ohmtable} for both zero and Meissner boundary conditions at the crust-core interface, and matching to an external vacuum at the surface. Note that for the case of a field confined in the star as in \S \ref{ch::equilpoltor} the solutions satisfy $\alpha\propto\beta$, and are equivalent to the ones shown in Table \ref{appendix::Ohmtable} for the toroidal field with zero boundary conditions.

\begin{table}
	\begin{center}
		\begin{tabular}{cc|rr|rr|rr}
			$n$&$l$&$\tau_P$&$\tau_{T,zbc}$&$\tau_{T,Mbc}$\\
			\hline
			1&1&  0.323 & 0.0995 & 0.376\\
			2&1& 0.0438 & 0.0252 & 0.0447\\
			1&2&  0.260 & 0.0964 & 0.330\\
			2&2& 0.0422 & 0.0250 & 0.0440
		\end{tabular}
	\end{center}
	\caption{First two radial modes for each of $l=1,2$, with $r_{min}=0.75R$. The $\tau_P$ are the values associated to the poloidal modes both for the cases of zero and Meissner boundary conditions, while $\tau_{T,zbc}$ and $\tau_{T,Mbc}$ are the values for the toroidal modes in the case of zero boundary conditions (for the subscript $zbc$) and Meissner boundary conditions (for the subscript $Mbc$). Values are given in units of $t_{Ohm}=L^2/\eta$}\label{appendix::Ohmtable}
\end{table}

\section{Comparison with the spectral code of Hollerbach (2000)}\label{appendix::holcom}
The most immediate way to test the validity of a code is to compare it against analytically known solutions. These are available for the case of pure Ohmic dissipation (see Appendix \ref{appendix::ohmmodes}), and our code properly reproduces these Ohmic eigenmodes. The same cannot be done for the Hall drift term; although solutions describing wave phenomena can be obtained for a constant background field \citep{golrei+92}, these are incompatible with our boundary conditions.

On the other hand, we had access to the spectral code developed by \citet{hol+00}, which was used in the simulations of \citet{holrud+02, holrud+04}. We used this code to run 4 test cases in order to compare with our results. These are particularly useful for a comparison since this code uses a spectral decomposition, a method radically different from our finite-difference approach. All tests were made using zero boundary conditions at the crust-core interface, and matching to an external vacuum field with no surface currents at the surface of the star.

\begin{figure}
\begin{center}
\begin{minipage}{0.5\columnwidth}
\includegraphics[width=\columnwidth]{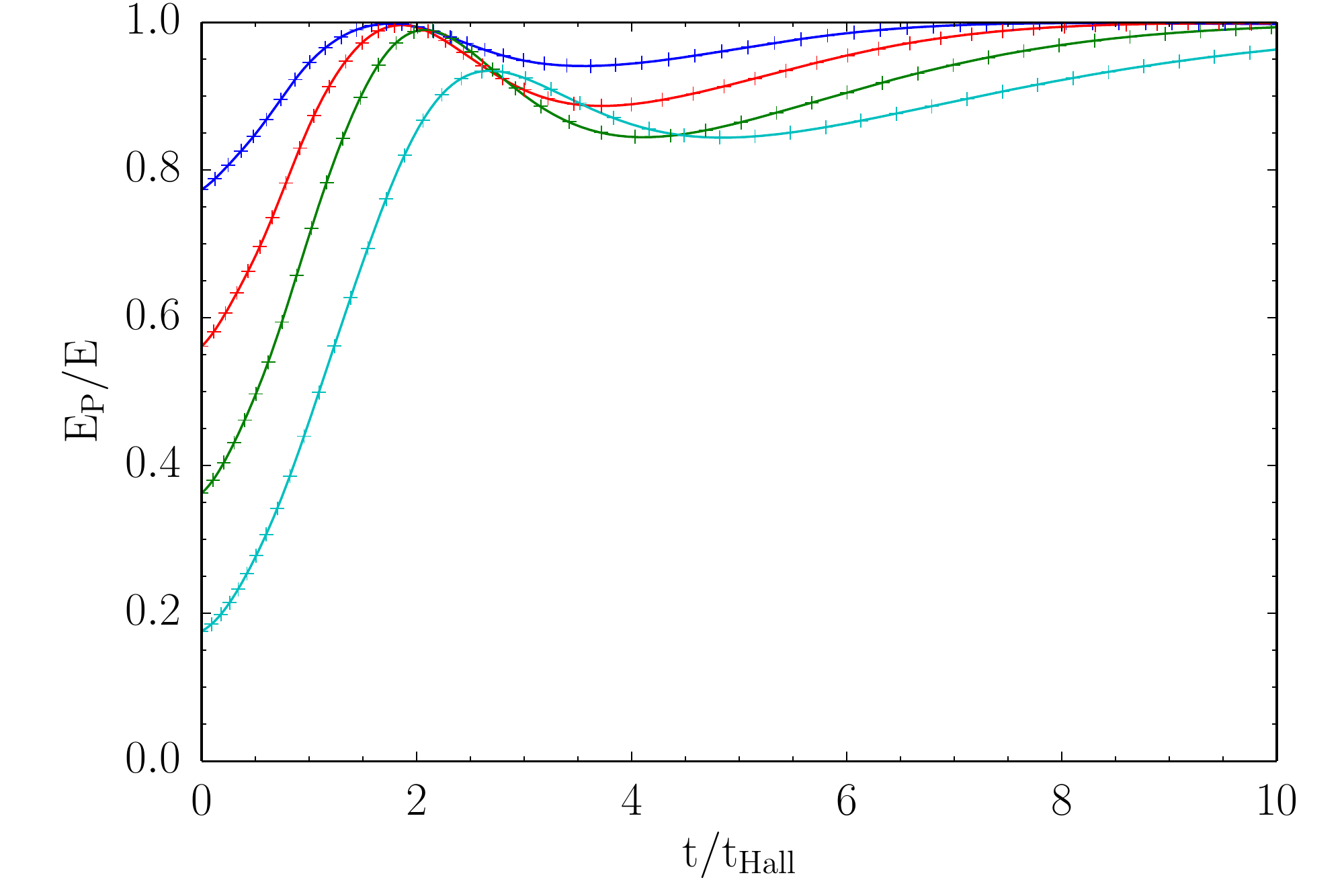}
\end{minipage}
\begin{minipage}{0.45\columnwidth}
\includegraphics[scale=0.3]{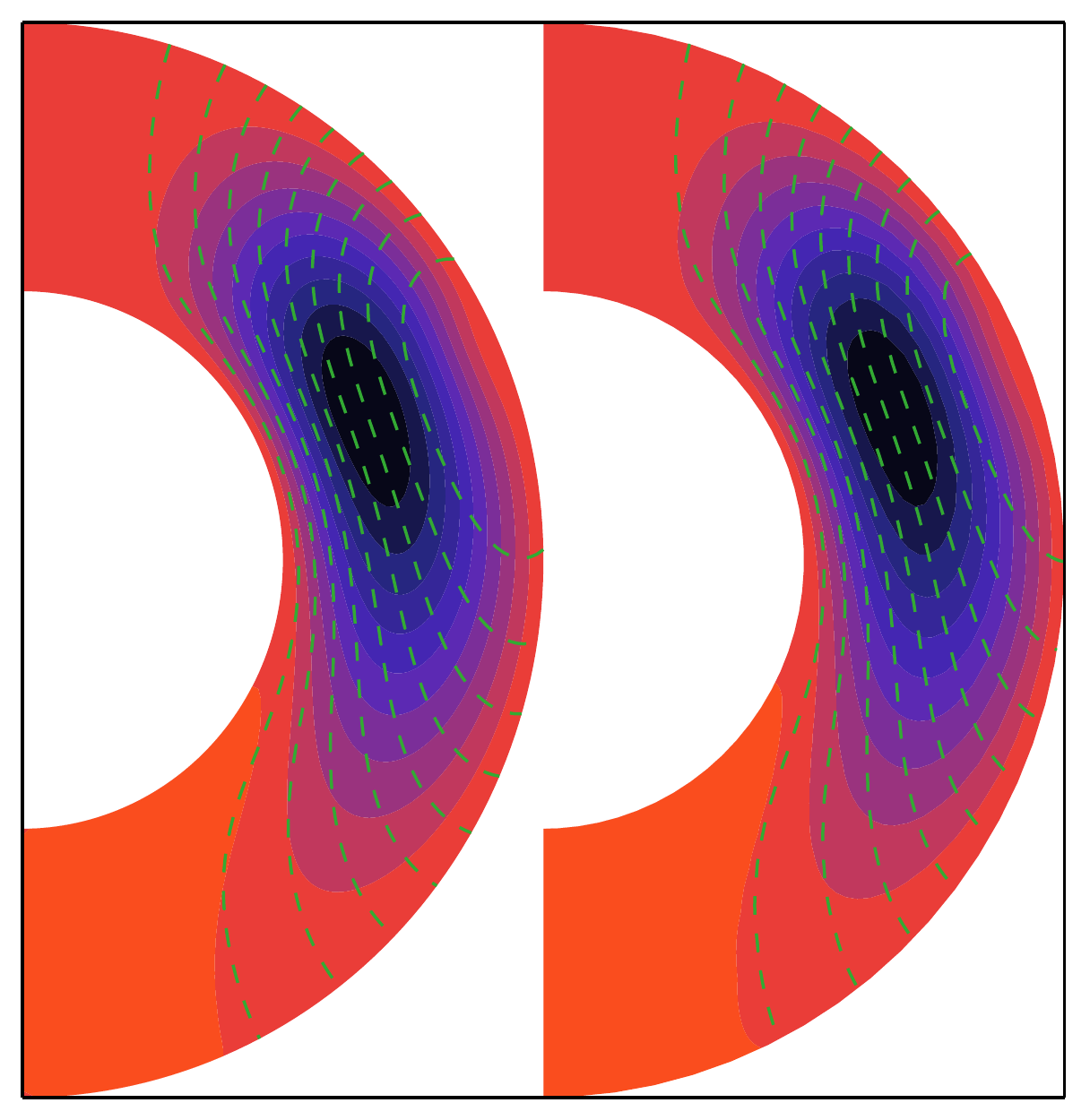}
\includegraphics[scale=0.3]{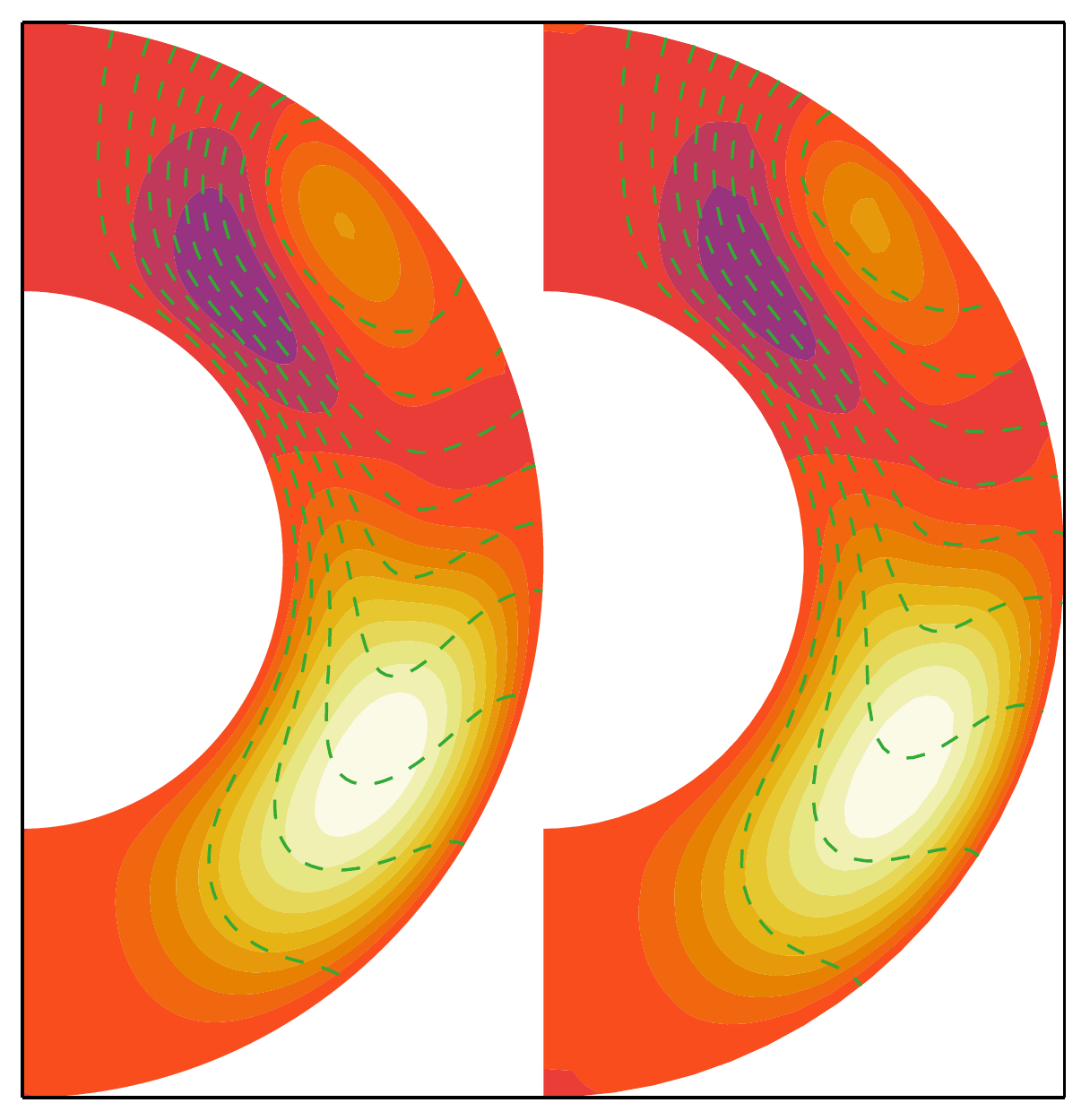}
\includegraphics[scale=0.8]{colorbar}
\end{minipage}
\end{center}
\caption{(left) Evolution of the ratio of poloidal to total magnetic energy $E_{P}/E$ as a function of time for the 4 test cases of Appendix \ref{appendix::holcom}, which are combinations of the fundamental poloidal and toroidal Ohm modes, with $R_B=25$. Solid lines are results given by Hollerbach's spectral code while crosses show the results of the finite-difference code developed for this work. (right) Snapshots of the evolution of the test case with lowest and highest initial $E_P/E$ ratio (left and right frames respectively). For both frames, the plot on the left is done with our finite-difference code while the one on the right is done with Hollerbach's spectral code. All plots correspond to $t/t_{Hall}=2.40$.}\label{num::hollplot}
\end{figure}

The resolution used when running these simulations with the spectral code was an expansion in terms of $20$ radial and $20$ latitudinal modes, which we compare against the same resolution used in our simulations, i.e. $40$ radial and $120$ latitudinal grid points with $L=24$ for the multipole expansion outside the star. Initial conditions were chosen using combinations of the fundamental poloidal and toroidal modes with $R_B=25$. The choice of a low resolution for the runs with the spectral code and a small value of $R_B$ where due to the long times required for each simulation with Hollerbach's code. Fig. \ref{num::hollplot} shows the evolution of the ratio of poloidal to total magnetic energy $E_P/E$, together with comparative snapshots of the structure of the field using both codes.

It can be seen that both the structure and the energetics are remarkably consistent between both codes. Since the implementation of the spectral code is significantly different from ours, the positive outcome of this comparison is indicative that our implementation is correct at least for the case of zero boundary conditions.

\end{document}